\documentclass[submission,Phys]{SciPost}

\usepackage[utf8]{inputenc} 
\usepackage[T1]{fontenc} 	
\usepackage[english]{babel} 

\usepackage[bitstream-charter]{mathdesign}
\usepackage{geometry} 		
\usepackage{amsmath} 		
\usepackage{amsthm} 		
\usepackage{mathtools} 		
\usepackage{float} 			
\usepackage{graphicx} 		
\usepackage{tabularx} 		
\usepackage{booktabs} 		
\usepackage{color, xcolor} 	
\usepackage{pdfpages} 		
\usepackage{extarrows} 		
\usepackage{multirow} 		
\usepackage{multicol} 		
\usepackage{caption} 		
\usepackage{subcaption} 	
\usepackage{enumitem} 		
\usepackage{setspace} 		
\usepackage{xspace} 		
\usepackage{ragged2e} 		
\usepackage{stackrel} 		
\usepackage{tikz} 			
\usepackage{braket} 		
\usepackage{bm} 			
\usepackage{tensor} 		
\usepackage{slashed} 		
\usepackage{lastpage} 		
\usepackage{cite} 			
\usepackage[normalem]{ulem} 
\usepackage{fontawesome} 	
\usepackage{tocloft} 		
\usepackage{titlesec} 		
\usepackage{doi} 			
\usepackage{hyperref} 		
\usepackage[most]{tcolorbox} 					
\usepackage[nameinlink, capitalize]{cleveref} 	
\usepackage[nottoc, notlot, notlof]{tocbibind} 	
\usepackage[ruled, vlined]{algorithm2e} 		
\usepackage{pifont}

\DeclareSymbolFont{usualmathcal}{OMS}{cmsy}{m}{n}
\DeclareSymbolFontAlphabet{\mathcal}{usualmathcal}

\SetArgSty{textnormal}
\SetKwComment{Comment}{{\small\#}~}{}
\SetCommentSty{mycommfont}

\setitemize{itemsep=0pt, parsep=0pt} 				
\setenumerate{itemsep=0pt, parsep=0pt} 				
\setitemize{itemsep=2pt,topsep=2pt,parsep=0pt,partopsep=0pt,leftmargin=*}
\setenumerate{itemsep=0pt,topsep=2pt,parsep=0pt,partopsep=0pt,labelindent=3pt,leftmargin=*}
\newcommand{\ie}{\textsl{i.e.}\;}

\newcommand{\qqquad}{\qquad\quad}
\newcommand{\qqqquad}{\qquad\qquad}

\newcommand\one{\leavevmode\hbox{\small1\normalsize\kern-.33em1}}
\newcommand{\tr}{\operatorname{Tr}}			

\newcommand{\ope}{\mathcal{O}}

\newcommand{\Dfb}{\mbox{$\raisebox{2mm}{\boldmath ${}^\leftrightarrow$}\hspace{-4mm} D$}}
\newcommand{\Dfba}{\mbox{$\raisebox{2mm}{\boldmath ${}^\leftrightarrow$}\hspace{-4mm} D^a$}}

\newcommand{\like}{\mathcal{L}} 	
\newcommand{\lag}{\mathscr{L}} 	

\newcommand{\arXiv}[2][]{%
	\ifthenelse{\equal{#1}{}}%
	{\href{http://arxiv.org/abs/#2}{arXiv:#2}}%
	{\href{http://arxiv.org/abs/#2}{arXiv:#2~[#1]}}}

\def\slashchar#1{\setbox0=\hbox{$#1$}           
   \dimen0=\wd0                                 
   \setbox1=\hbox{/} \dimen1=\wd1               
   \ifdim\dimen0>\dimen1                        
      \rlap{\hbox to \dimen0{\hfil/\hfil}}      
      #1                                        
   \else                                        
      \rlap{\hbox to \dimen1{\hfil$#1$\hfil}}   
      /                                         
   \fi}

\usepackage{units}
\usepackage{placeins}
\graphicspath{{./figs/}}                

\begin{document}

\begin{center}{\Large \textbf{
To Profile or To Marginalize -- A SMEFT Case Study
}}\end{center}

\begin{center}
  Ilaria Brivio\textsuperscript{1,2},
  Sebastian Bruggisser\textsuperscript{1,3},
  Nina Elmer\textsuperscript{1},
  Emma Geoffray\textsuperscript{1},
  Michel Luchmann\textsuperscript{1}, and 
  Tilman Plehn\textsuperscript{1}
\end{center}

\begin{center}
  {\bf 1} Institut f\"ur Theoretische Physik, Universit\"at Heidelberg, Germany\\
  {\bf 2} Physik Institut, University of Zurich, Switzerland \\
  {\bf 3} Department of Physics and Astronomy, Uppsala University, Sweden \\
\end{center}

\begin{center}
\today
\end{center}

\section*{Abstract}
{\bf Global SMEFT analyses have become a key interpretation framework
  for LHC physics, quantifying how well a large set of kinematic
  measurements agrees with the Standard Model. This agreement is
  encoded in measured Wilson coefficients and their uncertainties. A
  technical challenge of global analyses are correlations.  We
  compare, for the first time, results from a profile likelihood and a
  Bayesian marginalization for a given data set with a comprehensive
  uncertainty treatment. Using the validated Bayesian framework we
  analyse a series of new kinematic measurements. For the updated dataset
  we find and explain differences between the marginalization and
  profile likelihood treatments.}
         
\vspace{10pt}
\noindent\rule{\textwidth}{1pt}
\tableofcontents\thispagestyle{fancy}
\noindent\rule{\textwidth}{1pt}

\clearpage
\section{Introduction}
\label{sec:intro}

Higgs physics at the LHC~\cite{Dawson:2018dcd} perfectly illustrates a
deep tension in contemporary particle physics: on the one hand, the
existence of a fundamental Higgs boson is a direct consequence of
describing the electroweak gauge sector in terms of a quantum field
theory, specifically a renormalizable gauge theory. It looks like
Nature chose the simplest possible realization of the Higgs mechanism,
with one light scalar particle and an electroweak vacuum expectation
value (VEV) of unknown origin. On the other hand, puzzles like dark
matter or baryogenesis seem to point to non-minimal Higgs sectors for
convincing solutions based on renormalizable quantum field theory, but
without any LHC hint in these directions. The main goal of the LHC
Higgs program is to understand if Nature really took the opportunity
of a minimal electroweak and Higgs sector solving as many problems as
possible, or why she skipped this opportunity in favor of
theoretically less attractive alternatives. Or, more practically
speaking, we need to study as many Higgs properties as precisely as
possible.

Given the vast LHC dataset already after Run~2 and our fundamental
ignorance of the correct UV-completion of the Standard Model (SM), we
need to measure Higgs-related observables and express them in a
consistent, fundamental, and comprehensive theory framework. To
provide the necessary precision, this framework has to be defined
beyond leading order in perturbation theory, it needs to incorporate
kinematic information, and it should allow us to combine as many LHC
observables as possible. The effective field theory (EFT) extension of the Standard Model
(SMEFT)~\cite{Brivio:2017vri} fulfills precisely these three
requirements and defines a theoretical path to understanding the
entire LHC dataset in terms of a fundamental Lagrangian. Its main
shortcoming is the necessary truncation in the operator
dimensionality. The truncated SMEFT approximation will hardly describe
new physics appropriately, so SMEFT should really be viewed as a
systematic, conservative limit-setting tool.  Of course, this
practical aspect does not cut into the fundamental attractiveness of
an effective quantum field theory description of all LHC data.\medskip

While ATLAS and CMS have not published properly global SMEFT
analyses of the Higgs-electroweak or top sector,
there exists a range of phenomenological Higgs-gauge
analyses~\cite{Biekoetter:2018ypq,daSilvaAlmeida:2018iqo,Kraml:2019sis,vanBeek:2019evb,Dawson:2020oco,Almeida:2021asy},
top analyses~\cite{Brown:2018gzb,Hartland:2019bjb,Brivio:2019ius},
combinations of the two~\cite{Ellis:2020unq,Ethier:2021bye}, and
combinations with parton densities~\cite{Iranipour:2022iak,Carrazza_2019,Greljo_2021,Kassabov_2023}. These
analyses are typically based on experimentally preprocessed
information, including the full range of uncertainties. Given that by
assumption any SMEFT analysis will be centered around the
renormalizable SM-Lagrangian, the main focus of all global analyses is
the uncertainty treatment and the correlations between the different
operators. Technically, these two tasks tend to collide.  We can
choose a conservative uncertainty treatment based on profile
likelihoods and nuisance parameters, but it is much more
computing-efficient to treat correlations through covariance matrices
of marginalized Gaussian likelihoods~\cite{Bissmann:2019qcd}. The
SFitter
framework~\cite{Lafaye:2009vr,Klute:2012pu,Corbett:2015ksa,Butter:2016cvz,Biekoetter:2018ypq,Brivio:2019ius}
is unique in the sense that it has mostly been used for profile
likelihood analyses, but can provide marginalized limits equally
well~\cite{Lafaye:2007vs,Lafaye:2009vr}.

We make use of this flexibility and study, for the first time, the
difference between profiled and marginalized likelihoods of the same
global Run~2 dataset. In Sec.~\ref{sec:bayes} we find that for the
Higgs-electroweak dimension-6 operators and the given dataset the two
approaches agree well, so we can use the marginalized setup to treat
correlated measurements and uncertainties efficiently. Based on these
results, we include a range of recent Run~2 measurements from Higgs
studies as well as from exotics resonance searches, again with a focus
on a comprehensive and conservative uncertainty treatment, in
Sec.~\ref{sec:new}. Finally, we study the impact of these new
measurements and the inputs from a global top analysis in
Sec.~\ref{sec:fit} and find interesting differences between the
profiling and marginalization methods.

\section{SMEFT Lagrangian}
\label{sec:lag}

The SMEFT Lagrangian is based on the same field content, global, and
gauge symmetries as the SM. It includes interactions with
canonical dimension larger than four, organised in a systematic
expansion in the inverse powers of a new physics
scale~\cite{Weinberg:1978kz,Georgi:1984zwz,Donoghue:1992dd}.
Neglecting lepton number violation at dimension five, the leading
beyond-SM effects stem from dimension-six terms,
\begin{align}
\lag = \sum_x \frac{f_x}{\Lambda^2} \; \ope_x \;\;.
\label{eq:def_f}
\end{align}
There are 59 baryon-number conserving operators, barring flavor
structure~\cite{Leung:1984ni,Buchmuller:1985jz,Gonzalez-Garcia:1999ije,Grzadkowski:2010es,Passarino:2012cb}. We
use the operator basis of
Refs.~\cite{Corbett:2012ja,Biekoetter:2018ypq}, starting with a set of
$P$-even and $C$-even operators and then using the equations of motion
to define a basis without blind directions in the electroweak
precision data.  We neglect operators that cannot be studied at the
LHC yet, like those changing the triple-Higgs
vertex~\cite{DiVita:2017eyz,Goncalves:2018qas,Chang:2018uwu,Biekotter:2018jzu,Borowka:2018pxx}.
We also neglect operators which are too strongly constrained from
other LHC measurements to affect the Higgs-electroweak analysis, like
the ubiquitous triple-gluon operator
\begin{align}
  \ope_G =f_{ABC} G_{A \nu}^\rho G_{B \lambda}^\nu G_{C \rho}^\lambda \; ,
\end{align}
which is strongly constrained from multi-jet
production~\cite{Krauss:2016ely}.  In the bosonic sector the relevant
operators then are
\begin{alignat}{9}
\ope_{GG} &= \phi^\dagger \phi \; G_{\mu\nu}^a G^{a\mu\nu}  \quad 
&\ope_{WW} &= \phi^{\dagger} \; \hat{W}_{\mu \nu} \hat{W}^{\mu \nu} \; \phi  \quad 
&\ope_{BB} &= \phi^{\dagger} \; \hat{B}_{\mu \nu} \hat{B}^{\mu \nu} \; \phi 
\notag \\
\ope_W &= (D_{\mu} \phi)^{\dagger}  \hat{W}^{\mu \nu}  (D_{\nu} \phi) \quad 
& \ope_B &=  (D_{\mu} \phi)^{\dagger}  \hat{B}^{\mu \nu}  (D_{\nu} \phi) 
&\ope_{BW} &= \phi^\dagger \; \hat{B}_{\mu\nu} \hat{W}^{\mu\nu} \; \phi 
\notag \\
\ope_{\phi 1} &= (D_\mu \phi)^\dagger \; \phi \phi^\dagger \; (D^\mu \phi) \quad
&\ope_{\phi 2} &= \frac{1}{2} \partial^\mu ( \phi^\dagger \phi )
\partial_\mu ( \phi^\dagger \phi )
\notag \\
\ope_{3W} &= \tr \left( \hat{W}_{\mu \nu} \hat{W}^{\nu \rho} \hat{W}_\rho^\mu \right) \; ,
\label{eq:operators1}
\end{alignat}
where $\hat{B}_{\mu \nu} = i g' B_{\mu \nu}/2$ and $\hat{W}_{\mu\nu} =
i g\sigma^a W^a_{\mu\nu}/2$. The covariant derivative acting on the
Higgs doublet is $D_\mu = \partial_\mu+ i g' B_\mu/2 + i g
\sigma_a W^a_\mu/2$.\medskip

In addition to the purely bosonic operators, we also need to include
single-current operators modifying the Yukawa couplings,
\begin{align}
\ope_{e\phi,22} &= \phi^\dagger\phi \; \bar L_2 \phi e_{R,2} \qqquad   
&\ope_{e\phi,33} &= \phi^\dagger\phi \; \bar L_3 \phi e_{R,3} \notag \\
\ope_{u\phi,33} &= \phi^\dagger\phi  \; \bar Q_3 \tilde \phi u_{R,3} \qqquad  
&\ope_{d\phi,33} &= \phi^\dagger\phi \; \bar Q_3 \phi d_{R,3}  \; ,
\label{eq:operators2}
\end{align}
The main difference to earlier SFitter analyses is that we treat the
correction to the muon Yukawa $f_{\mu}$ as an independent
parameter, while previously it was tied to $f_{\tau}$ via an
approximate flavor symmetry. As LHC Run~2 found experimental evidence
for the Higgs coupling to muons, this approximation can now be
dropped.  However, when including the observed branching ratio to
muons, we will not be sensitive to the sign of the muon Yukawa, except
for the fact that such a sign flip is not consistent
with the SMEFT assumptions. Since the sign flip needs an interference term
to become visible, but there are no known interference effects.

Other single-current operators modify gauge and gauge-Higgs ($HVff$)
couplings~\cite{Ellis:2018gqa,daSilvaAlmeida:2018iqo,Zhang:2016zsp,Corbett:2017qgl,Baglio:2017bfe,Baglio:2018bkm,Alves:2018nof,Dawson:2018jlg},
\begin{align}
\ope_{\phi u}^{(1)} &=\phi^\dagger (i\,{\Dfb}_{\mu} \phi) (\bar u_{R}\gamma^\mu u_{R}) 
&
\ope_{\phi Q}^{(1)} &=\phi^\dagger (i\,{\Dfb}_{\mu} \phi) (\bar Q\gamma^\mu Q) 
\notag\\
\ope_{\phi d}^{(1)} &=\phi^\dagger (i\,{\Dfb}_{\mu} \phi) (\bar d_{R}\gamma^\mu d_{R})
&
\ope_{\phi Q}^{(3)} &=\phi^\dagger (i\,{\Dfba}_{\!\!\mu} \phi) \left(\bar Q\gamma^\mu \frac{\sigma_a}{2} Q\right)
\notag \\
\ope_{\phi e}^{(1)}&=\phi^\dagger (i\,{\Dfb}_{\mu} \phi) (\bar e_{R}\gamma^\mu e_{R}) \; .
\label{eq:ope_new2}
\end{align}
The four-lepton operator
\begin{align}
 \ope_{4L} &= (\bar{L}_1 \gamma_\mu L_2) \; (\bar{L}_2 \gamma^\mu L_1) 
\end{align}
induces a shift in the Fermi constant. For the operators in
Eq.\eqref{eq:ope_new2}, we maintain for simplicity a flavor symmetry,
and all currents are implicitly defined with diagonal flavor
indices. In this limit, the operators $\ope_{\phi L}^{(1)}, \ope_{\phi
  L}^{(3)}$, analogous to $\ope_{\phi Q}^{(1)}, \ope_{\phi Q}^{(3)}$,
are redundant with the bosonic set of Eq.\eqref{eq:operators1} via
equations of motion~\cite{Corbett:2012ja,Biekoetter:2018ypq}.

Dipole operators and $\ope_{\phi ud,ij}^{(1)} = \tilde \phi^\dagger
(i\,{D}_{\mu} \phi) (\bar u_{R,i}\gamma^\mu d_{R,j}) $ are neglected
for two reasons: the approximate flavor symmetry requires them to
scale with the SM Yukawa couplings and their interference with the SM
is always proportional to the fermion masses. Both factors suppress
their effects except for the top quark. The three dipole moments of
the top quark --- electric, magnetic and chromomagnetic --- are not
suppressed, so in this work we choose to retain the chromomagnetic
operator~\cite{Cirigliano:2009wk,Falkowski:2017pss,Alioli:2017ces,daSilvaAlmeida:2018iqo}
\begin{align}
  \ope_{tG} = i g_s (\bar{Q}_3 \sigma^{\mu \nu} T^A u_{R,3} ) \; \tilde{\phi} G^A_{\mu \nu} \; .
\end{align}
It affects the Higgs observables at the LHC significantly through the
loop-induced production process~\cite{Maltoni:2016yxb,Grazzini:2016paz,Deutschmann:2017qum,Ellis:2020unq,Ethier:2021bye}.\medskip

Our SMEFT Lagrangian is then defined as
\begin{align}
\lag_\text{eff} = \lag_\text{SM} 
&- \frac{\alpha_s }{8 \pi} \frac{f_{GG}}{\Lambda^2} \ope_{GG}  
 + \frac{f_{WW}}{\Lambda^2} \ope_{WW} 
 + \frac{f_{BB}}{\Lambda^2} \ope_{BB} 
 + \frac{f_{BW}}{\Lambda^2} \ope_{BW} \notag \\
&+ \frac{f_W}{\Lambda^2} \ope_W  
 + \frac{f_B}{\Lambda^2} \ope_B  
 + \frac{f_{\phi 1}}{\Lambda^2} \ope_{\phi 1} 
 + \frac{f_{\phi 2}}{\Lambda^2} \ope_{\phi 2}
 + \frac{f_{3W}}{\Lambda^2} \ope_{3W} \notag \\
&+ \frac{f_\mu m_\mu}{v \Lambda^2} \ope_{e\phi,22} 
 + \frac{f_\tau m_\tau}{v \Lambda^2} \ope_{e\phi,33} 
 + \frac{f_b m_b}{v \Lambda^2} \ope_{d\phi,33} 
 + \frac{f_t m_t}{v \Lambda^2} \ope_{u\phi,33} \notag \\
&+ \frac{f_{\phi Q}^{(1)}}{\Lambda^2} \ope_{\phi Q}^{(1)} 
 + \frac{f_{\phi d}^{(1)}}{\Lambda^2} \ope_{\phi d}^{(1)} 
 + \frac{f_{\phi u}^{(1)}}{\Lambda^2} \ope_{\phi u}^{(1)} 
 + \frac{f_{\phi e}^{(1)}}{\Lambda^2} \ope_{\phi e}^{(1)} 
 + \frac{f_{\phi Q}^{(3)}}{\Lambda^2} \ope_{\phi Q}^{(3)} 
 + \frac{f_{4L}}{\Lambda^2} \ope_{4L} \notag \\
&+ \frac{f_{tG}}{\Lambda^2} \ope_{tG}
 + \text{invisible decays}\;.
\label{eq:ourlag}
\end{align}
It contains 20 independent Wilson coefficients. The branching ratio of the Higgs to invisible final states, $\text{BR}_\text{inv}$,
  is treated as a free parameter, to account for potential Higgs
  decays to a dark matter agent.  For the global analysis it is
  convenient to work with the two orthogonal combinations
\begin{align}
 \ope_\pm = \frac{\ope_{WW} \pm \ope_{BB}}{2}
 \qquad \Rightarrow \qquad
 f_\pm = f_{WW} \pm f_{BB} \; .
\end{align}
The rotation is defined such that only $\ope_+$ contributes to the
$H\gamma\gamma$ vertex.\medskip

If we base our calculation on the Lagrangian like that given in
Eq.\eqref{eq:ourlag}, we strictly speaking need to supplement it with
a renormalization scheme or a renormalization condition.  For each
process, a reasonable assumption is that all Lagrangian parameters,
including the Wilson coefficients, are evaluated at the same
renormalization scale $\mu_R$. For the processes entering our global
analysis, an appropriate central scale choice is $\mu_R\simeq
m_H/2~...~m_H$.  To improve the precision beyond leading order, one
should eventually account for the renormalization group
evolution~\cite{Alonso:2013hga}, and evaluate the SMEFT predictions at
the energy scale appropriate for each process. This scale can vary for
instance across bins of a kinematic distribution.  In this work, all
SMEFT predictions are calculated at leading order, so we postpone an
in-depth analysis of renormalization group effects to a future work,
together with a systematic study of the impact of higher-order
corrections to inclusive Higgs production and decay rates.\medskip

The truncated Lagrangian of Eq.\eqref{eq:ourlag} as our
fundamental theory hypothesis needs to be put into context.  
The hypothesis based on a truncated Lagrangian is, strictly speaking,
not well defined once we include higher multiplicities of the dimension-6
operators in the amplitude. Therefore, the SMEFT analysis should be interpreted
as representing classes of
models~\cite{Dawson:2021xei,Dawson:2022cmu}, and the validity of the
SMEFT approach rests on the process-dependent assumption that in the
corresponding models no new particle is produced on its mass
shell~\cite{Biekotter:2016ecg}. While SMEFT is an excellent framework
to interpret global LHC analysis, possible anomalies need to be
interpreted by matching it to UV-complete
models~\cite{Anisha:2020ggj,DasBakshi:2020pbf,DasBakshi:2021xbl,Cepedello:2022pyx},
where for instance WBS signatures of corresponding models might
eventually require us to go beyond dimension-6
operators~\cite{Brass:2018hfw}.

If global SMEFT analyses should be interpreted as representing classes
of UV-complete models for a limited set of observables, we need to
consider the interplay between the SMEFT hypothesis and more
fundamental models.  Given the precision of the SMEFT analysis and its
field-theoretical advantages over the naive coupling analysis we can
and should perform this matching beyond leading
order~\cite{Brehmer:2015rna,Drozd:2015rsp,Fuentes-Martin:2016uol,Henning:2016lyp,Kramer:2019fwz,Cohen:2020fcu},
accounting for matching scale
uncertainties~\cite{Dawson:2020oco,Brivio:2021alv}, rather than
ignoring them at leading order. While this scale uncertainty clearly
does not cover all uncertainties induced by matching SMEFT limits to
UV-complete models, it also illustrates that such uncertainties exist
and have to be taken into account.

\section{Bayesian SFitter setup}
\label{sec:bayes}

Global SMEFT analyses are a key ingredient to a more general analysis
strategy at the LHC, which is to test theory predictions based on
perturbative quantum field theory using the full kinematic range of
the complete set of LHC measurements. It is worth stressing that SMEFT
analyses are currently the only way to systematically probe kinematic
LHC measurements beyond resonance searches. They come with two
assumptions which greatly simplify the actual analyses
\begin{enumerate}
\item experimentally, we know that our SMEFT analysis is not confronted with
  established anomalies; those should be discussed using
  properly defined BSM models; 
\item theoretically, SMEFT can only describe small deviations from the
  Standard model, otherwise the dimensional expansion in
  Eq.\eqref{eq:def_f} is not valid.
\end{enumerate}
While global SMEFT analyses with a truncated Lagrangian can translate
kinematic measurements into fundamental parameters, these two aspects
imply that their outcome will be limit-setting.  For our analysis this
means that we already know that the global maximum of the SMEFT
likelihood lies around the SM-limit $f_x/\Lambda^2 \to 0$.  The exact
position of the most likely parameter point is of limited interest,
the main task of the global analysis is to determine the uncertainty
on the values of the Wilson coefficients or, more in general, the
finite preferred region in the multi-dimensional SMEFT parameter
space.\medskip

In this spirit, the goal of the SFitter framework is to enable an
independent interpretation of experimental inputs, without relying on
pre-processed information and including a comprehensive treatment of
statistical, systematic, and theory
uncertainties~\cite{Lafaye:2009vr,Klute:2012pu,Corbett:2015ksa}.  The
SFitter methodology relies on the construction of a likelihood
function in which these uncertainties can be described by nuisance
parameters.  In all previous SFitter analyses, nuisance parameters are
profiled over. The resulting profile likelihood is then profiled over
the parameters of interest, to extract one- and two-dimensional limits
on the Wilson coefficients.  An alternative, Bayesian treatment is
based on marginalising over nuisance parameters and parameters of
interest. It has been adopted in several SMEFT
analyses~\cite{Dumont:2013wma,Fichet:2015xla,DeBlas:2019ehy,Bissmann:2019qcd,deBlas:2021wap}
and simplifies greatly the treatment of correlated uncertainties.  The
goal of this work is to perform an apples-to-apples comparison between
a profiled and a marginalized likelihood, employing exactly the same
data and uncertainties inputs in both cases.

\subsubsection*{Marginal likelihood}

Since marginalization is new in SFitter, we provide a brief
description of the main features. The corresponding profile likelihood
treatment is discussed in detail in
Refs.~\cite{Lafaye:2009vr,Klute:2012pu,Butter:2016cvz,Biekoetter:2018ypq,Brivio:2019ius}.
The first step of a global analysis is the construction of the fully
exclusive likelihood $\like_\text{excl}$, which is a function of the
parameters of interest $f_x$ and of a set of nuisance parameters
$\theta_i$. This $\like_\text{excl}$ is defined with the following
uncertainty treatment: (i) statistical uncertainties are included via
a Poisson distribution, in some cases approximated using a Gaussian
whenever this stabilizes the numerical evaluation; (ii) systematic
uncertainties are assumed to be Gaussian, organized in 31 categories,
such that uncertainties within the same category are fully correlated
through a covariance matrix or through nuisance parameters.
Systematics which do not fit into any of the 31 categories are assumed
to be uncorrelated; (iii) theory uncertainties are modelled as flat
distributions. Whenever theory uncertainties need to be correlated we
use an explicit nuisance parameter.

For a Bayesian analysis we first marginalize over or integrate out the
nuisance parameters.  This yields the marginal likelihood
$\like_\text{marg}$, for one counting measurement and one parameter
illustrated by
\begin{align}
  \like_\text{marg}(f_x) = \int d\theta\, \like_\text{excl}(f_x,\theta) =
  \int d\theta \; \text{Pois}(d | m(f_x,\theta)) \; p(\theta) \; .
\label{eq:bayesian_1D}
\end{align}
Here $d$ stands for the measured number of events, $m$ is the model
(theory) prediction, $\theta$~is a nuisance parameter and $p(\theta)$
the distribution over the nuisance parameter which, in the Bayesian
context, defines the prior. In SFitter, nuisance priors are either
Gaussian or flat. Computing $\like_\text{marg}$ in SFitter starts with
the marginalization procedure over the nuisance parameters, so we omit
the dependence on $f_x$ for now.\medskip

SFitter provides several options to define the statistical model of a
measurement, including a simplified Gaussian likelihood where
uncertainties add in quadrature. A more sophisticated and reliable
framework starts with a typical LHC measurement as an independent
counting experiment, which is modelled by a Poisson distribution.
Systematic uncertainties or theory uncertainties then define the
completely exclusive likelihood for one measurement
\begin{align}
  \like_\text{excl} (\theta)
  = \text{Pois}(d | m(\theta_1, \theta_2, ..., b)) \;
  p(b) \;  \prod_i p(\theta_i) \; .
\label{eq:bayesian_likelihood}
\end{align}
Here $d$ is the measured number of events, $b$ the background
estimate, and $m$ the model prediction, that is a function of the
nuisance parameters $\theta_i$. The distributions $p(b)$ and
$p(\theta_i)$ incorporate our knowledge about these quantities. In
general, they can be extracted from auxiliary measurements,
simulations, or other possible sources. However, because tracking
hundreds of different reference measurements is beyond the scope of
SFitter, we simply assume $p(\theta_i)$ to be Gaussian for systematic
uncertainties and flat or uniform for theory uncertainties,
\begin{align}
  p(\theta_i) =
  \begin{cases}
    \mathcal{N}_{0,\sigma_i}(\theta_{\text{syst}, i}) &\qqquad \text{systematics} \\
    \mathcal{F}_{0,\sigma_i}(\theta_{\text{theo}, i}) &\qqquad \text{theory} \; .
  \end{cases}
\end{align}
In this step we assume that all prior distributions for
$\theta_\text{syst}$ and $\theta_\text{theo}$ are centered around
zero, with given half-widths $\sigma$.

For $p(b)$, SFitter provides several choices: for measurements where
$b$ is extracted from a single control region (CR) measurement we use
\begin{align}
p(b) = \text{Pois} (b_\text{CR} | b k) \; ,
\label{eq:background_prior}
\end{align}
where $k$ is an interpolation factor between CR and signal region,
$b_\text{CR}$ is the measured number of events in the control region,
and $b$ is the expected number of background events in the signal
region. For measurements with several control regions or with
simulated backgrounds we assume the combined $p(b)$ to be a
Gaussian. Systematic uncertainties on the background measurement can
also be included, and are assumed to be fully correlated with the
uncertainties on the signal region within the same category.\medskip

Typically, the dependence of the theory prediction $m$ on the nuisance
parameters in Eq.\eqref{eq:bayesian_likelihood} is not spelled out or
extremely complex to determine. To simplify this task, we assume a
leading linear dependence on assumed-to-be small uncertainties
\begin{align}
  m \approx s + b + \theta_{\text{theo}, 1} + \theta_{\text{theo}, 2} + \cdots + \theta_{\text{syst}, 1} + \theta_{\text{syst}, 2} + \cdots
  \equiv s + b + \theta_\text{tot} \;.
\label{eq:m_linearized}
\end{align}
where $s$ is the expected number of signal events. The exclusive
likelihood of Eq.\eqref{eq:bayesian_likelihood} can then be written as
\begin{align}
  \like_\text{excl}(\theta) \approx
  \text{Pois}(d | s + b + \Sigma \theta_{\text{theo}, j} + \Sigma \theta_{\text{syst}, i})
  \; p(b)
  \; \prod_j \mathcal{F}_{0,\sigma_j}(\theta_{\text{theo}, j})
  \; \prod_i \mathcal{N}_{0,\sigma_i}(\theta_{\text{syst}, i})  \; ,
  \label{eq:bayesian_likelihood_before_integral}
\end{align}
The marginal likelihood for a single measurement is then
constructed by integrating over all nuisance parameters,
\begin{align}
  \like_\text{marg}
  &=
  \int \prod_j d\theta_{\text{theo}, j}
  \int \prod_i d\theta_{\text{syst}, i}
  \int d b \;
  \like_\text{excl}(\theta) 
  \notag \\
  &=
  \int \prod_j d\theta_{\text{theo}, j} \mathcal{F}_{0,\sigma_j}(\theta_{\text{theo}, j})
  \int \prod_i d\theta_{\text{syst}, i} \mathcal{N}_{0,\sigma_i}(\theta_{\text{syst}, i})
  \notag \\
  & \qqqquad \qqqquad \times \int d b \;
    \text{Pois}(d | s + b +\Sigma \theta_{\text{theo}, j} + \Sigma \theta_{\text{syst}, i})
    \; p(b) \; .
\label{eq:likelihood_last_integral1}
\end{align} 
The integration over $b$ can be performed analytically if 
$p(b)$ is a Poisson distribution. In this case, the convolution $\mathcal{P}(d|s+\theta_{\text{tot}})$ of
$p(b)$ and Pois($d|m$) gives a so-called Poisson-Gamma model, as
Eq.\eqref{eq:background_prior} is a special case of the Gamma
distribution,
\begin{align}
  \like_\text{marg}
  &=
  \int \prod_j d\theta_{\text{theo}, j} \mathcal{F}_{0,\sigma_j}(\theta_{\text{theo}, j})
  \int \prod_i d\theta_{\text{syst}, i} \mathcal{N}_{0,\sigma_i}(\theta_{\text{syst}, i}) \times
  \mathcal{P}(d | s + \theta_\text{tot}) \; .
\end{align} 
We use $\theta_\text{tot}$ as defined in Eq.\eqref{eq:m_linearized}.
To solve the remaining integrals over the nuisance parameters we
replace one of the integrals, for instance $\theta_{\text{syst}, 1}$
with $(\theta_\text{tot} -\Sigma_{i\neq 1} \theta_{\text{syst}, i})$,
\begin{align}
  \like_\text{marg}
  =
  \int d \theta_\text{tot}
  &\;
  \mathcal{P}(d | s + \theta_\text{tot})
  \notag\\
  &
  \times
  \underbrace{\int \prod_j d\theta_{\text{theo}, j} \mathcal{F}_{0,\sigma_j}(\theta_{\text{theo}, j})
  \int \prod_{i \ne 1} d\theta_{\text{syst}, i} \mathcal{N}_{0,\sigma_i}(\theta_{\text{syst}, i})
  \mathcal{N}_{0,\sigma_1} ( \theta_{\text{syst},1})
  }_\text{solved analytically}  \; .
\label{eq:likelihood_last_integral2}
\end{align} 
Assuming only Gaussian plus at most three flat priors, all
$\theta$-convolutions except for one can be performed
analytically. The corresponding closed formulas are implemented in
SFitter, speeding up the marginalization.  The remaining 1-dimensional
integral in Eq.\eqref{eq:likelihood_last_integral2} is solved
numerically with Simpson's method.\medskip

\begin{figure}[t]
  \centering
  \includegraphics[width=0.48\textwidth,page=1]{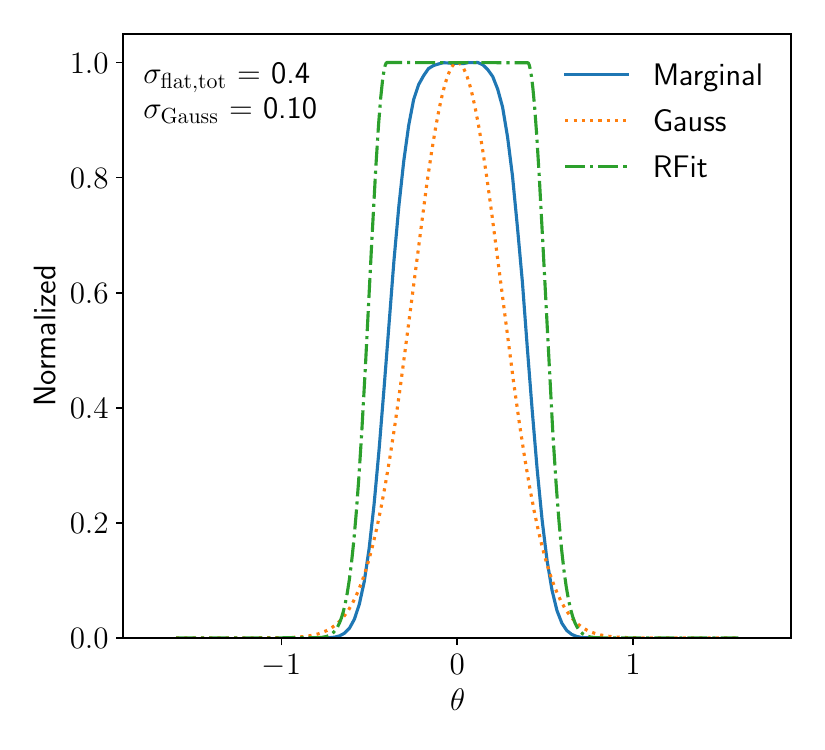}
  \includegraphics[width=0.48\textwidth,page=2]{comparison_rfit_vs_marginal.pdf}
  \caption{Marginalized and profiled likelihoods from the convolution of a Gaussian
distribution with one (left) and three (right) flat ones. In the right plot, each flat uncertainty is $\sigma_{\text{flat, i}}=0.4/3.$ The total flat uncertainty is $\sigma_{\text{flat,tot}}=0.4$ in both panels, such that the RFit scheme results are identical in the two cases (green dot-dashed curve). On the other hand, the Gaussian obtained adding half-widths in quadrature (orange curve) differs.}
  \label{fig:one_vs_many_th_uncertainties}
\end{figure}

Marginalizing over nuisance parameters and profiling over them will
not give the same marginalized likelihood.  Only for statistical
uncertainties described by Poisson statistics and Gaussian
systematics, the two lead to the same marginalized result in the limit
of large enough statistics. Differences appear when we use flat theory
uncertainties. For a Bayesian marginalization the central limit
theorem ensures that the final posterior will be approximately
Gaussian. Using a profile likelihood, two uncorrelated flat
uncertainties add linearly, while a combination of flat and Gaussian
uncertainties give the well-known RFit
prescription~\cite{Hocker:2001xe}.
Figure~\ref{fig:one_vs_many_th_uncertainties} shows, as an
illustration, the distributions obtained combining one Gaussian with
one (left) or three (right) flat nuisance parameters. We see that
the profile likelihood or RFit result maintains a flat core and is independent of the number of theory nuisances, while 
the marginalized result varies and is very close to a Gaussian in the right panel.

\subsubsection*{Combining channels}

Unlike probabilities, likelihoods of a set of measurements can simply
be multiplied. This means we can generalize
Eqs.\eqref{eq:bayesian_likelihood}
and~\eqref{eq:bayesian_likelihood_before_integral} to a set of $N$
measurements by replacing
\begin{alignat}{7}
  & \text{Pois}(d | m) p(b)
  \quad &&\longrightarrow \quad
  &&\prod_k \text{Pois}(d_k | m_k) p(b_k) \notag \\
  & \mathcal{N}_{0,\sigma_i} (\theta_{\text{syst},i}) 
  \quad &&\longrightarrow \quad 
  && \mathcal{N}_{\vec 0,\Sigma_i} (\vec \theta_{\text{syst},i}) \notag \\
  &\mathcal{F}_{0,\sigma_j} (\theta_{\text{theo},j}) 
  \quad &&\longrightarrow \quad
  &&\prod_k \mathcal{F}_{0,\sigma_{kj}}(\theta_{\text{theo}, kj}) \; ,
\end{alignat}
with 
\begin{align}
  m_k \approx s_k + b_k + \sum_i \theta_{\text{syst},ki} + \sum_j \theta_{\text{theo}, kj}
  \equiv s_k + b_k + \theta_{\text{tot},k} \; .
\end{align}
Here we assume that the theory uncertainties are uncorrelated, while
the systematics can be correlated, so we need to introduce an
$N$-dimensional Gaussian with the covariance matrices $\Sigma_i$
encoding the correlations between uncertainties of category $i$ entering different measurements $k$. We use either uncorrelated or fully correlated
systematics.

When we compute the marginal likelihood in analogy to
Eq.\eqref{eq:likelihood_last_integral1} the only non-trivial aspect are
the correlated systematic uncertainties including the covariance
matrix. However, the convolution of $N$-dimensional Gaussians still
leads to one $N$-dimensional Gaussian, where the combined covariance
matrix is the sum of the individual covariance matrices.  This means,
in the last step of Eq.\eqref{eq:likelihood_last_integral2} we are now
left with an $N$-dimensional integral over $\theta_{\text{tot},k}$,
correlated through the covariance matrix appearing in the distribution
of the systematic nuisance parameters.

In SFitter, this integral is solved by approximating it with the
Laplace method. This is computationally efficient and works well for
cases where most of the probability is concentrated around one
mode. This is the case when the nuisance parameters are Gaussians or
flat. We can then write
\begin{align}
\int dx^n f(x) = \int d x^n e^{\log f(x)}\; ,
\end{align}
and assume that $f(x)$ has a maximum at $x=x_0$. 
Then one can expand $\log f(x)$ up to second
order around $x_0$ as
\begin{align}
\log f(x) \approx \log f(x_0) + \underbrace{\frac{\partial}{\partial x} \log f(x_0)}_{= 0} (x - x_0) +  \underbrace{\frac{\partial^2}{\partial x_i x_j} \log f(x_0)}_{= F_{ij}(x_0)} (x - x_{0})_i \; (x - x_{0})_j + ...
\label{eq:laplace}
\end{align}
such that the integral is approximated by
\begin{align}
\int dx^n f(x) \approx f(x_0) \; \sqrt{\dfrac{(2\pi)^n}{\det F(x_0)}} \; .
\label{eq:laplace_2}
\end{align}
Note that $f(x)$ is given by the exclusive likelihood, with the
maximum at $f(x_0)$ kept through profiling but not through
marginalization.  The matrix $F(x_0)$ is the Hessian of the
log-likelihood at the maximum, \ie the Fisher information matrix in
the space of the nuisance parameters.  In SFitter, $x_0$ is extracted
with an analytic expression, approximating the Poisson distribution in
Eq.\eqref{eq:bayesian_likelihood} with a Gaussian.  The resulting
error is compensated by keeping a finite first derivative in
Eq.\eqref{eq:laplace}, which in turn requires us to modify
Eq.\eqref{eq:laplace_2} by introducing an additional term depending on
the first derivative of the log-likelihood. Both the first and second
derivatives can be computed numerically. All these approximations in
evaluating the exclusive and marginal likelihoods have been checked by
evaluating the exclusive likelihood using Markov chains.

\subsubsection*{Validation}

\begin{figure}[t]
\centering
\includegraphics[width=0.95\textwidth]{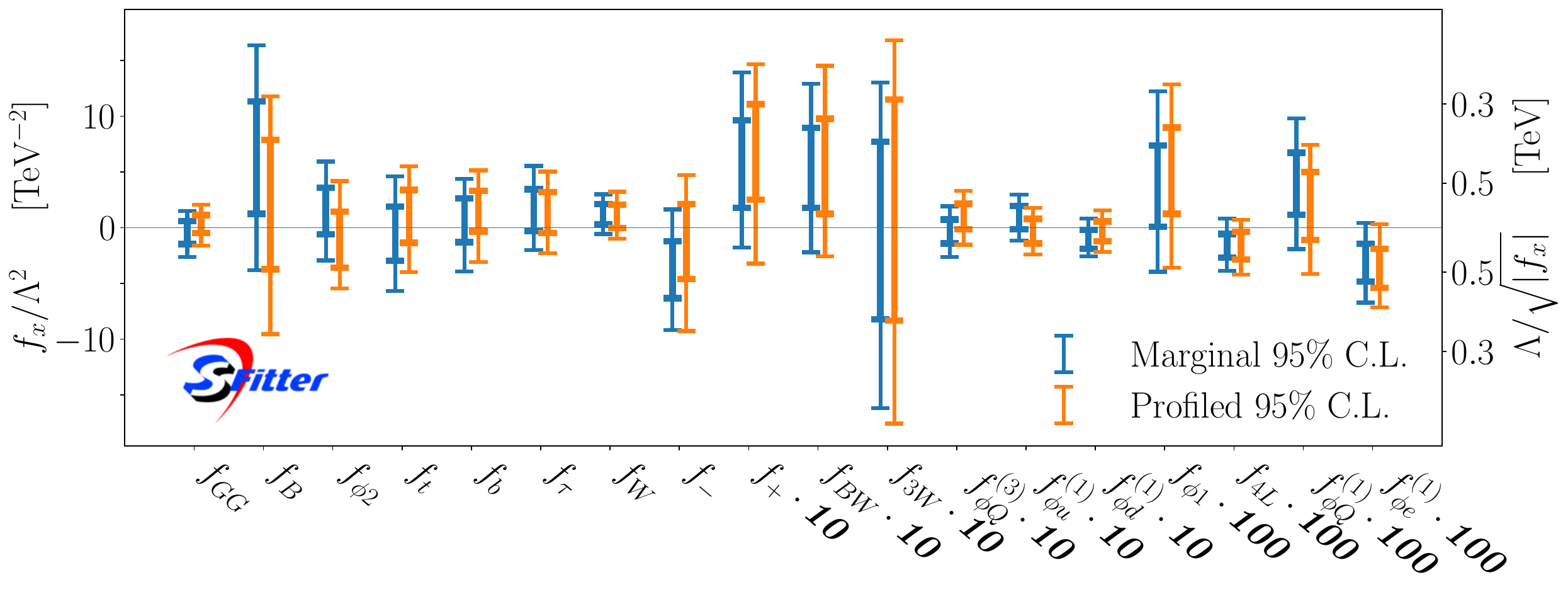}
\caption{68\% and 95\% confidence intervals from profile likelihoods
  and Bayesian marginalization. The dataset is the same as in
  Ref.\cite{Biekoetter:2018ypq}.}
\label{fig:comparison_freq_vs_bayesian_summary}
\end{figure}

We can validate the implementation of the Bayesian marginalization
over nuisance parameters and Wilson coefficients starting from the
fully exclusive likelihood using the operator basis and dataset of
Ref.~\cite{Biekoetter:2018ypq}. The SMEFT Lagrangian is given in
Eq.\eqref{eq:ourlag}, but without the muon Yukawa, the top-gluon
coupling $\ope_{tG}$, and the invisible branching ratio of the Higgs.  For the direct
comparison we construct the marginal likelihood by profiling or
marginalizing over all nuisance parameters and Wilson
coefficients. We then extract the posterior probability and 68\% and
95\% confidence intervals. Unless otherwise specified, we assume flat,
wide priors for all Wilson coefficients.  This choice minimizes the
impact of the prior on the final result, and we have verified that our
priors on the Wilson coefficients indeed fulfill this condition.  In
Fig.~\ref{fig:comparison_freq_vs_bayesian_summary}, we show the 68\%
and 95\%CL limits from the corresponding 18-dimensional operator
analysis. We see that the results of the two methods are in good
agreement.\medskip

\begin{figure}[t]
  \includegraphics[width=0.33\textwidth, page= 3]{18d_without_new_meas_bayesian_vs_freq}
  \includegraphics[width=0.33\textwidth, page= 1]{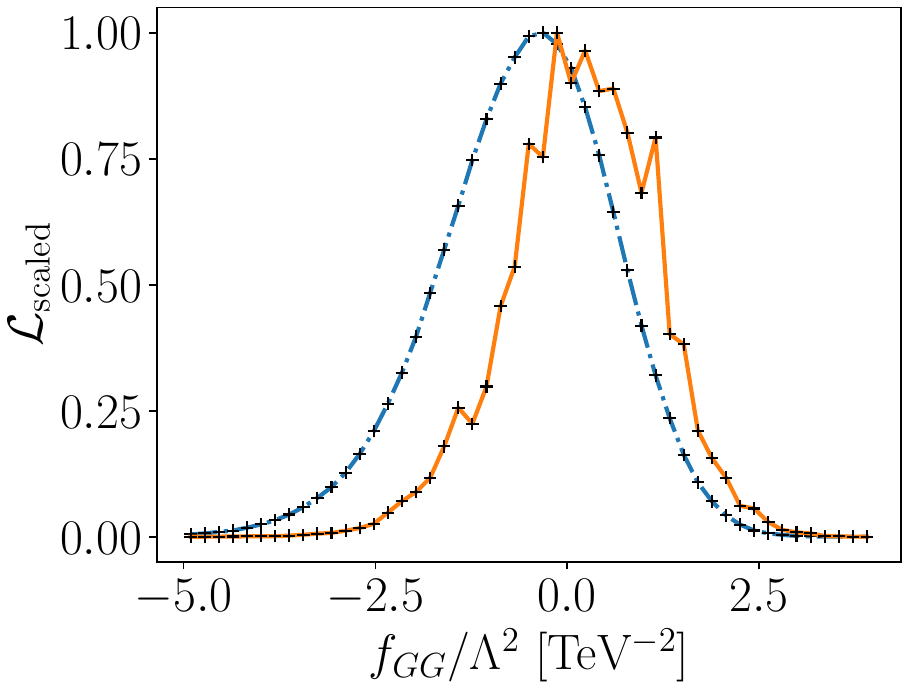}
  \includegraphics[width=0.33\textwidth, page=17]{18d_without_new_meas_bayesian_vs_freq}
  \caption{Profile likelihoods vs marginalized likelihood for a set of
    Wilson coefficients. The two curves are scaled such that the
    maximum values are at $\like_\text{scaled} = 1$.}
\label{fig:comparison_freq_vs_bayesian}
\end{figure}

Going beyond confidence intervals, we can look at the distributions of
the 1-dimensional profile likelihoods or marginalized probabilities.
We show three examples in
Fig.~\ref{fig:comparison_freq_vs_bayesian}. Because the analysis
relies on actual LHC data, the central values are not at zero Wilson
coefficients. The well-measured Wilson coefficient $f_W$ shows no
difference between the profile and the marginalized results. For
$f_{GG}$, we see a slight deviation in the central values, within one
standard deviation and therefore not statistically significant. This
effect points to the theory and pdf uncertainties, which we assume to
be flat, and which therefore allow the central value to move freely
for the profile likelihood approach, while the marginalization leads
to a well-defined maximum when combining two individually flat
likelihood distributions. In
Fig.~\ref{fig:comparison_freq_vs_bayesian_summary} we see that this
difference only has a slight effect on the lower boundary when we
extract 95\%CL limits on $f_{GG}$. Finally, we see a similar effect
for $f_-$, even though this measurement depends on several different
LHC channels. According to
Fig.~\ref{fig:comparison_freq_vs_bayesian_summary} this is one of the
largest and still not significant differences between the two methods.
We use Monte Carlo Markov chains to scan the parameter space. To construct the profile likelihood we need to find a singular point, the maximum, for each bin. Because of the high dimensionality of our fit this requires us to sample a large amount of points to get a decent estimate of this maximum. Even after sampling 100M points the results can still look very spiky. In contrast, marginalization results into smoother outcomes because it does not depend on only one point per bin. Additionally marginalization smooths out statistical fluctuation, which the profile likelihood does not.

\begin{figure}[b!]
  \includegraphics[width=0.33\textwidth, page=3]{18d_without_new_meas_bayesian_prof_vs_marg}
  \includegraphics[width=0.33\textwidth, page=1]{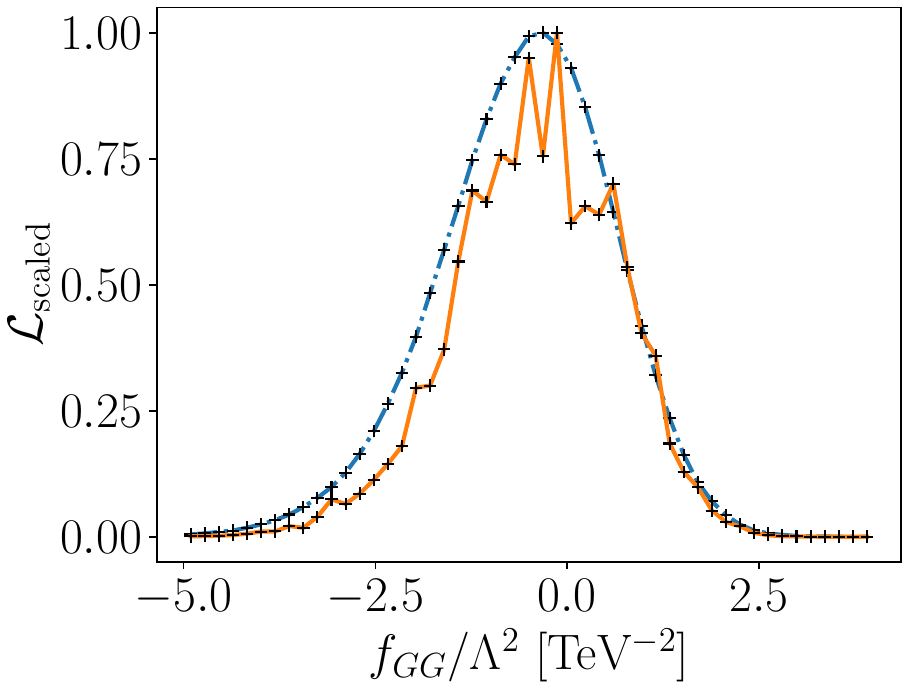}
  \includegraphics[width=0.33\textwidth, page=17]{18d_without_new_meas_bayesian_prof_vs_marg}
  \caption{Likelihoods profiled vs marginalized over the Wilson
    coefficients $f_x$, but always marginalized over all nuisance
    parameters $\theta$. We show the same Wilson coefficients as in
    Fig.~\ref{fig:comparison_freq_vs_bayesian}.}
\label{fig:bayesian_prof_vs_marg}
\end{figure}

The source of the differences in
Fig.~\ref{fig:comparison_freq_vs_bayesian} can be traced back to
whether the uncertainty-related nuisance parameters are marginalized
or profiled. Fig.~\ref{fig:bayesian_prof_vs_marg} shows that, once the
uncertainty treatment is fixed, the results are independent of whether
the Wilson coefficients are marginalized or profiled over.\medskip

\begin{figure}[t]
\includegraphics[width=0.33\textwidth, page= 1]{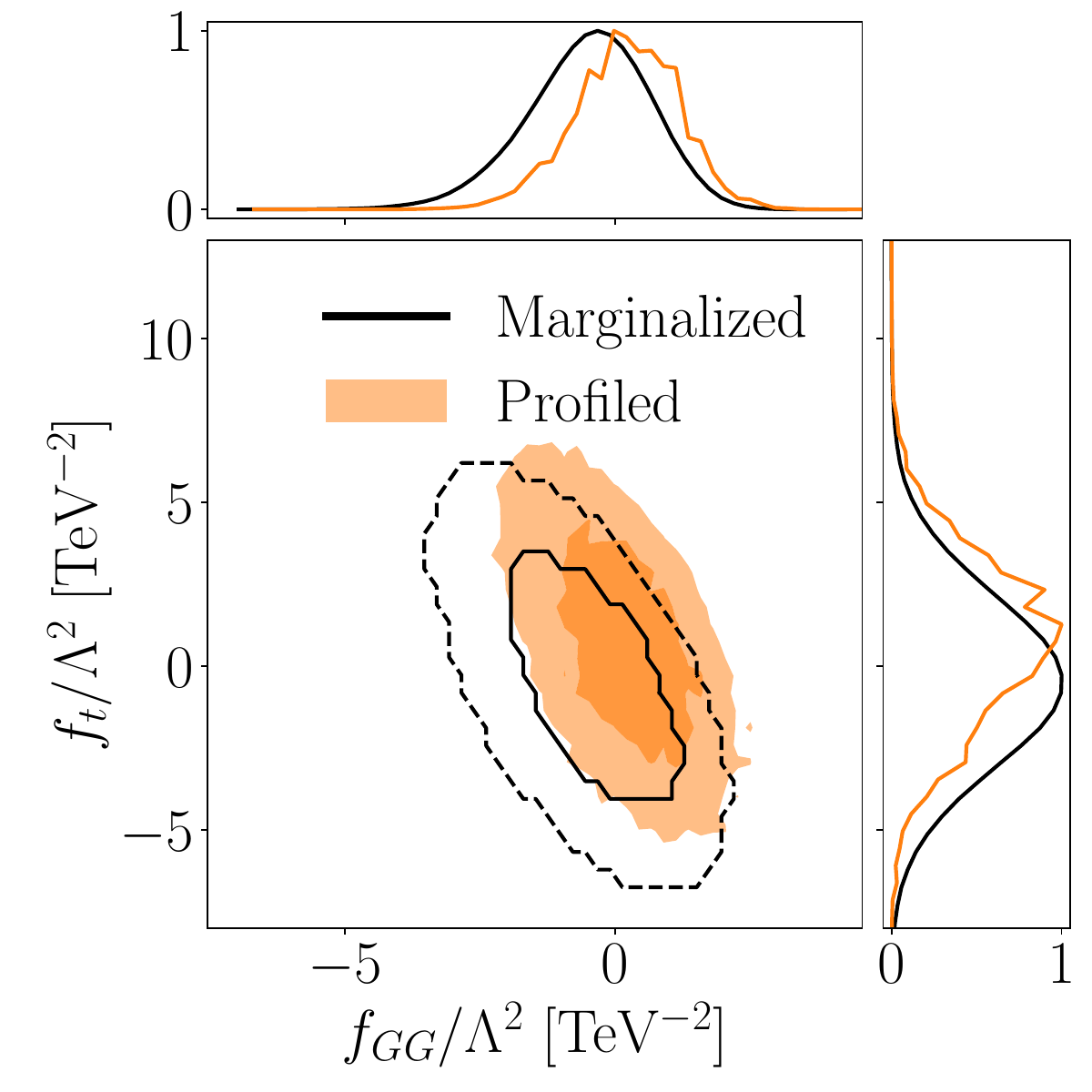}
\includegraphics[width=0.33\textwidth, page= 2]{18d_without_new_meas_2d_correlations_selection_bayesian_vs_frequentist_v2} 
\includegraphics[width=0.33\textwidth, page=3]{18d_without_new_meas_2d_correlations_selection_bayesian_vs_frequentist_v2}
\caption{Comparison of 2-dimensional correlations of profiled
  and marginalized likelihoods.}
\label{fig:comparison_freq_vs_bayesian_2d_v2}
\end{figure}

Next, we check 2-dimensional profiled and marginalized likelihoods.
Figure~\ref{fig:comparison_freq_vs_bayesian_2d_v2} shows three
examples involving the same Wilson coefficients as in
Fig.~\ref{fig:comparison_freq_vs_bayesian}. First, we see that there
exists an anti-correlation between $f_{GG}$ and $f_t$, the modified
top Yukawa also affecting the loop-induced production process $gg \to
H$. This suggests that a slightly high rate measurement can be
accommodated by adjusting either of the two Wilson
coefficients. Because the uncertainty on this measurement includes
sizeable theory and pdf contributions, the same difference between the
two methods can be seen for each of the two Wilson coefficients
individually and for their correlation. Another instructive example is
the correlation between $f_W$, determined from kinematic
distributions, and $f_{\phi 2}$ leading to a shift in the Higgs wave
function. Here the difference only appears in $f_{\phi 2}$, the
parameter extracted from total rates and especially sensitive to
theory uncertainties. Finally, we show the correlation between $f_-$
and $f_{\phi Q}^{(3)}$ and observe the usual correlation from the
sizeable range of kinematic di-boson
measurements~\cite{Brehmer:2019gmn}.\medskip
 
\begin{figure}[b!]
\includegraphics[width=0.33\textwidth, page=18]{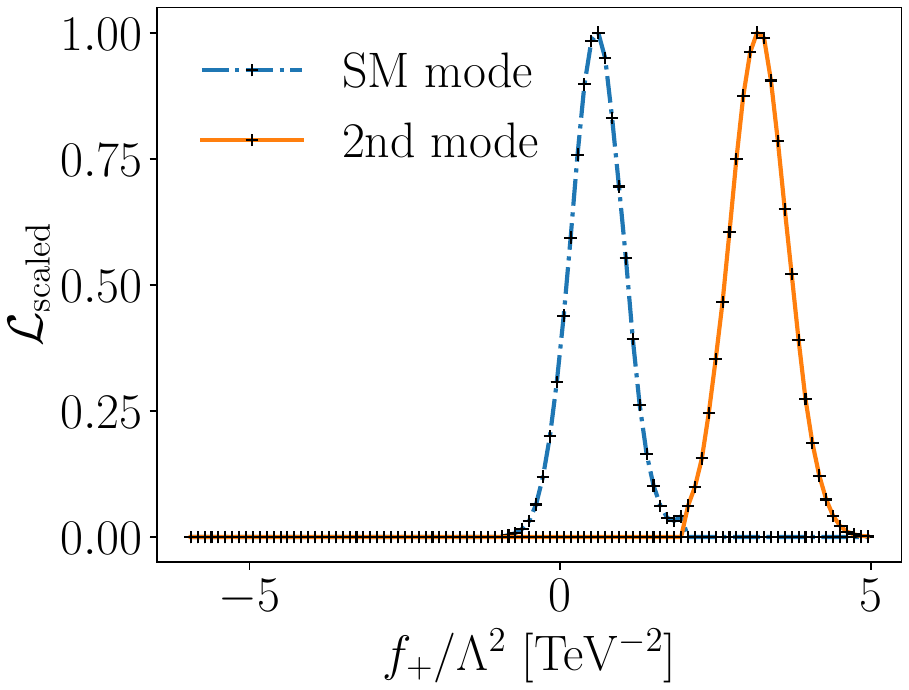}
\includegraphics[width=0.33\textwidth, page=17]{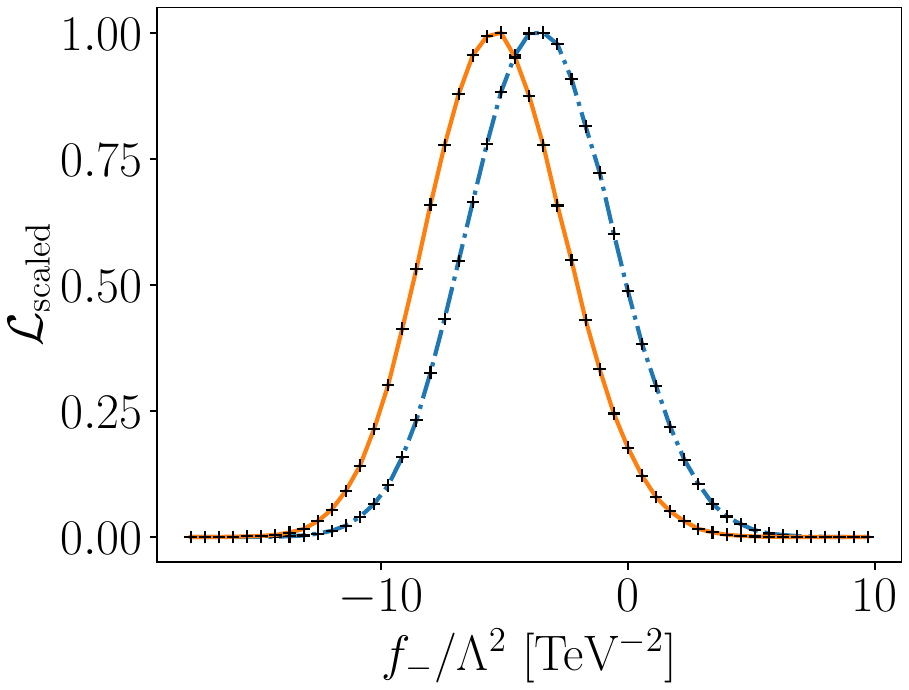}
\includegraphics[width=0.33\textwidth, page=2]{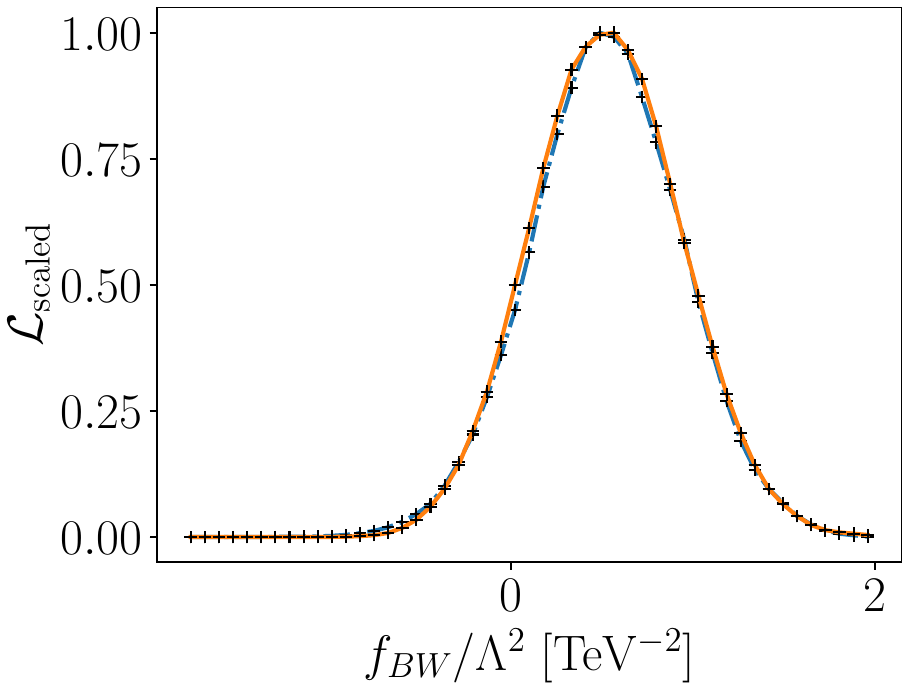}
\caption{Marginalized likelihoods for the SM-like and the second mode in
  $f_+$, again for the 18-dimensional analysis.}
\label{fig:second_mode_1dplots}
\end{figure}

Finally, we can check for alternative maxima in the likelihood and
find that $f_+$ is the only Wilson coefficient exhibiting a
non-trivial second mode. This can be understood from the $f_+$ vs
$f_-$ plane.  By a numerical accident, the SMEFT corrections to all
Higgs production and decay processes vanish in the SM-maximum and also
close to the point $f_-/\Lambda^2 = -3$ and $f_+/\Lambda^2 = 2.7$.  The
only measurement which breaks this degeneracy is $H\to Z\gamma$, with
limited statistical power. In the $f_+$ axis, the position of the
maximum is fully determined by $H\to \gamma\gamma$, which is measured
precisely enough to resolve the two modes, while in the $f_-$ axis the
constraints cannot distinguish the second maximum from the SM point.

Given the consistency condition of the SMEFT approach, we should not
compare the two modes at face value, even though the Bayesian setup
would allow for this.  On the other hand, we need to confirm that this
choice of modes does not affect other parameters in a significant
manner once it is embedded in the 18-dimensional space.  In
Fig.~\ref{fig:second_mode_1dplots} we show what happens if we restrict
our parameter analysis to either the SM-mode or the second mode. To
this end we run Markov chains mapping out both modes and then separate
the samples through the condition $f_+/\Lambda^2 \lessgtr 2$.  We see
that choosing the second mode in $f_+$ has a small effect on $f_-$,
pushing the best-fit closer to $f_- = -3$, but none of the other Wilson
coefficients is affected.  We also confirmed that both modes are of
equal height by choosing a Breit-Wigner proposal function, which
ensures that the Markov chains can move large distances, helping each
individual chain to jump between both modes.

\subsubsection*{Uncertainties and correlations}

\begin{figure}[t]
\includegraphics[width=0.33\textwidth, page=1]{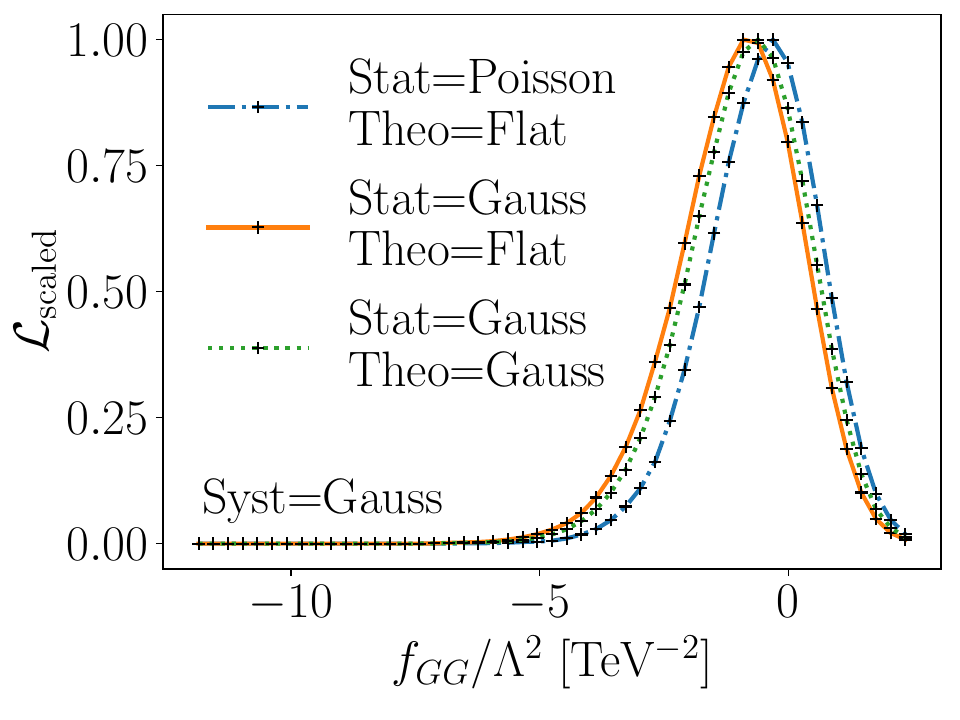}
\includegraphics[width=0.33\textwidth, page=17]{18d_without_new_meas_bayesian_different_settings}
\includegraphics[width=0.33\textwidth, page=12]{18d_without_new_meas_bayesian_different_settings}
\caption{Marginalized likelihoods for different uncertainty
  modeling. The SFitter default is a Poisson likelihood with flat
  theory uncertainties and Gaussian systematics (blue dot-dashed).}
\label{fig:bayes_diff_settings}
\end{figure}

\begin{figure}[b!]
\includegraphics[width=0.33\textwidth, page=1]{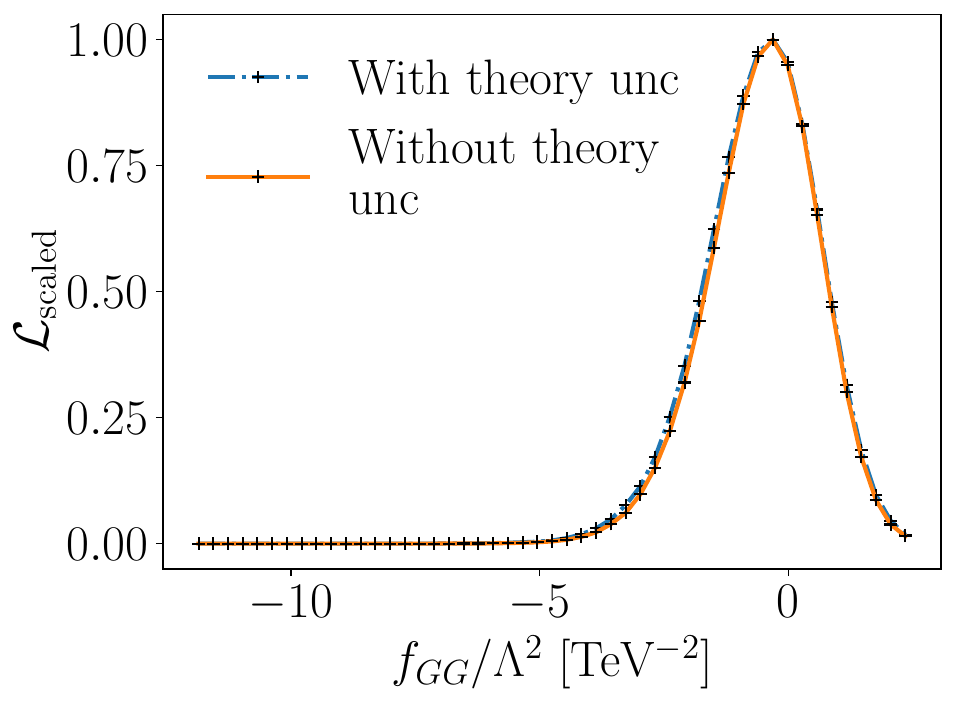}
\includegraphics[width=0.33\textwidth, page=17]{18d_without_new_meas_bayesian_no_theory_unc}
\includegraphics[width=0.33\textwidth, page=12]{18d_without_new_meas_bayesian_no_theory_unc}
\caption{Marginalized likelihoods with and without theory uncertainties.}
\label{fig:bayes_notheo}
\end{figure}

After confirming that the slight differences between the profile and
marginalization approaches are related to the treatment of
uncertainties, we can check the impact of the SFitter-specific
uncertainty treatment. By default, and as explained earlier, we
construct the exclusive likelihood with flat theory uncertainties and
Gaussian systematics.  By switching all uncertainties to Gaussian
distributions we construct the completely Gaussian likelihood shown in
Fig.~\ref{fig:bayes_diff_settings}. If we marginalize over the
different uncertainties, the central limit theorem guarantees that for
enough different uncertainties the results will be identical. The
exact level of agreement between different uncertainty models depends on
the dataset and the size of the individual uncertainties and cannot be
generalized. For instance, sizeable differences will appear when an
outlier measurement generates a tension in the global analysis. Such a
tension can be accommodated more easily using a single flat
uncertainty with its reduced cost in the likelihood value.\medskip

Because the main difference between profiling and marginalizing over
uncertainties appears for the flat theory uncertainties, the
results from Fig.~\ref{fig:bayes_diff_settings} motivate the question
how relevant the theory uncertainties really are for the Run~2
dataset analyzed in Ref.~\cite{Biekoetter:2018ypq}. We show three 1-dimensional
likelihoods in Fig.~\ref{fig:bayes_notheo} and indeed find that after
marginalizing over all nuisance parameters and over all other Wilson
coefficients the theory uncertainties do not play any visible
role. Obviously, this statement is dependent on a given dataset, on
the operators we are looking at, and on the assumed uncertainties, and
it clearly does not generalize to all global Run~2 analyses.\medskip

\begin{figure}[t]
\includegraphics[width=0.33\textwidth, page=1]{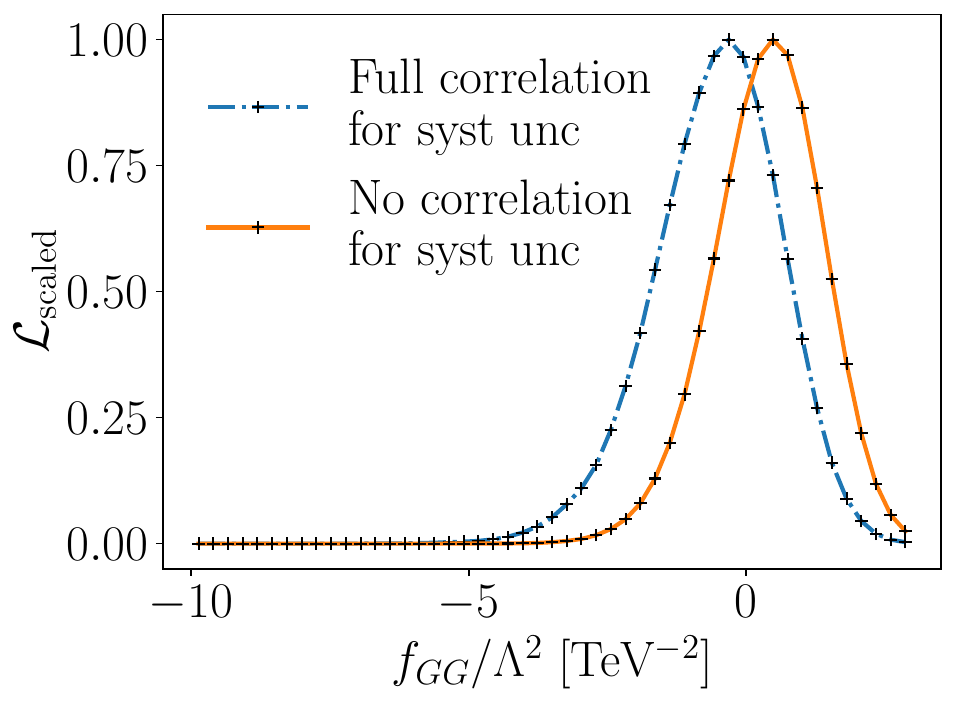}
\includegraphics[width=0.33\textwidth, page=17]{18d_without_new_meas_bayesian_turn_off_correlation}
\includegraphics[width=0.33\textwidth, page=4]{18d_without_new_meas_bayesian_turn_off_correlation}
\includegraphics[width=0.33\textwidth, page=6]{18d_without_new_meas_bayesian_turn_off_correlation}
\includegraphics[width=0.33\textwidth, page=13]{18d_without_new_meas_bayesian_turn_off_correlation}
\includegraphics[width=0.33\textwidth, page=14]{18d_without_new_meas_bayesian_turn_off_correlation}
\caption{Marginalized likelihoods with and without correlations
  between systematic uncertainties of the same category.}
\label{fig:turn_off_correlations}
\end{figure}

The last effect we need to study is the impact of correlations between
the different uncertainties.  In Fig.~\ref{fig:turn_off_correlations}
we show what happens with the 1-dimensional marginalized likelihoods
when we switch off all correlations between systematic uncertainties
of the same kind. We see that the correlations have a much larger
impact than anything else we have studied in this section. While the
size of the uncertainties do not change much, the central values
essentially vary freely within one standard deviation. An analogous
effect was observed in Ref.~\cite{Bissmann:2019qcd}. We cannot
emphasize enough that all statements about the validity of different
approximations do not generalize to new, incoming measurements, as we
will see in the following section. However, something that will not
change is the key relevance of correlations as indicated by
Fig.~\ref{fig:turn_off_correlations}.
 
\section{Updated dataset}
\label{sec:new}

After the detailed comparison of a profile likelihood and Bayesian
SFitter approach we can, in principle, apply the numerically simpler
Bayesian approach to update the SMEFT analysis of the
Higgs-electroweak sector with a series of new Run~2 results. As a
first step, we introduce the set of new kinematic measurements
entering the updated SFitter analysis. We focus on an improved
treatment of correlated uncertainties.

\subsection{WW resonance search}
\label{sec:new_vv}

\begin{figure}[t]
  \includegraphics[width=.48\textwidth]{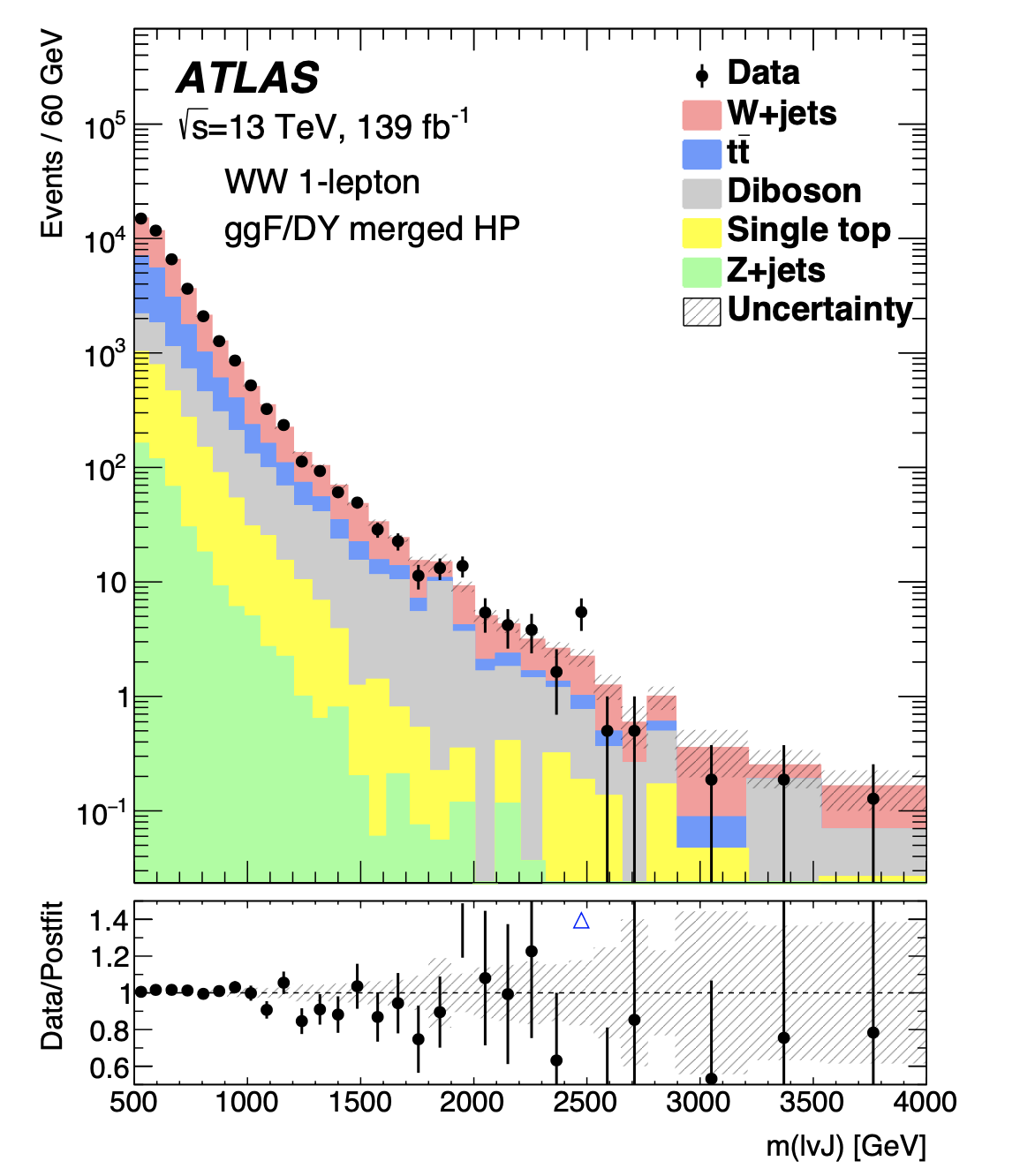}
  \includegraphics[width=.453\textwidth]{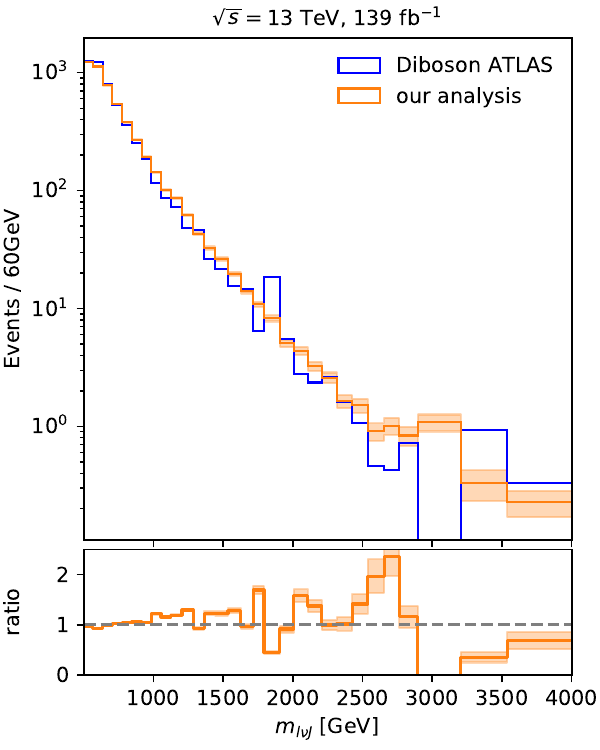}
  \caption{Left: measured $m_{VV}$ distribution~\cite{Aad:2020ddw}.
    Right: comparison between ATLAS results and our SM background
    estimate.  The orange band shows the statistical uncertainty from
    the Monte Carlo generation.}
  \label{fig:calib_VV}
\end{figure}

Once we notice that especially boosted kinematics with large momentum
transfer through Higgs interactions play a key role in SMEFT
analyses~\cite{Brehmer:2016nyr,Brehmer:2019gmn}, it is clear that the
reinterpretation of $VH$ and $VV$ resonance searches should be
extremely useful for a global SMEFT
analysis~\cite{Butter:2016cvz,Biekoetter:2018ypq}. To the best of our
knowledge, SFitter is currently the only global analysis framework
which includes these kinds of signatures.

First, we add the ATLAS search for resonances in the semi-leptonic $VV$
final state~\cite{Aad:2020ddw}, as briefly discussed in
Ref.~\cite{Brivio:2021alv}.  We only use the $WW$ 1-lepton category in
the merged Drell-Yan and gluon-fusion high-purity signal region,
\begin{align}
  pp \to W^+ W^-
  \to \ell^+ \nu_\ell \; j j + \ell^- \bar{\nu}_\ell \; j j \; .
\end{align}
Our signal consists of $W^+W^-$ production modified by SMEFT
operators. We neglect SMEFT effects in the leading $W$+jets and $t\bar
t$ backgrounds. We include all other $W_{\ell \nu} V_{jj}$ and
$Z_{\ell \ell} V_{jj}$ channels as SM-backgrounds and verified that
SMEFT corrections to the other di-boson channels are sufficiently
suppressed by the analysis setup.

The signal is simulated using Madgraph~\cite{Alwall:2014hca},
Pythia~\cite{Sjostrand:2014zea}, FastJet~\cite{Cacciari:2011ma}, and
Delphes~\cite{deFavereau:2013fsa} with the standard ATLAS card at
leading order and in the SM and requiring the lepton pair to come from
an intermediate on-shell $W^\pm$. The hadronic $W$-decay is simulated
using Pythia.  Fat jets are identified using the default
categorization in Delphes and ignoring the cut on the $D_2$ variable.
The complete SM-rate is compared to the left panel of
Fig.~\ref{fig:calib_VV}, taken from Ref.~\cite{Aad:2020ddw}.  We
reproduce the event selection based on the analysis cuts listed in
Tab.~2 of Ref.~\cite{Aad:2020ddw}. No re-calibration of energy scales
or fat-jet invariant mass windows is required, but we adjust the
histogram entries by a factor $1.606$ to match the ATLAS normalization
of the di-boson background and accommodate efficiencies and
higher-order corrections~\cite{Campbell:2011bn}.  In the right panel
of Fig.~\ref{fig:calib_VV} we show the final $m_{WW}$ distribution
obtained with this procedure.  Finally, we extract the statistical and
systematic uncertainty from the ATLAS analysis, as shown in the lower
panel in Fig.~\ref{fig:calib_VV}.  Whenever backgrounds are estimated
from control regions, the Gaussian systematic uncertainties are
smaller than the Poisson-shaped statistical uncertainties in the
signal region.

\begin{figure}[t]
\centering
\includegraphics[width=.48\textwidth]{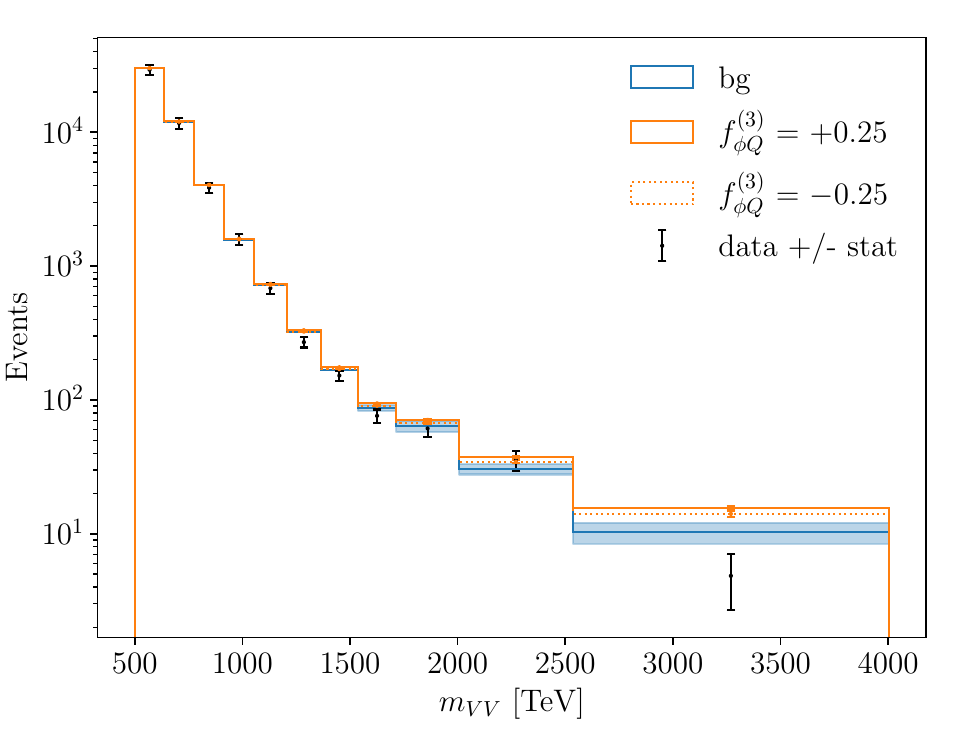}
\includegraphics[width=.48\textwidth]{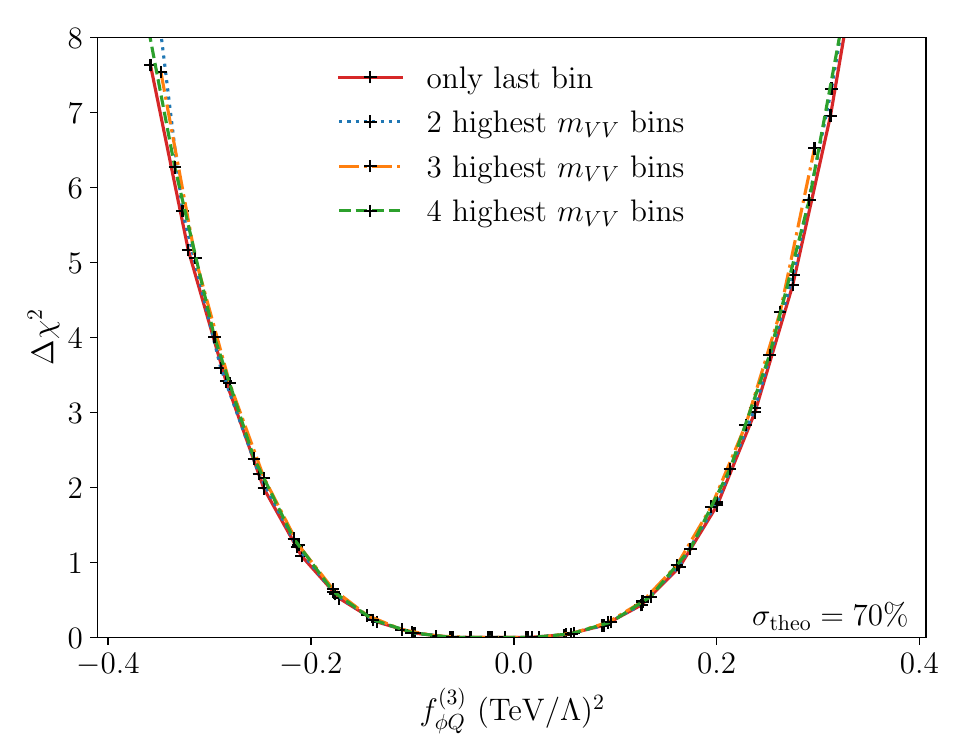}
    \caption{Left: re-binned $m_{WW}$ distribution for the
      semi-leptonic $WW$ analysis implemented in SFitter.  We show the
      complete continuum background, including
  statistical and systematic uncertainties, and the effect of a finite Wilson
      coefficient $f^{(3)}_{\phi Q}$. Right: toy analysis for the same
      Wilson coefficient using different numbers of bins.}
  \label{fig:impl_vv_chq3}
\end{figure}

To include the $VV$ channel in our SMEFT analysis we re-bin the
original distribution such that we have a minimum of five observed
events per bin. The kinematic distribution we use in SFitter is shown
in the left panel of Fig.~\ref{fig:impl_vv_chq3}. Here all statistical
uncertainties are treated as uncorrelated and added in quadrature, the
same for the systematic background uncertainties linked to Monte Carlo
statistics, while other systematic uncertainties are conservatively
treated as fully correlated and consequently added linearly.  Finally,
we add a 80\% theory uncertainty on the signal predictions in all bins
and assuming no correlation among them. Of this 70\% account for the
uncertainties in our SMEFT Monte Carlo predictions and 10\% for
$V$+jets and single-top modeling.

In the right panel of Fig.~\ref{fig:impl_vv_chq3} we show the limit in
terms of the Gauss-equivalent
\begin{align}
  \Delta \chi^2
  = \chi^2 - \chi^2_\text{min}
  = - 2 \log \like + 2 \log \like_\text{max} \; ,
\end{align}
extracted from different bins of the measured $m_{WW}$
distribution. We see that the likelihood maximum slightly deviates
from the SM point $f_{\phi Q}^{(3)} = 0$, and the last bin completely
dominates the likelihood distribution. This is expected for
momentum-enhanced operators which modify the tails of momentum
distributions, as systematically analyzed in
Ref.~\cite{Brehmer:2019gmn}. We will discuss the effect of the
under-fluctuation in the last bin in more detail in
Sec.~\ref{sec:fit_bayes}.

\subsection{WH resonance search}
\label{sec:new_wh}

\begin{figure}[t]
  \includegraphics[width=0.45\textwidth]{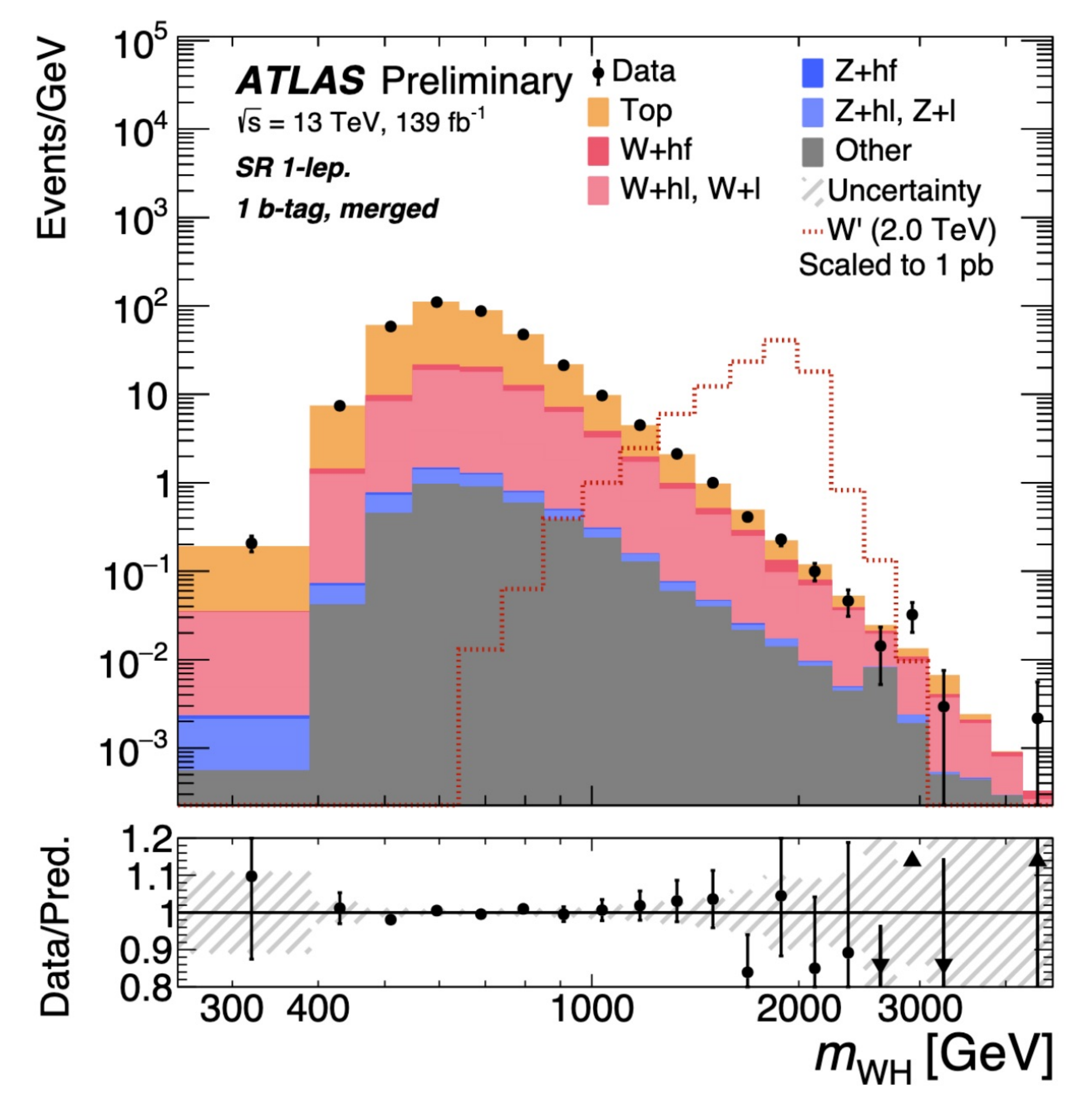} 
  \hspace*{0.05\textwidth}
  \includegraphics[width=0.45\textwidth]{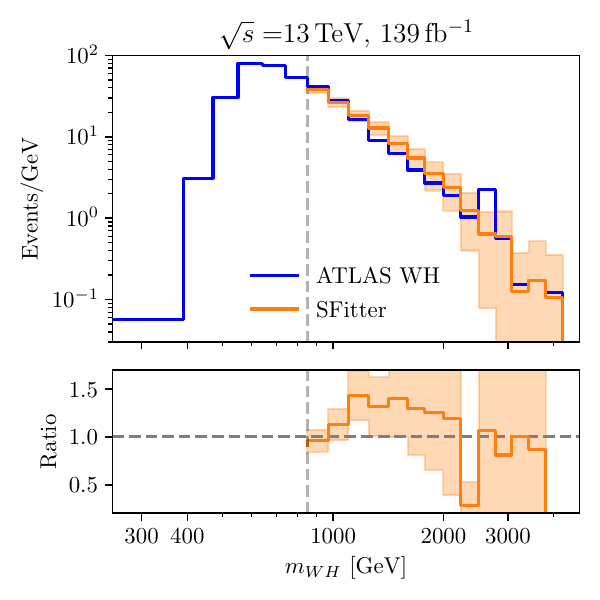}
  \caption{Left: measured $m_{WH}$
    distribution~\cite{ATLAS-CONF-2021-014}. Right: comparison between the
    the ATLAS results and our SM background estimate.  The orange band
    shows the statistical uncertainty from the Monte Carlo
    generation.}
\label{fig:Atlas_comp}
\end{figure}

\begin{figure}[b!]
  \centering
  \includegraphics[width=0.49\textwidth]{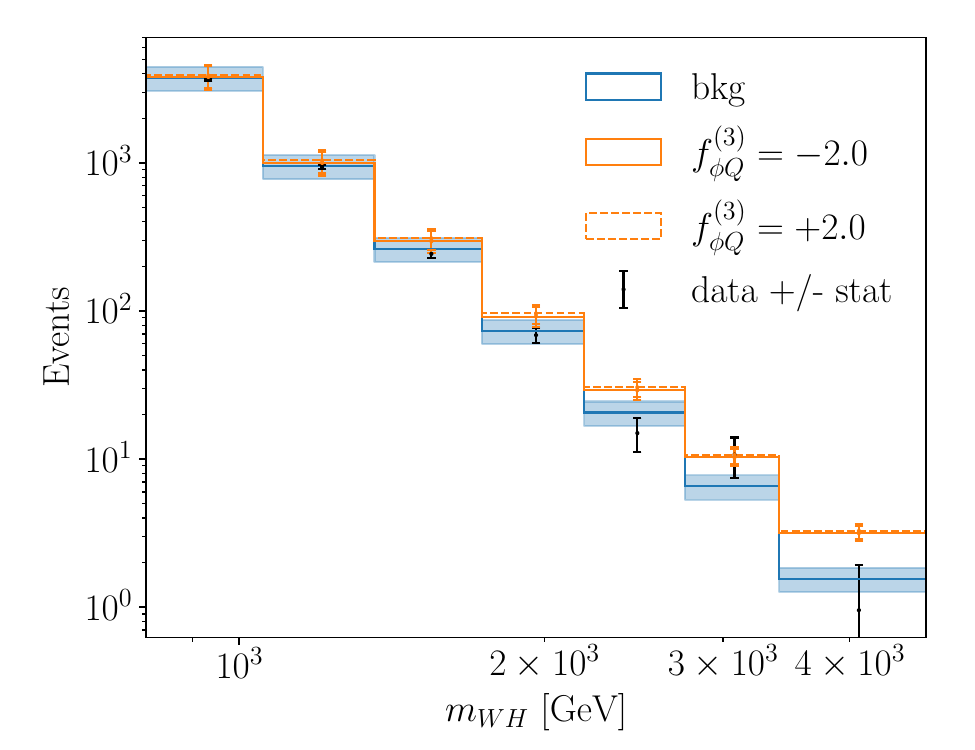}
  \caption{Re-binned $m_{WH}$ distribution implemented in SFitter,
    including statistical and systematic uncertainties. We show the
      complete continuum background and the effect of a finite Wilson
    coefficient $f^{(3)}_{\phi Q}$.}
\label{fig:wh_rebin}
\end{figure}

Complementing the dataset of Ref.~\cite{Biekoetter:2018ypq} we include
two new resonance searches, one described in
Ref.~\cite{Brivio:2021alv} and another ATLAS analysis looking for
\begin{align}
  pp \to WH \to \ell \bar{\nu}_\ell \; b \bar{b}
\end{align}
at high invariant masses~\cite{ATLAS-CONF-2021-014}. We focus on $WH$ production with one
$b$-tag, because it includes the best kinematic measurement at high
$m_{VH}$. This analysis applies cuts on the $WH$ topology and requires
exactly one single-$b$-tagged fat jet. In the merged category the
$b$-tags are part of a fat jet.

We generate di-boson events for the combined di-boson channels
with lepton-hadron decays
\begin{align}
  pp \to W_{\ell \nu} W_{jj}, \; 
  W_{\ell\nu} Z_{jj}, \; 
  Z_{\ell \ell} W_{jj}, \;
  Z_{\ell \ell} Z_{jj} \; ,
\end{align}
again using the Madgraph-Pythia-FastJet-Delphes chain with the
standard ATLAS card at leading order.  They can be compared to the
grey di-boson background in the left panel of
Fig.~\ref{fig:Atlas_comp}, including the $b$-tagging and corresponding
mis-tagging. After adjusting the $m_{WH}$-independent efficiency
factor we find the agreement illustrated in the right panel of
Fig.~\ref{fig:Atlas_comp}. We apply the same efficiency factor for the
$W H$ signal and then use the reweighting module in Madgraph to
estimate the SMEFT rates. The $W$-decay to electrons or muons is
included through Madgraph, while the Higgs decay to $b\bar{b}$ pairs
is simulated by Pythia. We neglect SMEFT corrections to the $t\bar{t}$
and $W/Z$+jets backgrounds, assuming that the targeted phase space region
favors the Higgs signal. Having to make this assumption is
unfortunate, but we emphasize that the number of experimental
measurements should prevent us from falling for SMEFT corrections
canceling between the different signals and backgrounds.

\begin{figure}[t]
  \includegraphics[width=0.32\textwidth,page=3]{theo_comp}
  \includegraphics[width=0.32\textwidth]{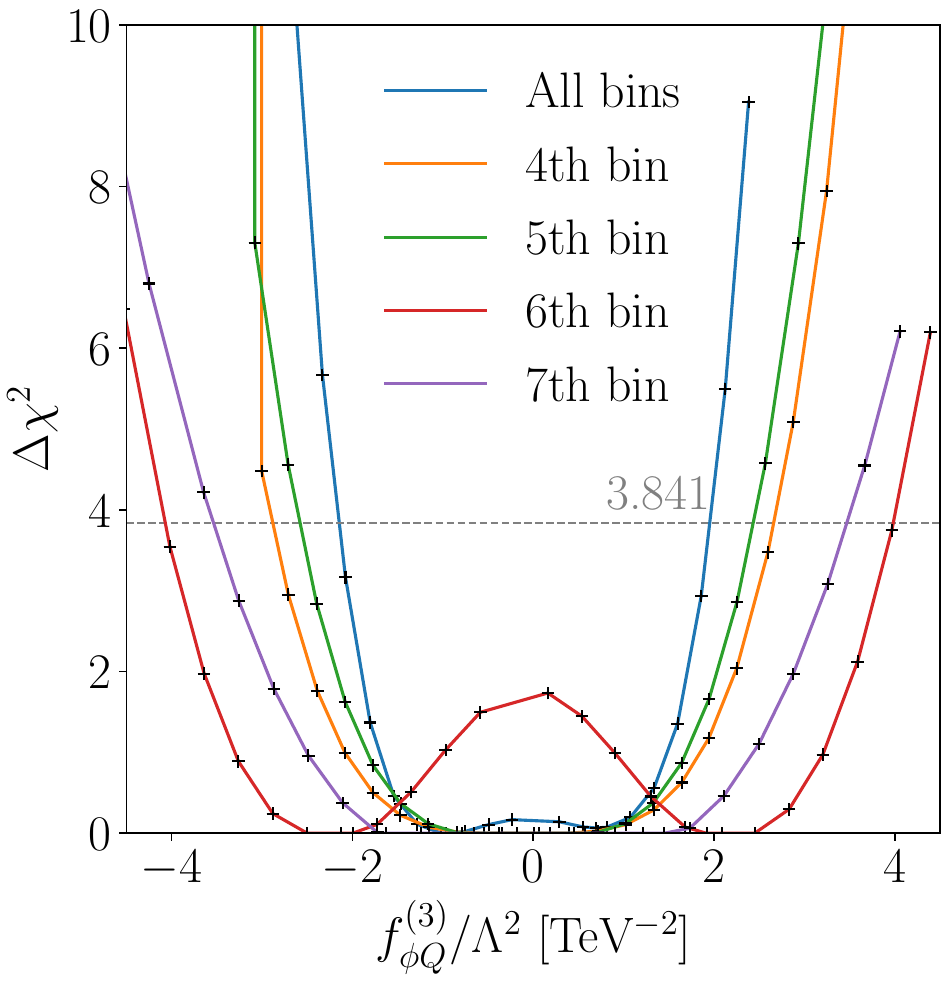} 
  \includegraphics[width=0.32\textwidth, page=3]{WH_2022_underfluctuations}
  \caption{Log-likelihood for a 3-parameter analysis of the $WH$
    search as a function of $f^{(3)}_{\phi Q}$. We vary the theory
    uncertainties and their correlation (left), the number of bins
    included with uncorrelated theory uncertainties for a
    1-dimensional analysis (center), and the treatment of
    under-fluctuations (right).}
\label{fig:nuisance_comp}
\end{figure}

To define a meaningful measurement for our global analysis we have to
merge bins of the original distribution such that at least three
observed events appear per bin.  In Fig.~\ref{fig:wh_rebin}, we show
the actually implemented distribution for the complete SM
background and including a finite Wilson coefficient $f^{(3)}_{\phi
  Q}$. For each bin we include a statistical uncertainty following a
Poisson distribution and a Gaussian systematic uncertainty, as
reported by ATLAS.  In addition, we include a $13\%$ theory
uncertainty also reported by ATLAS and a theory uncertainty between
$1\%$ and $4\%$ per bin from our SMEFT predictions, but neglecting
correlation between various bins.

We can check some of our assumptions on the way we model theory
uncertainties from a three-parameter analysis with $f^{(3)}_{\phi Q}$,
$f_{W}$ and $f_{WW}$.  Neglecting the correlations in the theory
uncertainties is justified by the left panel of
Fig.~\ref{fig:nuisance_comp}.  It shows the Gauss-equivalent
$\Delta \chi^2$ for varying the theory uncertainties with different
correlations; the orange and green lines represent a $10\%$ and $30\%$
theory uncertainty, fully correlated. The green line shows results
without theory uncertainty, and the red line assumes our SMEFT theory
uncertainty without correlations. These results are very close to each
other, so we can ignore correlations in the theory uncertainties from
the EFT prediction.

The central panel compares constraints from the 3-parameter analysis
from the entire $m_{WH}$ distribution and only including one bin at a
time. The limit improves sharply when the 4th and 5th bins are
included. This can be understood from Fig.~\ref{fig:wh_rebin}, where
both of these bins show significant under-fluctuations. In the right
panel of Fig.~\ref{fig:nuisance_comp} we show that by removing
under-fluctuations from the global analysis by setting all measured values to the
number of events expected from the SM we lose constraining
power. Again, demonstrating that our analysis strongly benefits from
under-fluctuations.

\subsection{ZH resonance search}
\label{sec:new_zh}

\begin{figure}[t]
  \includegraphics[width=0.45\textwidth]{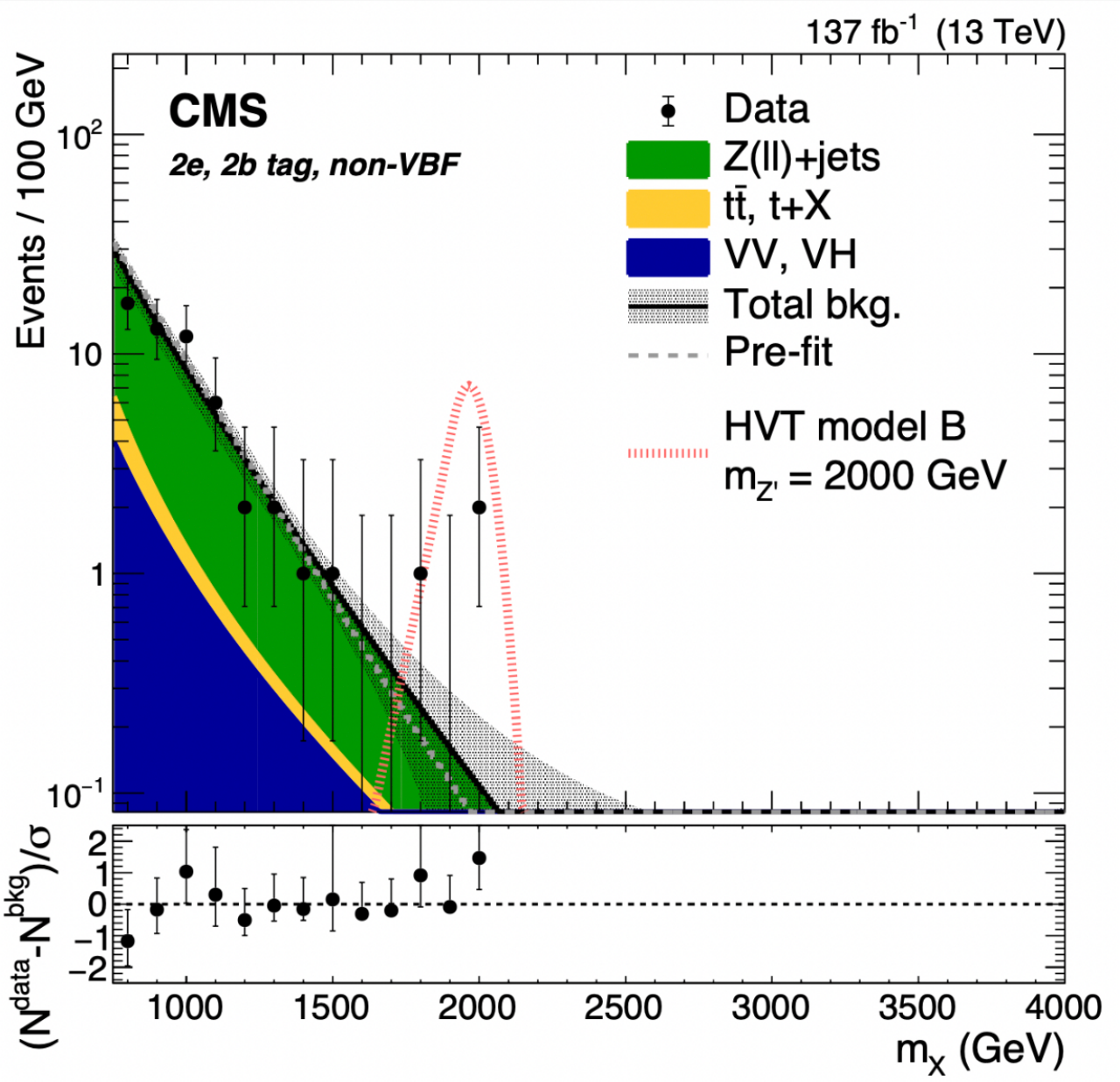}
  \hspace*{0.05\textwidth}\includegraphics[width=0.45\textwidth]{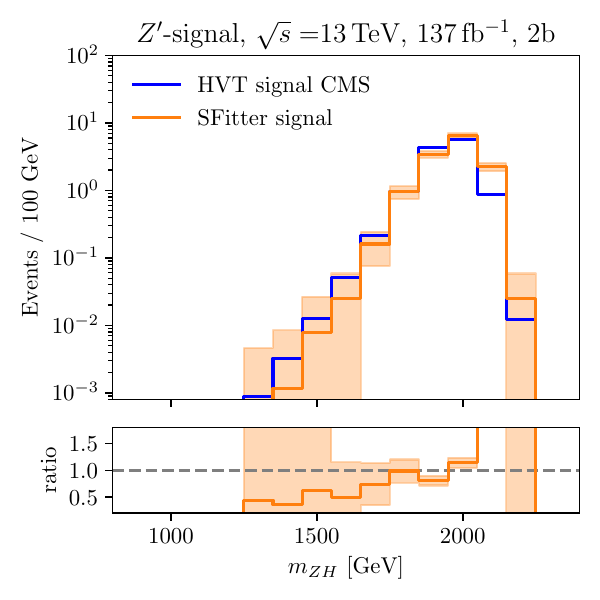}
\includegraphics[width=0.45\textwidth]{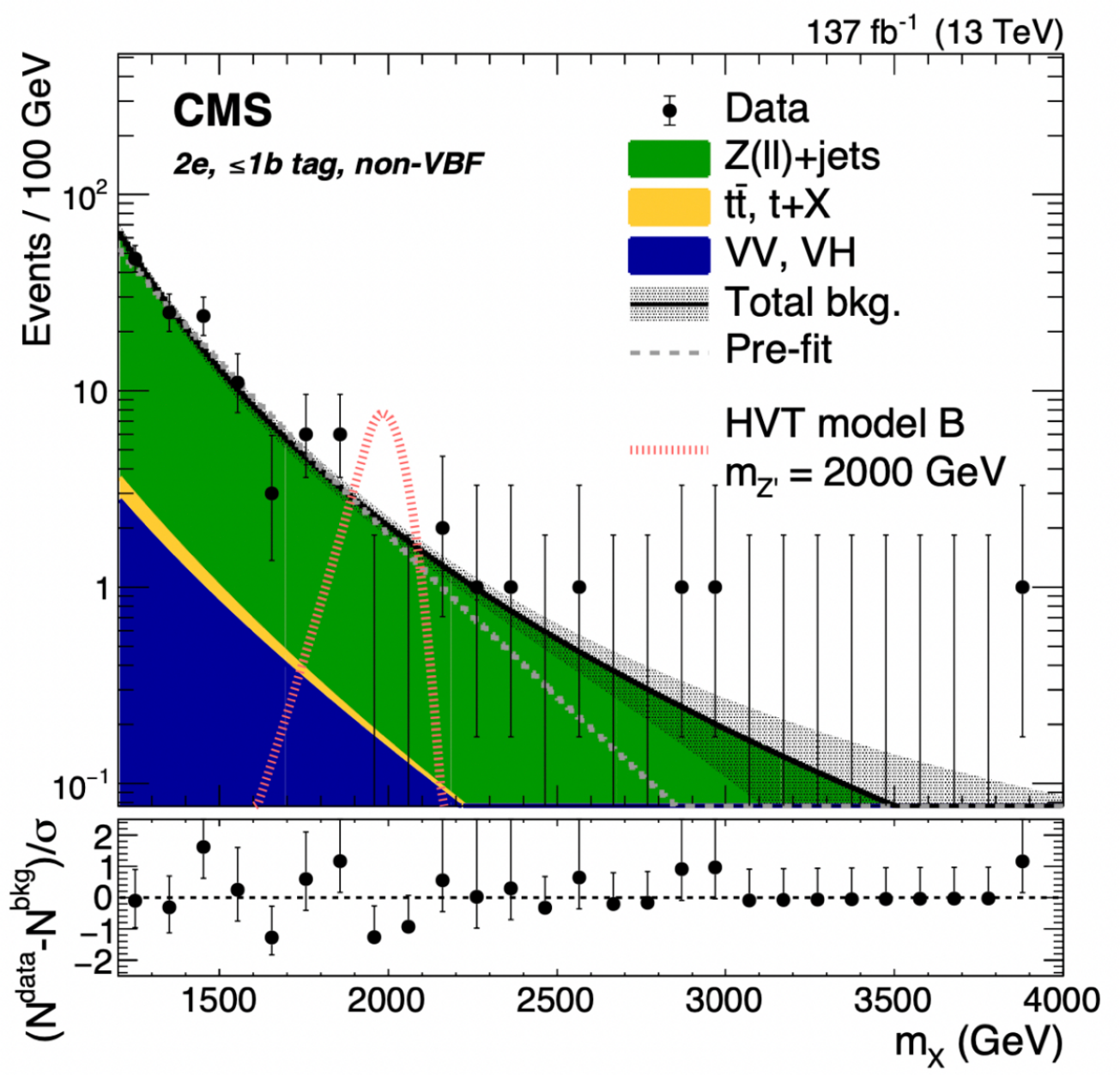}
\includegraphics[width=0.45\textwidth]{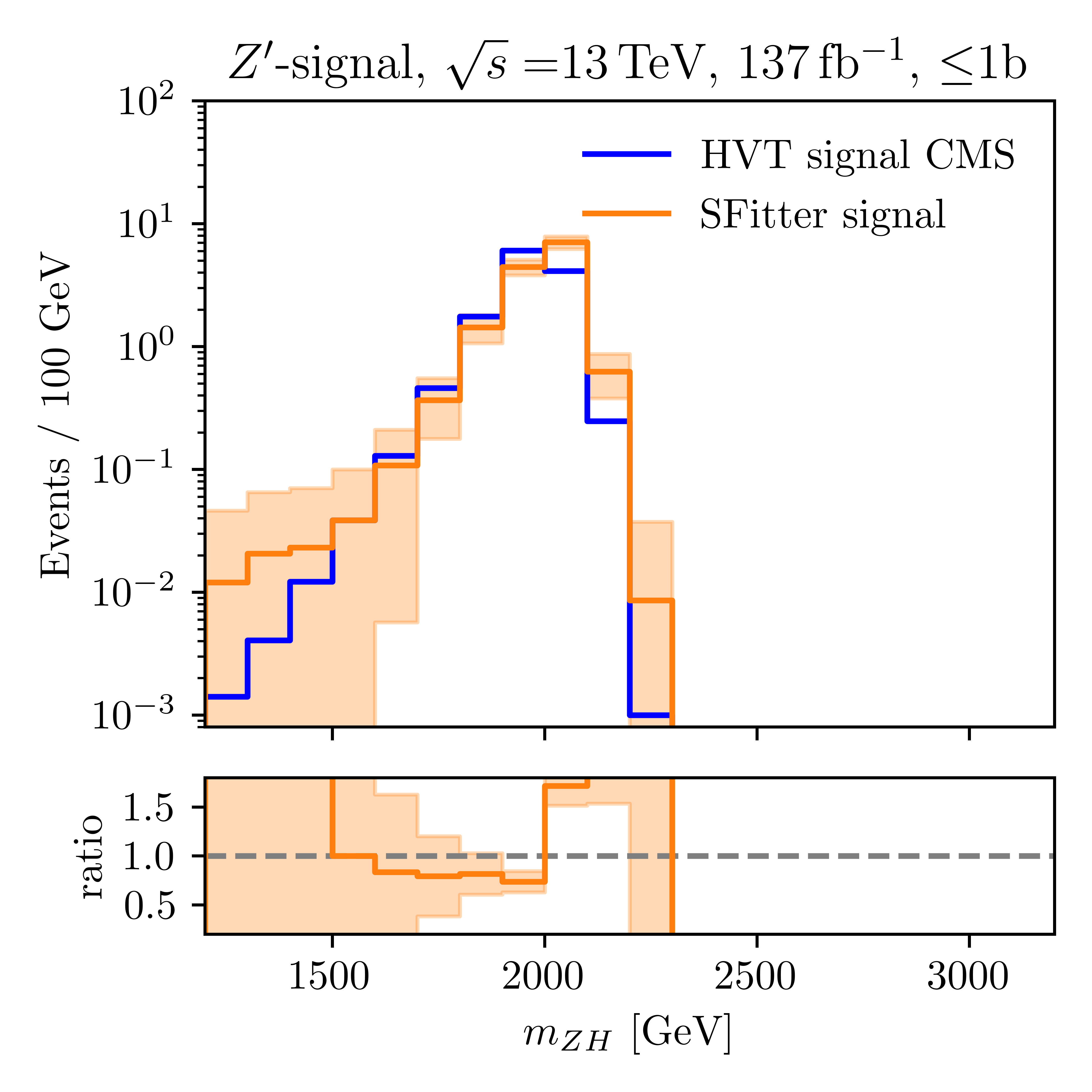}\hspace{5em}
\caption{Left: measured $m_{ZH}$ distributions for the two $b$-tagging
  categories~\cite{CMS:2021fyk}. Right: comparison between the
  $Z^{\prime}$ signal quoted by CMS and our estimate.  The orange
  bands show the statistical uncertainty from the Monte Carlo
  generation.}
\label{fig:CMS_comp}
\end{figure}

The second boosted $VH$ analysis we re-interpret in terms of SMEFT is
a CMS resonance search in the process~\cite{CMS:2021fyk}
\begin{align}
  pp \to ZH \to e^+ e^- \; b \bar{b} \; .
\end{align}
We include the non-VBF category with $\leq$1 $b$-tags and with two
$b$-tags.  We find that the two-$b$ category is more constraining than
the $\leq 1b$ category. This can happens because 
the relative size of the SMEFT correction prefers this category.
To determine the number of $b$-tags in an event, we look at the
corresponding fat jet and the number of $b$-quarks inside the jet.

We validate our analysis simulating events for $Z'$ peak in the heavy vector triplet model (HVT), that is used by CMS to illustrate a
possible signal,
\begin{align}
  pp \to Z' \to
  Z_{\ell \ell} H_{bb} \; .
\end{align}
This signal has the advantage that it is localized in $m_{ZH}$ and
simulated at leading order using Madgraph, which means it is easier to
use for calibration than a continuum background.  Again, we use
Madgraph, Pythia, FastJet, and Delphes with the standard CMS card at
leading order.  The combined sample is then compared to the
HVT peak shown in Fig.~\ref{fig:CMS_comp}.  We extract the
experimental efficiencies after scaling the invariant mass by the same
factor 1.05 for both categories.  The right panels in
Fig.~\ref{fig:CMS_comp} show the simulated $Z'$ signal for the two
categories, compared with the quoted CMS distributions.

The SMEFT signal in the $ZH$ channels is then computed using the same
efficiencies and the reweighting module in Madgraph. The $Z$-decays
are included in the Madgraph simulation, while the Higgs decays are
simulated in Pythia. As before, we ignore SMEFT effects on the
$t\bar{t}$ background.

\begin{figure}
\includegraphics[width=0.49\textwidth]{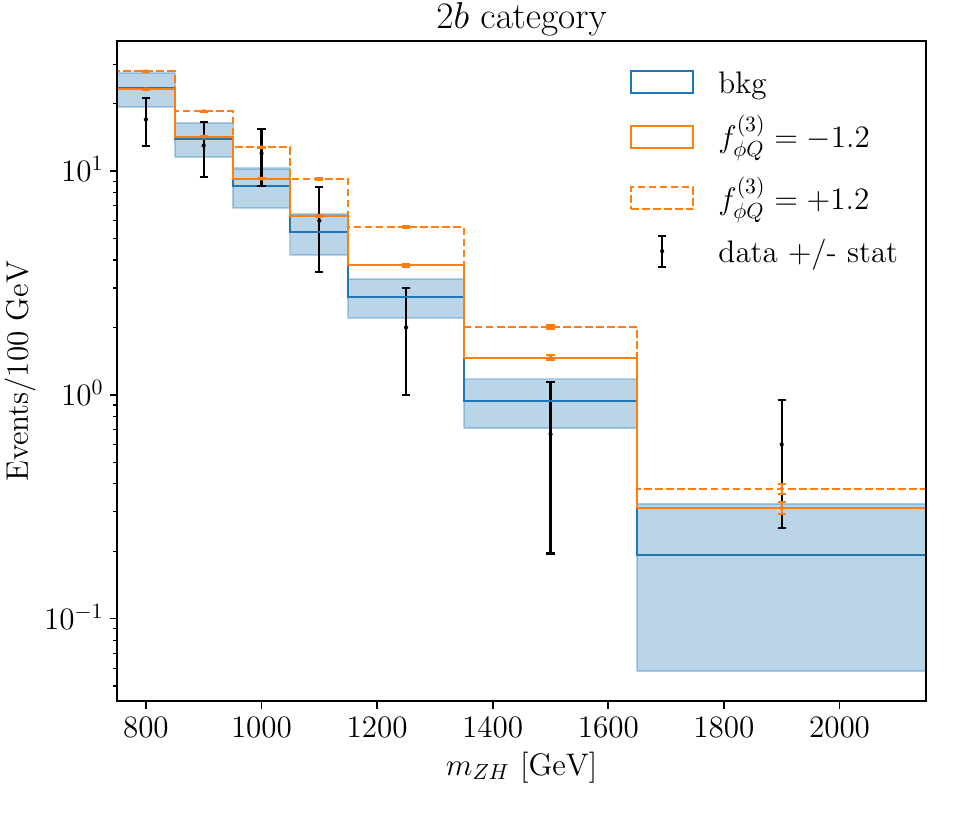}
\includegraphics[width=0.49\textwidth]{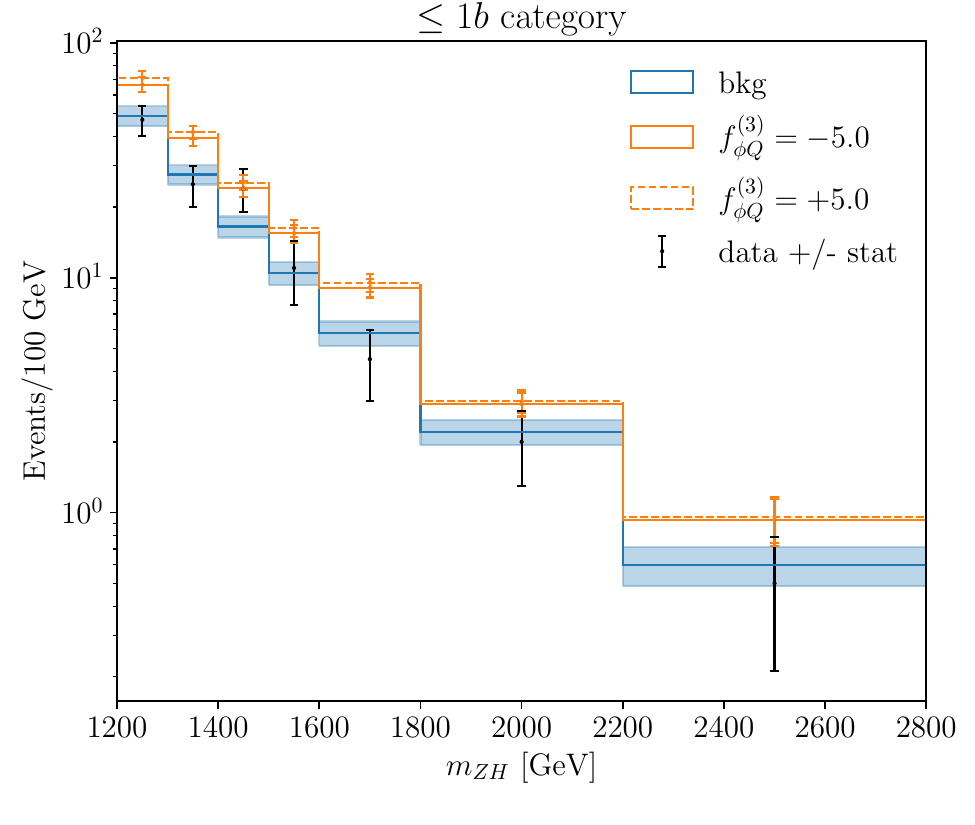}
\caption{Re-binned $m_{ZH}$ distributions for the $2b$ category (left)
  and the $\leq 1b$ category implemented in SFitter, including
  statistical and systematic uncertainties. We show the complete
  continuum background and the effect of a finite Wilson
  coefficient $f^{(3)}_{\phi Q}$.}
\label{fig:CMS_SFitter_results}
\end{figure}

Also for the CMS $ZH$ channel we need to re-bin the $m_{ZH}$
distribution to define a meaningful set of measurements, now with at
least two events per bin and separately for the two categories. The
results are shown in Fig.~\ref{fig:CMS_SFitter_results}.  For each bin
we include the systematic and statistical uncertainties from
Ref~\cite{CMS:2021fyk}. In addition, we include different theory uncertainties
per bin from the SMEFT prediction and event generation in Madgraph. As
discussed in detail for the ATLAS $WH$ analysis, we neglect the
correlation between bins.

\subsection{Boosted Higgs production}
\label{sec:new_boosted}

\begin{figure}[t]
\includegraphics[width=0.49\textwidth]{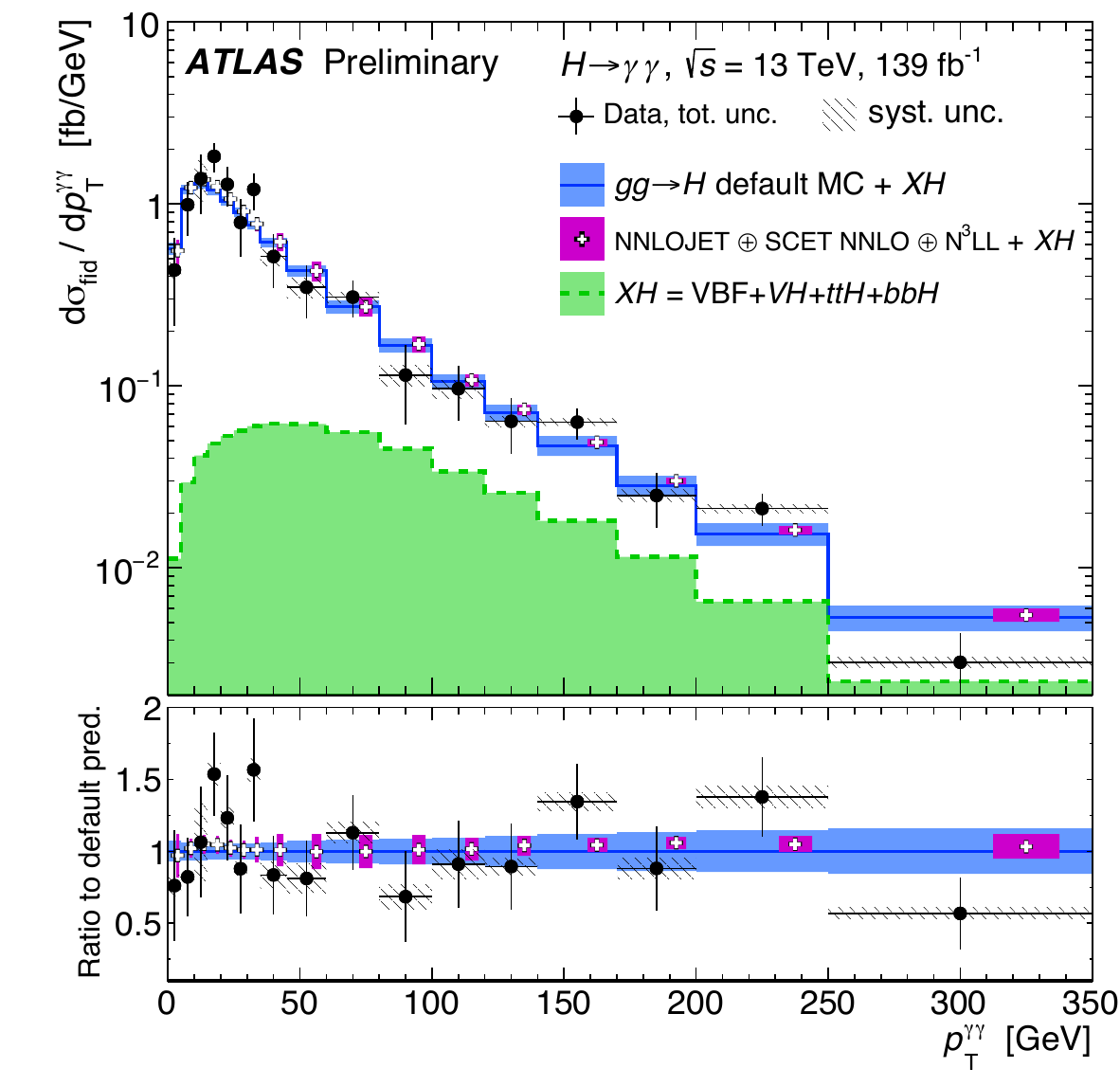}
\includegraphics[width=0.49\textwidth]{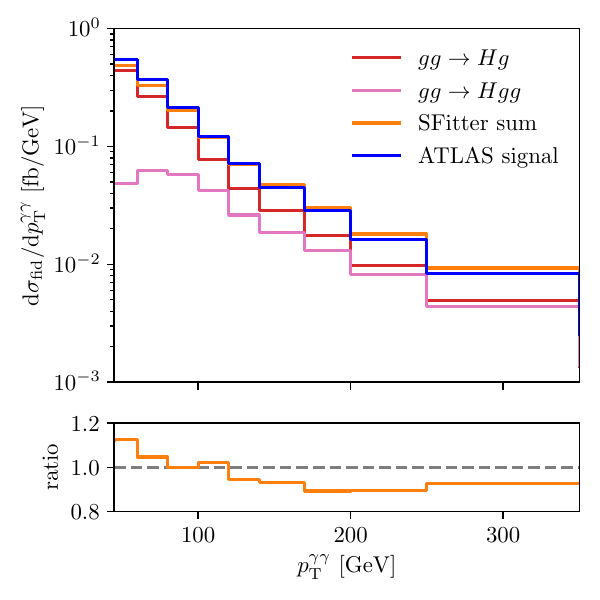}
\caption{Left: measured
  $\text{d}\sigma_\text{fid}/\text{d}p_\text{T}^{\gamma\gamma}$
  distribution~\cite{ATLAS:2019jst}. Right: comparison between the the
  ATLAS distribution and our SM estimate summing contributions from
  $gg\to Hg$ and $gg\to Hgg$.}
\label{fig:pTH_Comp}
\end{figure}

Boosted Higgs production, in association with one or more
hard jets,
\begin{align}
  pp \to H j(j) \; ,
\end{align}
has been known to distinguish between a top-induced Higgs-gluon-gluon
coupling and the corresponding dimension-6 operator for a long
time~\cite{Ellis:1987xu,Baur:1989cm}. It has therefore been suggested
as a channel to measure the dimension-6 Wilson coefficient $f_{GG}$ in
the presence of a modified top Yukawa coupling
$f_t$~\cite{Banfi:2013yoa,Azatov:2013xha,Harlander:2013oja,Grojean:2013nya},
where it competes with channels like the off-shell Higgs
production~\cite{Buschmann:2014twa,Buschmann:2014sia}.  In the SFitter
Higgs analysis it can be added to the set of measurements to provide
complementary information to the total Higgs production rate.  We take
the measurement of the Higgs $p_T$ distribution in the $\gamma\gamma$
channel by ATLAS~\cite{ATLAS:2019jst}.

The main contribution to boosted Higgs production comes from the
partonic channel $gg\to Hg$, with subleading corrections from $gg\to
Hgg$. This allows us to include SMEFT corrections to $gg\to Hg$
only. They can be separated into rescalings of the top Yukawa
coupling, for instance via $\ope_{u\phi,33}$, corrections to the
top-gluon coupling from $\ope_{tG}$, and the effective Higgs-gluon
interaction induced by $\ope_{GG}$.

Because these effective vertices enter also $t\bar{t}H$ production,
these operators lead to a non-trivial interplay in the global
analysis.  Moreover, as discussed in Sec.~\ref{sec:new_top} below,
$f_{tG}$ is well-constrained by top pair production $pp\to
t\bar{t}$. In fact, it constitutes the most significant contact
between global top and Higgs
analyses\cite{Ellis:2020unq,Ethier:2021bye}.\medskip

We calibrate the boosted Higgs analysis simulating the SM signal for
the partonic sub-channels $gg\to Hg$ and $gg\to Hgg$ using
Madgraph. The gluon-initiated channels are simulated at 1-loop, while
the quark-initiated one at tree level. For the one-loop simulation we
use a fixed renormalization scale $\mu_R=m_H$. This setup is also used
for the SMEFT simulations.  Figure~\ref{fig:pTH_Comp} shows the
comparison between our simulation and the SM signal estimate
provided by ATLAS.  We use the same binning as in the original
distribution, but omit the bins with $p_{T,\gamma \gamma} < 45$~GeV.\medskip

The simulation of SMEFT effects is tackled with different methods. The
effect of a shifted top Yukawa is just a rescaling of the SM cross
section, that can be easily computed analytically,
\begin{align}
 \frac{\sigma_\text{SMEFT}}{\sigma_\text{SM}} = \left(1 
  - \frac{f_t}{\sqrt2}\; \frac{v^2}{\Lambda^2}\right)^2\; .
\end{align}

Second, $\ope_{tG}$ also enters the top loops, but induces a different
Lorentz structure compared to the SM amplitude. Its contributions are
simulated independently using SMEFT\@@NLO~\cite{Degrande:2020evl} in
Madgraph. In the event generation, the EFT operator is renormalized at
$\mu_\text{EFT}=\mu_R =m_H$.

Finally, $\ope_{GG}$ enters at the tree level. Because the pure
interference between tree and loop diagrams cannot be generated
directly in Madgraph, we choose to simulate both the linear and the
squared term with a modified {\tt loop\_sm} UFO model, where the
point-like Higgs-gluon vertices are mimicked by sending the bottom
quark mass and Yukawa coupling to 15~TeV. We verified that any value
larger than 10~TeV gives equivalent results. This way the simulation
is formally at one loop for all terms. The results of this
approximation were cross-checked against the analytic results in
Refs.~\cite{Ellis:1987xu,Baur:1989cm} for the interference and against
the tree-level simulation for the pure square.

The mixed quadratic terms, \ie the interferences between two
operators, can be computed analytically for the combination of
$\ope_{tG}$ or $\ope_{GG}$ with a shifted Yukawa coupling.  The
combination of $f_{tG}$ and $f_{GG}$ needs to be simulated
independently, in our case using using SMEFT\@@NLO and the reweighting
module in Madgraph.

\begin{figure}[t]
\centering
\includegraphics[width=0.49\textwidth]{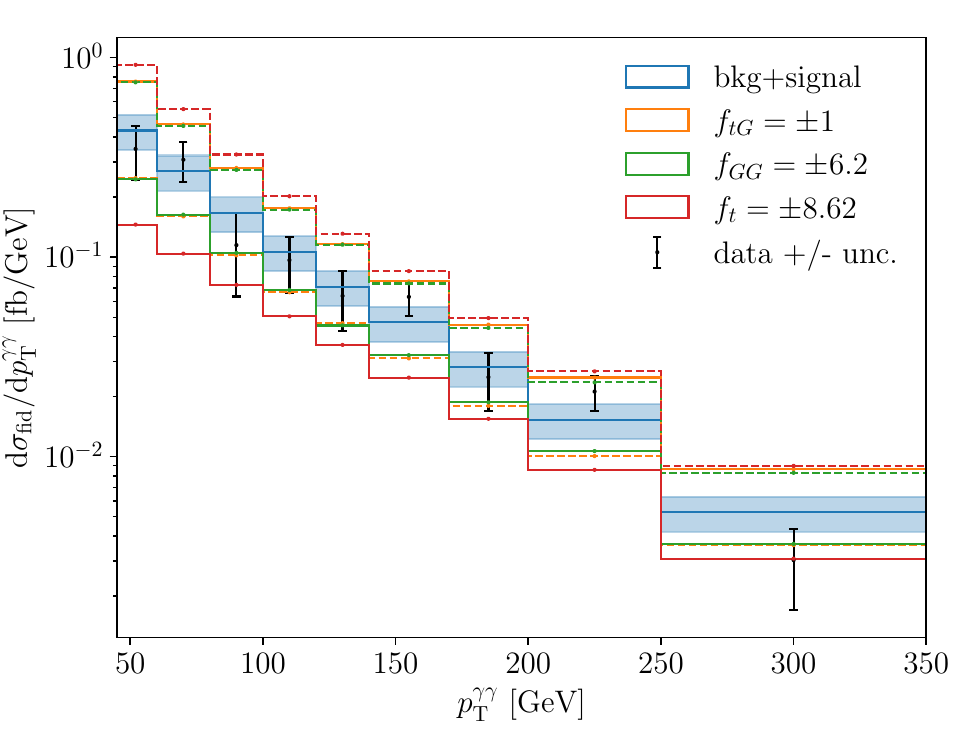}
\caption{Reconstructed $p_{T,H}$ distribution implemented in SFitter,
  including statistical and systematic uncertainties as well as
  additional uncertainties on our prediction. We show the complete
  continuum of signal and background and the effect of three finite
  Wilson coefficients $f_{t}$, $f_{t G}$ and $f_{GG}$. The negative
  values are represented by dashed lines and the positive values
  by solid lines.}
\label{fig:pTH_SFitter_results}
\end{figure}

In Fig.\ref{fig:pTH_SFitter_results} we show the impact of four
relevant SMEFT coefficients on the kinematic distribution we implement
in SFitter.  For each bin we include the systematic and statistical
uncertainties from Ref~\cite{ATLAS:2019jst}, as well as an additional
$20\%$ theory uncertainty reflecting the scale uncertainty on the
SMEFT prediction.

\subsection{From the top}
\label{sec:new_top}

\begin{figure}[b!]
\centering
\includegraphics[width=0.45\textwidth]{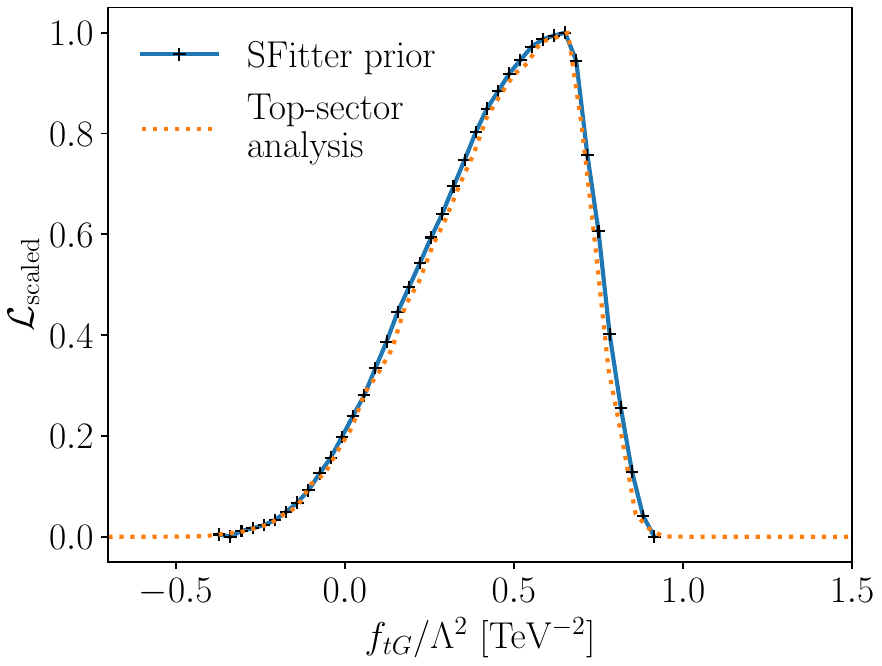}
\caption{Profile likelihood for $f_{tG}$ from the SFitter top-sector
  analysis~\cite{Brivio:2019ius}.}
\label{fig:ctg_prior}
\end{figure}

From the combined top-Higgs
analyses~\cite{Ellis:2020unq,Ethier:2021bye} we know that the
Higgs-gauge sector and the top sector cannot be treated completely
independently. The two operators
\begin{align}
  \ope_{u\phi,33} &= \phi^\dagger\phi  \; \bar Q_3 \tilde \phi u_{R,3} 
  \qquad \text{and} \qquad 
  \ope_{tG} = ig_s (\bar{Q}_3 \sigma^{\mu \nu} T^A u_{R,3} ) \; \tilde{\phi} G^A_{\mu \nu} 
\end{align}
contribute to top pair and associated $t\bar{t}H$ production and are,
at the same time, crucial to interpret gluon-fusion Higgs production,
together with the Higgs-related operator $\ope_{GG}$, as discussed
above. By the definition of top-sector and Higgs-sector SMEFT analyses
in SFitter, $\ope_{tG}$ is covered by the top analysis, while we keep
$\ope_{u\phi,33}$ as part of the Higgs analysis, together with a
complete treatment of $t\bar{t}H$ production. This means we can
include the limits on $f_{tG}$ from the dedicated SFitter analysis of
the top sector~\cite{Brivio:2019ius} using its 1-dimensional profile
likelihood.  We implement these constraints as an external measurement
or prior. The corresponding profile likelihood is shown in
Fig.~\ref{fig:ctg_prior}. It consists of 100 data points which are
dense enough that we can linearly interpolate between them.

We choose the range in $f_{tG}$ to cover extremely small
log-likelihoods, to avoid numerical issues in the combined
analysis. Still, while it is very unlikely to occur, we also want to
describe points outside of this range, so we extrapolate the
log-likelihood further with two quadratic fits; one fitted to negative
Wilson coefficients and one fitted to positive Wilson coefficient. A
quadratic fit in this context means exponentially suppressed Gaussian
tails.

\subsection{Rates and signal strengths}
\label{sec:new_rate}

\begin{table}[b!]
\centering
\begin{tabular}{cl|cc} 
 \toprule
 Production & Decay & ATLAS & CMS \\ 
 \midrule
 All & $H\rightarrow\gamma\gamma$ & \cite{arxiv.2207.00348} & \cite{Sirunyan_2021} \\
 \hline
 $ZH$ & $H\rightarrow\text{inv}$ & \cite{Aad_2022} & \cite{Sirunyan:2727805} \\
 VBF (ggF, $VH$) & $H\rightarrow\text{inv}$ & \cite{arxiv:2202.07953} & \\
 VBF (ggF, $ZH$, $t\bar{t}H$) & $H\rightarrow\text{inv}$ & & \cite{CMS-HIG-20-003} \\
 \hline
 All & $H\rightarrow\tau\tau$ & \cite{arxiv.2201.08269} & \\
 $VH$ & $H\rightarrow\tau\tau$ & & \cite{arxiv.2204.12957} \\
 \hline
 ggF, VBF & $H\rightarrow WW$ & \cite{ATLAS-CONF-2021-014} & \\ 
 ggF, VBF, $VH$ & $H\rightarrow WW$ & & \cite{arxiv.2206.09466} \\
 \hline
 $WH$, $ZH$ & $H\rightarrow b\bar{b}$ & \cite{Aad:2723187} & \\ 
  \hline
 ggF, VBF ($VH$, $ttH$) & $H\rightarrow \mu\mu$ & & \cite{Tumasyan_2021} \\ 
  \bottomrule
\end{tabular}
\caption{List of the new Run~2 Higgs measurements included in this
  analysis, we denote $V = W,Z$.}
\label{tab:HiggsMeas}
\end{table}

In addition to the new kinematic measurements above, we update the set
of Higgs rate measurements of Ref.~\cite{Biekoetter:2018ypq}, adding
those listed in Tab.~\ref{tab:HiggsMeas}. The two $H \to \tau \tau$
and three out of four $H \to \text{inv}$ measurements are completely
new constraints, while the others update results included in our
previous analysis.  The first column indicates which production
channels were implemented in SFitter. We do not always use all the
channels covered in a given ATLAS or CMS paper, if some of them are
clearly subleading or some of them appear impossible to implement in
the necessary details. Production channels in parentheses are
numerically subleading, but were retained nevertheless.

The systematic and statistical uncertainties of the new measurements
are typically smaller compared to the older ones. On the other hand,
we attempt a more comprehensive and conservative estimate of the
theory uncertainties, given the available information.  In
Ref.~\cite{Biekoetter:2018ypq} we typically discarded many theory
uncertainties on the signal quoted in the actual papers and replaced
them with the leading uncertainty on the complete signal prediction
from the HXSWG~\cite{HXSWGReport,HXSWGWebXS,HXSWGWebBR}, added
linearly as expected for uncorrelated flat uncertainties combined by
profiling.  In our new, comprehensive treatment, all theory
uncertainties quoted by the analyses are retained. We include them
separately and combine them. In addition, we include the uncertainties
reported by the HXSWG~\cite{HXSWGReport,HXSWGWebXS,HXSWGWebBR} as the
uncertainty on our SFitter prediction, again split by contribution and
ready to be profiled over or marginalized.\medskip

We illustrate the implementation procedure in some more detail only
for the recent Run-2 $H\rightarrow WW$ analysis by
CMS~\cite{arxiv.2206.09466}.  Among the results presented, we
implement the four signal strength measurements. Because they are
reported for individual production modes (and not only in the STXS
binning), they can be directly compared to the known expressions for
Higgs production rates in the SMEFT, without re-deriving. These have
been long implemented in SFitter for the main Higgs production
channels (ggF, VBF, $WH$, $ZH$, $ttH$) and decays ($b\bar{b}$, $WW$,
$gg$, $\tau\tau$, $ZZ$, $\gamma\gamma$, $Z\gamma$, $\mu\mu$).  A
re-derivation of the SMEFT expression can also be avoided in cases
where the final results are not given for specific production
channels, but the expected signal contribution from each production
channel is provided.\medskip

The key ingredient to SFitter is a detailed breakdown of all
uncertainties.  This is crucial in order to obtain the best possible
approximation of the full experimental likelihood.  For
Ref.~\cite{arxiv.2206.09466} we consider different uncertainties for
each production channel, that are reported in the paper and in the
corresponding HepData entry.

The statistical uncertainty is taken from the experimental paper,
symmetrized and implemented as Poisson or Gaussian distribution.  For
experimental systematics, SFitter provides 31 predefined categories of
Gaussian uncertainties, correlated across measurements and, where
appropriate, across experiments. All uncertainties belonging to the
same category are added in quadrature.  The categories used to
implement the CMS analysis cover luminosity, detector effects, lepton
reconstruction, and b-tagging. Detector effects combine the jet energy
scale and resolution uncertainties, as well as the missing transverse
momentum scale uncertainty.  Whenever the experimental papers quote
significant uncertainties that do not fit any predefined category, we
add them as an uncorrelated Gaussians, but this is not the case for
the analysis of Ref.~\cite{arxiv.2206.09466}.

Theoretical uncertainties are typically implemented with flat
uncorrelated likelihoods. One exception is the Monte Carlo statistics
uncertainty, which we usually treat as an uncorrelated Gaussian.  The
CMS analysis quotes five theoretical uncertainties, that are all
introduced independently.  In addition, we have six theoretical
uncertainties on the SFitter prediction: three on the production rate
and three on the decay branching ratio, following the HXSWG
prescription~\cite{HXSWGReport,HXSWGWebXS,HXSWGWebBR}.

As a final step we compare the systematic uncertainties quoted on the
final result with the sum of the uncertainties implemented in
SFitter. If we are missing information for example on the
correlations, our implementation might not be conservative, so we
introduce an additional uncorrelated Gaussian uncertainty to
compensate. This happens for the CMS reference analysis in the $ZH$
channel.  For this measurement we implement two uncorrelated
Gaussian uncertainties, three correlated Gaussian uncertainties, plus the
eleven flat uncertainties.
 
\section{Global SFitter analysis}
\label{sec:fit}

After validating the marginalization technique in SFitter and
introducing a set of promising new observables, we can provide the
final global analysis of the Higgs and electroweak sector after Run~2,
including the leading link to the top sector.  To be conservative, we
will compare all our results with a profile likelihood treatment. We
will find and explain differences of the two methods facing the same
extended dataset.

\subsection{Marginalization vs profiling complications}
\label{sec:fit_bayes}

While in Sec.~\ref{sec:bayes} we have found that for the dataset of
Ref.~\cite{Biekoetter:2018ypq} the marginalization and profiling
approaches lead to, essentially, identical results, one analysis
implemented in SFitter as part of Ref.~\cite{Brivio:2021alv} actually
leads to significant differences.  The data driving this separation of
profiling and marginalization is the $m_{WW}$ distribution measured by
ATLAS~\cite{Aad:2020tps}, shown in the left panel of
Fig.~\ref{fig:impl_vv_chq3}. It has the unique feature of a sizeable
under-fluctuation in the last bin.

Such an under-fluctuation is challenging to accommodate in the SMEFT. First,
under-fluctuations can only be explained by operators with large
interference terms, where the Wilson coefficients have to be carefully
tuned to be large enough to explain a sizeable effect and small enough
to not be dominated by dimension-6 squared contributions. Second, a
localized under-fluctuation in only one bin of one kinematic
distribution requires a subtle balance of several Wilson coefficients, to
control all other bins in all other di-boson and $VH$ channels.

\begin{figure}[t]
  \includegraphics[width=.33\textwidth]{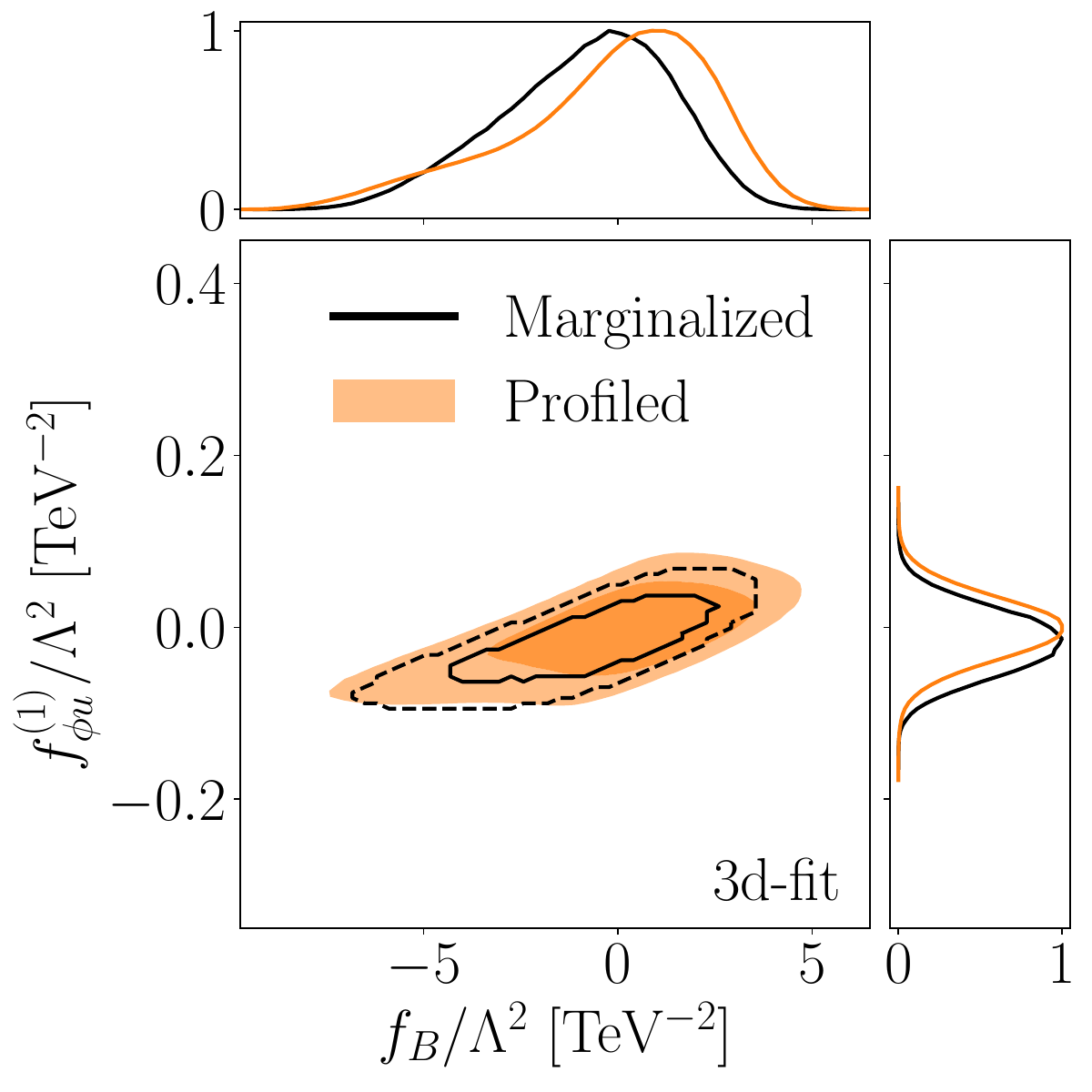}
  \includegraphics[width=.33\textwidth]{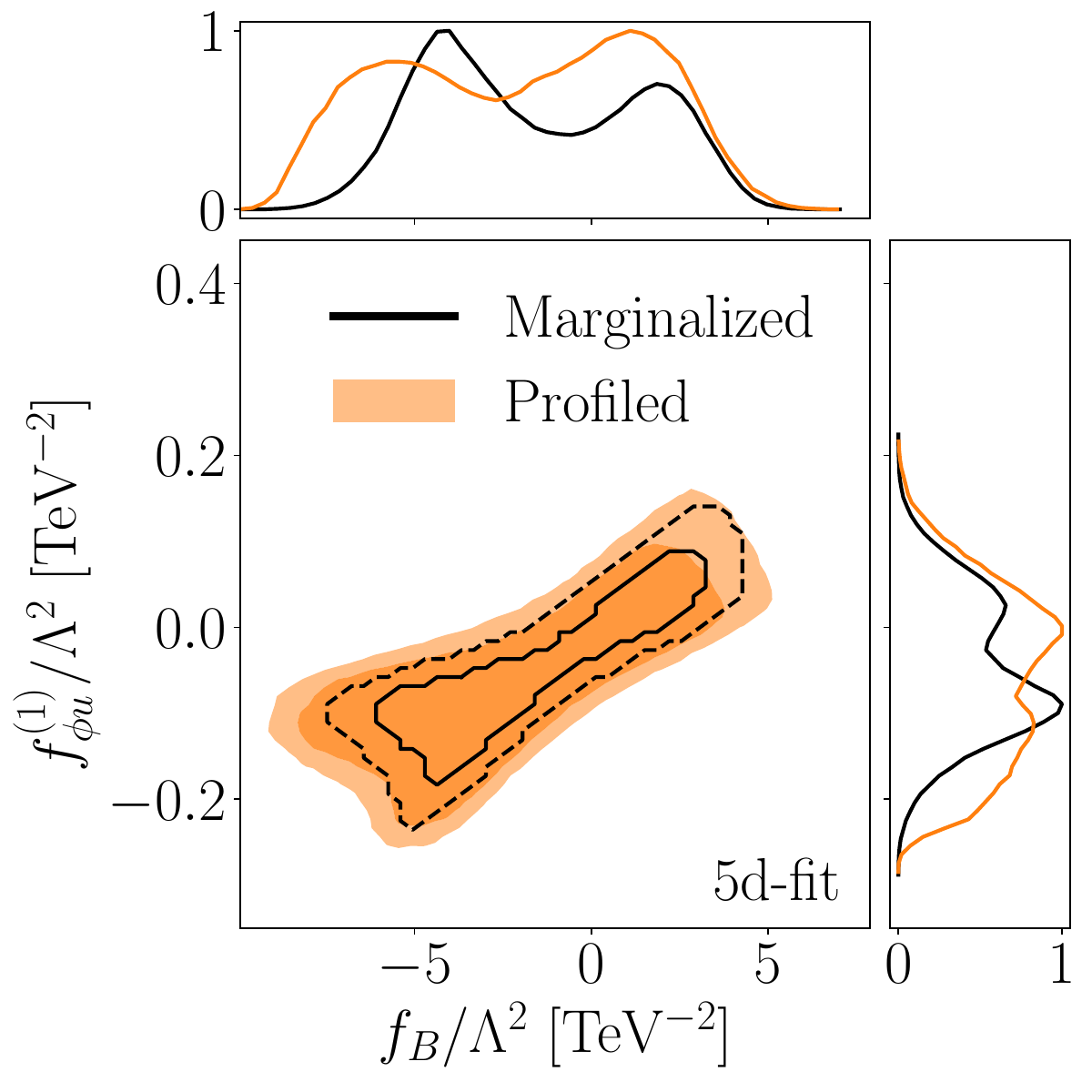}
  \includegraphics[width=.33\textwidth]{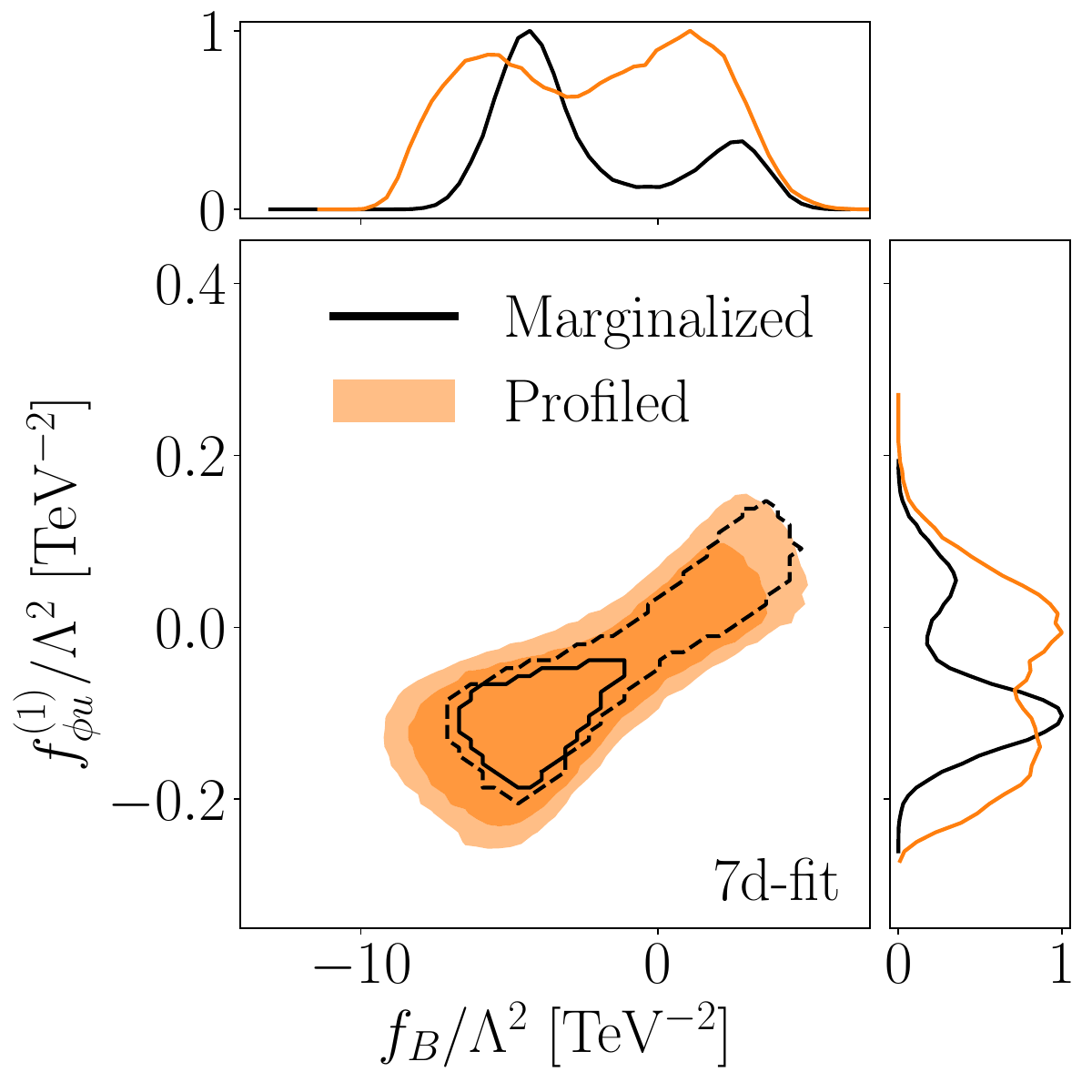}
  \caption{SFitter analysis with different SMEFT models describing the
    full Run~2 dataset, including the boosted $WW$ production.}
  \label{fig:doubleMin_2d}
\end{figure}

In Fig.~\ref{fig:doubleMin_2d} we show low-dimensional analyses of the full
Run~2 dataset including the $WW$ kinematics shown in
Fig.~\ref{fig:impl_vv_chq3}, constraining three, five and seven Wilson
coefficients.  For the three parameters $\{ f_B, f_{\phi u}^{(1)}, f_W
\}$ we see that the maximum of the likelihood is perfectly compatible
with the SM. The reason is that the SMEFT model is not flexible enough
to accommodate the under-fluctuation, so we only encounter the issue
when we look at the value of the likelihood in the maximum. Adding
first $\{ f^{(1)}_{\phi Q}, f^{(3)}_{\phi Q}\}$ and then $\{ f^{(1)}_{\phi d}, f_{3W} \}$ to the SMEFT model
allows us to accommodate the under-fluctuation, leading to a second
likelihood maximum.

When we compare the two likelihood maxima, differences between the
profiling and the marginalization appear. By definition, the
profile likelihood identifies the most likely parameter point, which
according to Fig.~\ref{fig:doubleMin_2d} is close to the SM point, $f_B
\approx 0 \approx f_{\phi u}^{(1)}$. This does not change when we
increase the operator basis or expressivity of the SMEFT model. The
marginalization adds volume effects in the space of Wilson
coefficients, and they increasingly prefer the non-SM maximum once the
SMEFT model is flexible enough to explain the
under-fluctuation. Consequently, the marginalized analysis proceeds to
challenge the SM in favor of an alternative SMEFT parameter
point.

\subsection{Full analysis}
\label{sec:fit_full}


\begin{figure}[t]
  \includegraphics[width=.33\textwidth, page=4]{21d_fit_with_everything_bayes}
  \includegraphics[width=.33\textwidth, page=5]{21d_fit_with_everything_bayes}
  \includegraphics[width=.33\textwidth, page=15]{21d_fit_with_everything_bayes}\\
  \includegraphics[width=.33\textwidth, page=21]{21d_fit_with_everything_bayes}
  \includegraphics[width=.33\textwidth, page=20]{21d_fit_with_everything_bayes}
  \includegraphics[width=.33\textwidth, page=1]{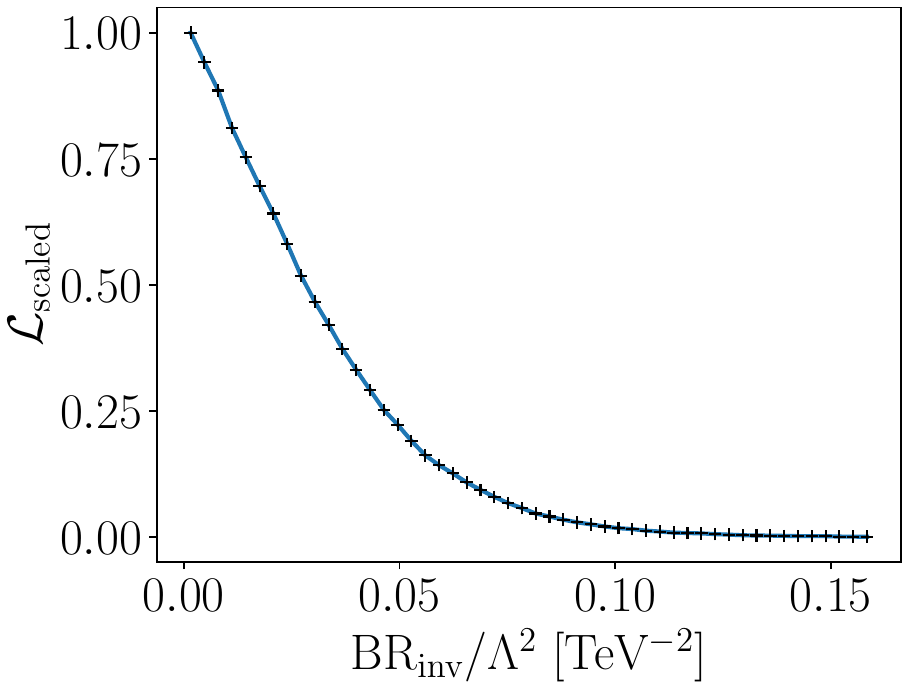}
  \includegraphics[width=.33\textwidth, page=2]{21d_fit_with_everything_bayes}
  \includegraphics[width=.33\textwidth, page=9]{21d_fit_with_everything_bayes}
  \includegraphics[width=.33\textwidth, page=18]{21d_fit_with_everything_bayes}\\
  \caption{Set of marginalized likelihoods for the 21-dimensional
    SFitter analysis including the full set of measurements.}
  \label{fig:full_1d}
\end{figure}

\begin{figure}[b!]
  \includegraphics[width=.33\textwidth, page=1]{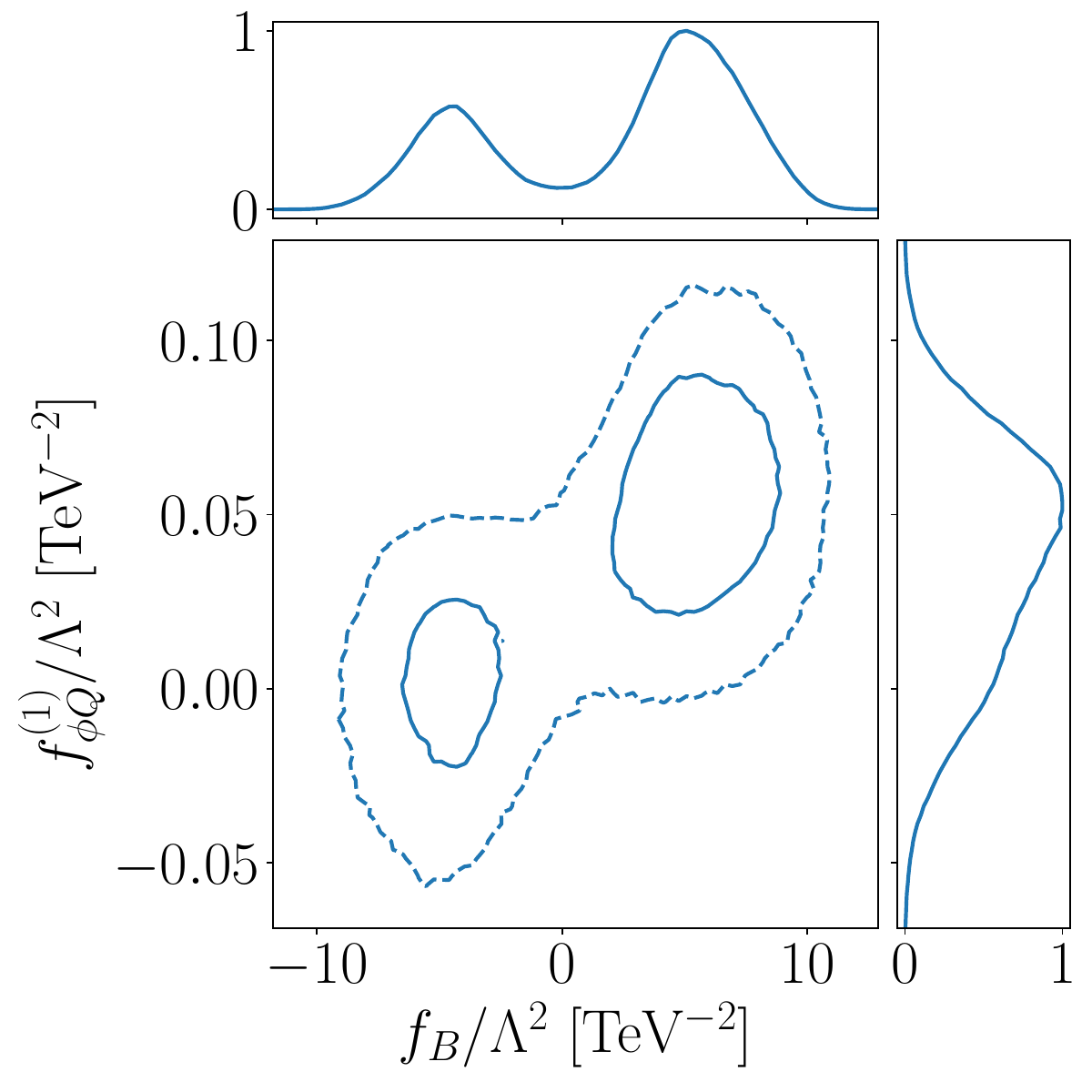}
  \includegraphics[width=.33\textwidth, page=2]{21d_fit_bayes_relevant_correlations_v2}
  \includegraphics[width=.33\textwidth, page=3]{21d_fit_bayes_relevant_correlations_v2}
  \caption{Set of marginalized correlations for the 21-dimensional
    SFitter analysis including the full set of measurements. The solid and dashed lines show $\Delta\chi^2=2$ and 7 respectively.}
\label{fig:full_2d}
\end{figure}

After identifying and understanding the issue with marginalized
likelihoods for the updated dataset we now perform the full,
21-dimensional parameters analysis on all available data. The theory
framework is defined by the Lagrangian in Eq.\eqref{eq:ourlag}. The
dataset consists of all measurements from
Ref.~\cite{Biekoetter:2018ypq}, combined with the new and updated
channels described in Sec.~\ref{sec:new}. We will discuss the standard
profile likelihood results below, in a first step we focus on the
marginalization. In Fig.~\ref{fig:full_1d} we show a set of
1-dimensional marginalized likelihoods. In the first row we show three
Wilson coefficients affected by the under-fluctuation in $m_{WW}$, as
discussed in the previous Sec.~\ref{sec:fit_bayes}. While the
marginalized likelihood for $f_W$ follows a standard single-mode
distribution, those for $f_B$ and $f_{\phi u}^{(1)}$, for example,
show two distinct modes accommodating the observed under-fluctuation.

In the second row we show the alternative maximum in $f_+$ we already
observed for the dataset from Ref.~\cite{Biekoetter:2018ypq} and which
we discuss in Fig.~\ref{fig:second_mode_1dplots} of
Sec.~\ref{sec:bayes}. For the final SFitter result we will remove the
second maximum as an expansion around the wrong SMEFT limit. We also
see that the invisible Higgs width is strongly constrained, even after
we account for a modified Higgs production process rather than
assuming SM Higgs production combined with the exotic invisible Higgs
decay.

In the last row we show the effect of including $\ope_{tG}$ in the
Higgs analysis.  Comparing the limit on $f_{tG}$ to its prior in
Fig.~\ref{fig:ctg_prior} we see that this parameter gains essentially
nothing from the Higgs measurements, but it will broaden the limits on
the correlated parameter $f_{GG}$ affecting gluon-fusion Higgs
production.\medskip

\begin{figure}[t]
  \includegraphics[width=.33\textwidth, page=5]{21d_fit_with_and_without_WW_res_search_bayes}
  \includegraphics[width=.33\textwidth, page=15]{21d_fit_with_and_without_WW_res_search_bayes}
  \includegraphics[width=.33\textwidth, page=1]{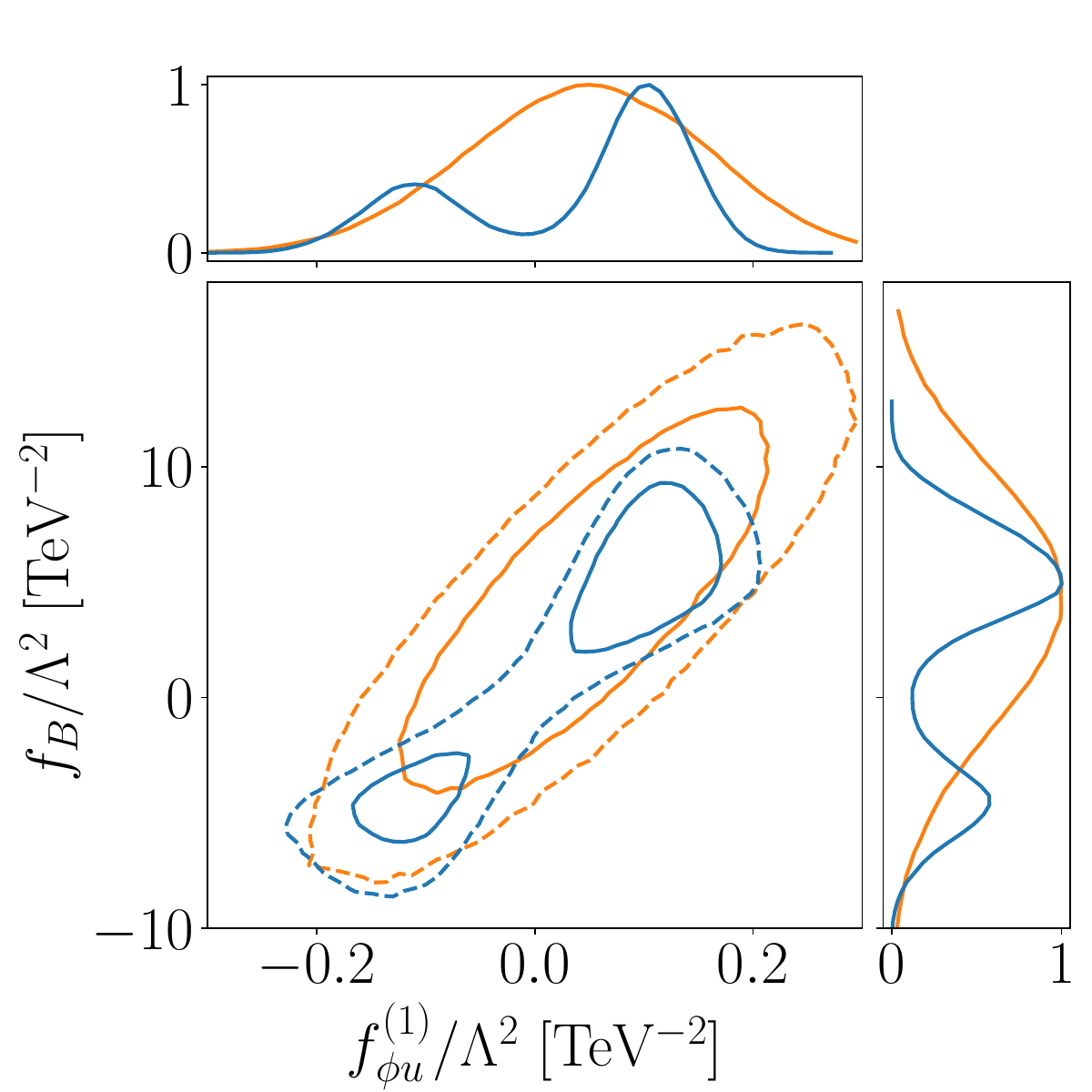} \\
  \includegraphics[width=.33\textwidth, page=4]{21d_fit_with_and_without_WW_res_search_bayes}
  \includegraphics[width=.33\textwidth, page=16]{21d_fit_with_and_without_WW_res_search_bayes}
  \includegraphics[width=.33\textwidth, page=8]{21d_fit_with_and_without_WW_res_search_bayes}
  \caption{Set of marginalized likelihoods for the 21-dimensional
    SFitter analysis with and without the ATLAS $WW$ resonance
    search altogether.}
  \label{fig:full_no_ww}
\end{figure}

To follow up on the discussion of Fig.~\ref{fig:doubleMin_2d} we show a
more complete set of 2-dimensional marginalized likelihoods related to
the $m_{WW}$ under-fluctuation in Fig.~\ref{fig:full_2d}. In the full
analysis the correlation does not just affect $f_{\phi u}^{(1)}$, but
the full range of gauge-fermion operators. This is expected from the
argument that we need to carefully tune many Wilson coefficients to
accommodate a deviation in a single di-boson process in a single bin of
the high-invariant-mass distribution. As mentioned before, the
apparent signal for physics beyond the Standard Model is an artifact
of the marginalization and its volume effects, and cannot be
reproduced with the profile likelihood. Note that this does not mean
the marginalization is wrong or wrongly done, this difference just
reflects the two methods asking different questions.

To study the impact of the critical $WW$-resonance analysis on our
global analysis we show a set of marginalized likelihoods with and
without this analysis, \ie with and without the entire $m_{WW}$
distribution. Obviously, removing this distribution also removes the
secondary maximum structure, as we immediately see in
Fig.~\ref{fig:full_no_ww}. Removing the entire distribution replaces
the marginalized likelihoods for $f_{B}$ and $f_{\phi u}^{(1)}$ by
their broad envelopes, still correlated, but without the distinctive
maxima. For $f_W$ the additional observable has limited impact, for
$f_{\phi d}^{(1)}$ is leads to a smaller uncertainties combined with a
shifted maximum, and for $f_{3W}$ the
$WW$-analysis provides key information.

\begin{figure}[b!]
  \includegraphics[width=.33\textwidth, page=5]{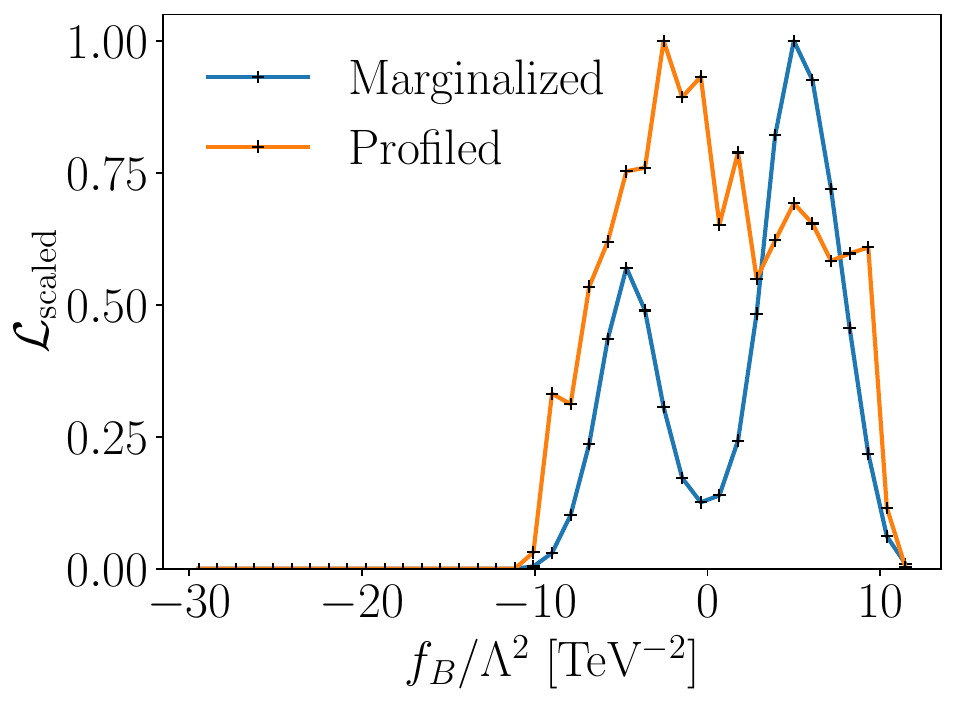}
  \includegraphics[width=.33\textwidth, page=15]{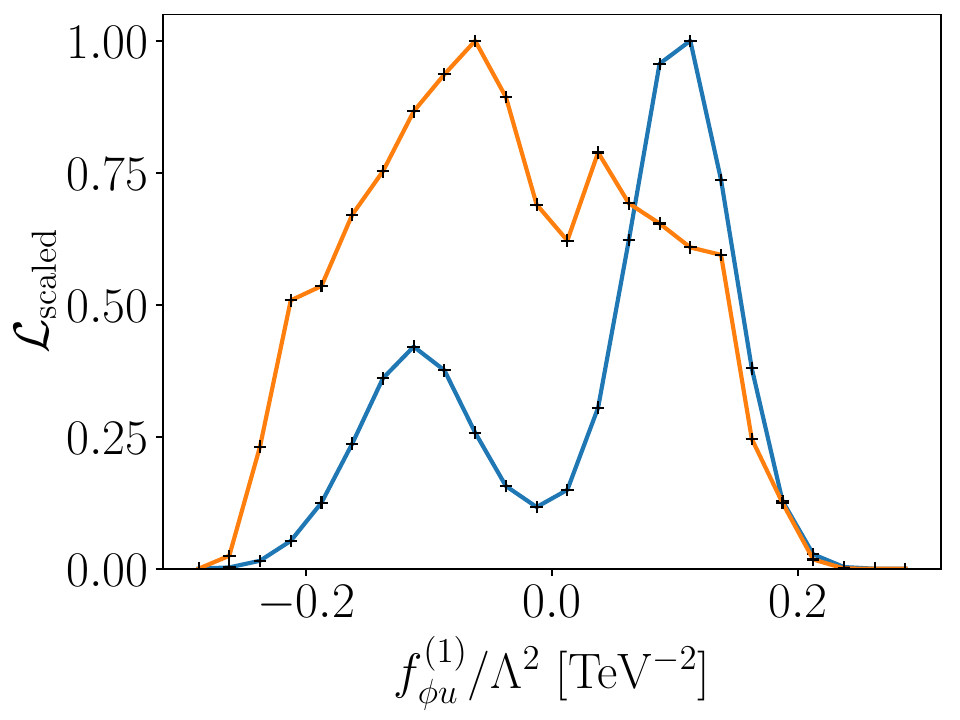}
  \includegraphics[width=.33\textwidth, page=4]{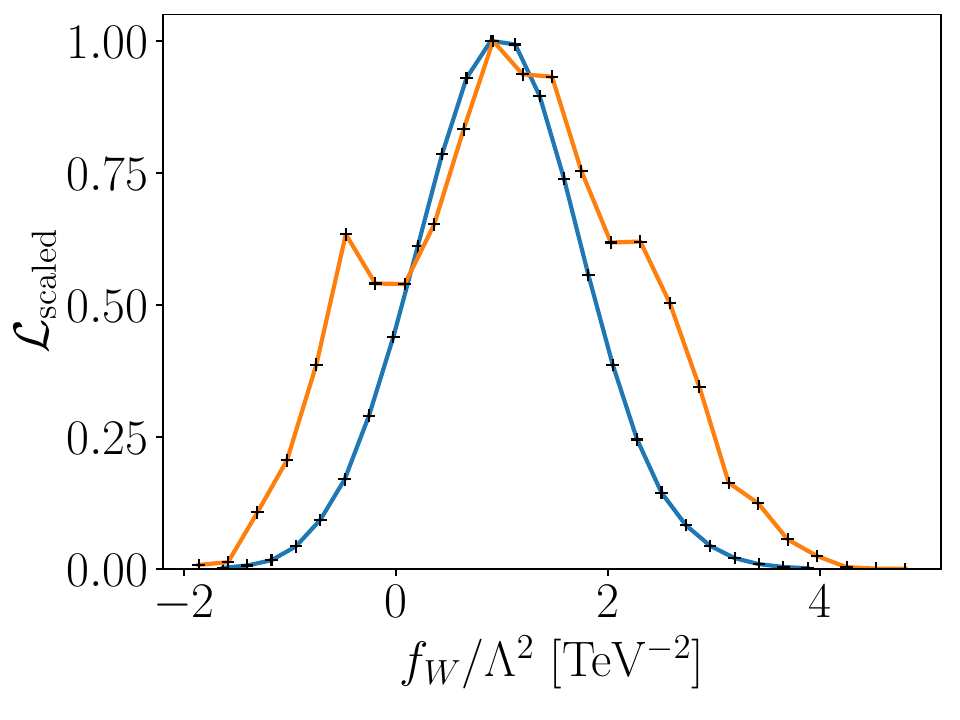}
  \caption{Set of marginalized and profiled likelihoods for the
    21-dimensional SFitter analysis with the ATLAS $WW$ resonance
    search.}
  \label{fig:full_ww_profile}
\end{figure}

\begin{figure}[t]
  \includegraphics[width=.33\textwidth, page=2]{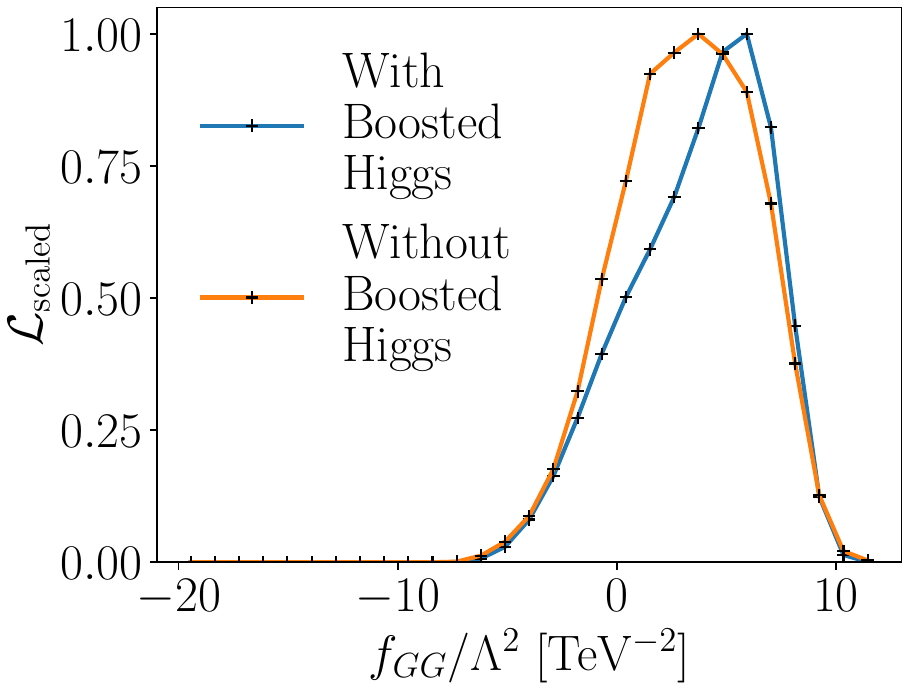}
  \includegraphics[width=.33\textwidth, page=9]{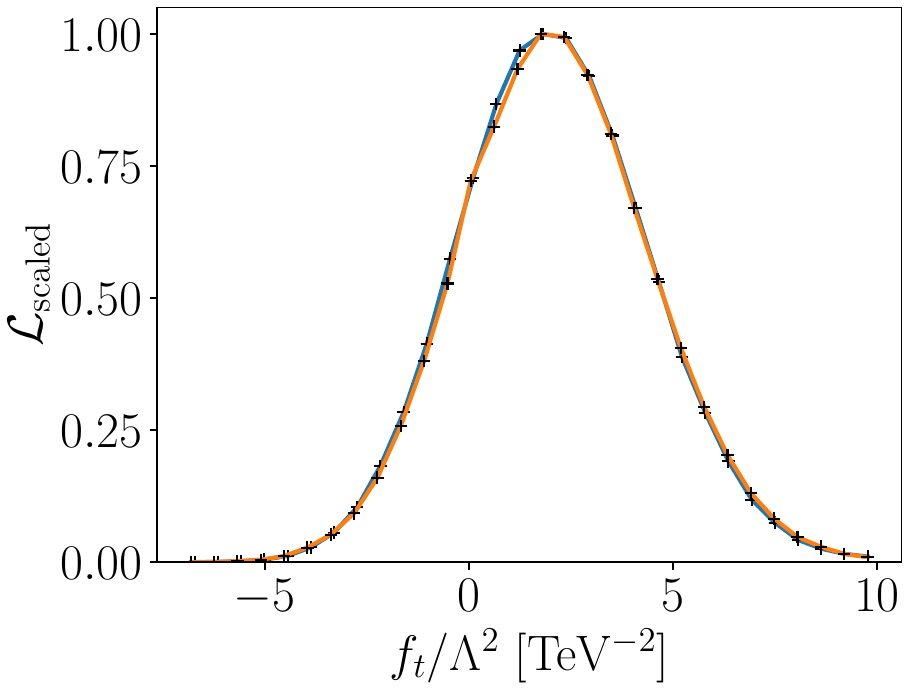}
  \includegraphics[width=.33\textwidth, page=18]{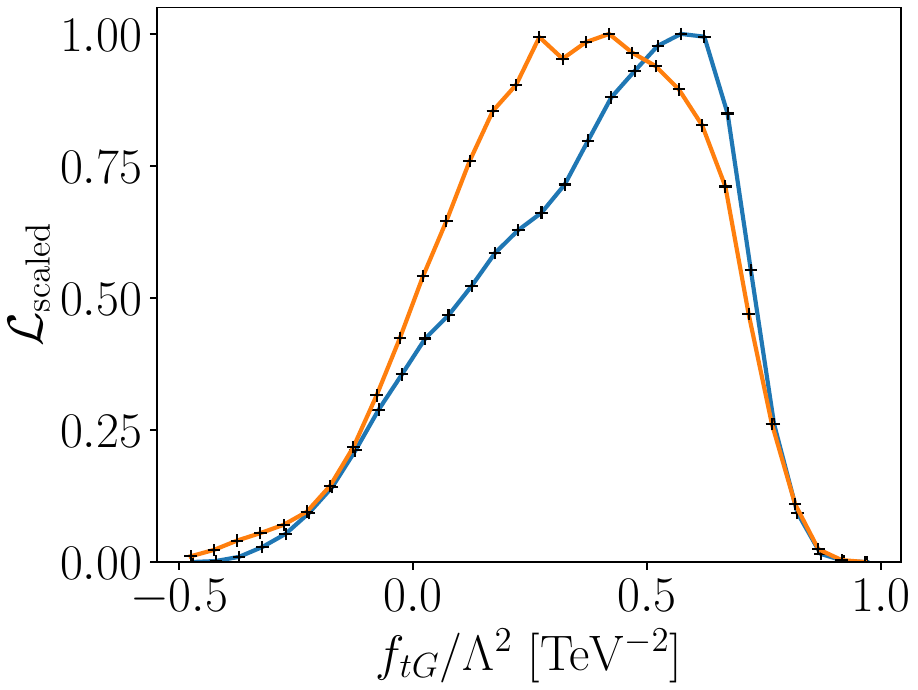}
  \caption{Set of marginalized likelihoods for the 21-dimensional
  SFitter analysis with and without the boosted Higgs analysis.}
  \label{fig:full_tG}
\end{figure}

Finally, in Fig.~\ref{fig:full_ww_profile} we compare the
1-dimensional marginalized likelihoods with the corresponding profile
likelihoods for a set of Wilson coefficients. For $f_B$ and $f_{\phi
  u}^{(1)}$ we see the difference in the treatment of the secondary
likelihood maximum, while $f_W$ serves as an example for the many
parameters where the two methods give the same results, as discussed
in detail in Sec.~\ref{sec:bayes} and
Fig.~\ref{fig:comparison_freq_vs_bayesian_summary}. Indeed, the
results from the two methods only disagree when the likelihoods
develop secondary maxima.\medskip

Moving on with the effects observed in Fig.~\ref{fig:full_1d} we can
look at the top-Higgs sector with $f_{GG}$, $f_t$, and the added
$f_{tG}$. These three Wilson coefficients are constrained by the Higgs
production in gluon fusion, associated top-Higgs production, and top
pair production through the prior shown in
Fig.~\ref{fig:ctg_prior}. We have already seen that this prior is
practically identical to the final outcome in
Fig.~\ref{fig:full_1d}. Nevertheless, we can ask what the impact of
the boosted Higgs production process is, given that it should provide
a second measurement of the three Wilson coefficients with different
relative weights. In Fig.~\ref{fig:full_tG} we show the results of the
21-dimensional SFitter analysis with and without the new boosted Higgs
measurement introduced in Sec.~\ref{sec:new_boosted}. Unfortunately,
the likelihood distributions are similar, corresponding
to our expectation from the limited statistics of this measurements
and the limited range in $p_{T,H}$, where significant differences can
only be expected for $p_{T,H} > 250$~GeV~\cite{Buschmann:2014twa}, and
even for this kinematic range it is not clear how well the measurement
separates effects from $f_{GG}$ and $f_{tG}$, while the $f_t$
measurement is completely dominated by $t\bar{t}H$ production.

\begin{figure}[b!]
  \centering
  \includegraphics[width=.33\textwidth, page=2]{21d_vs_20d_fit_ftg_bayes}
  \includegraphics[width=.33\textwidth]{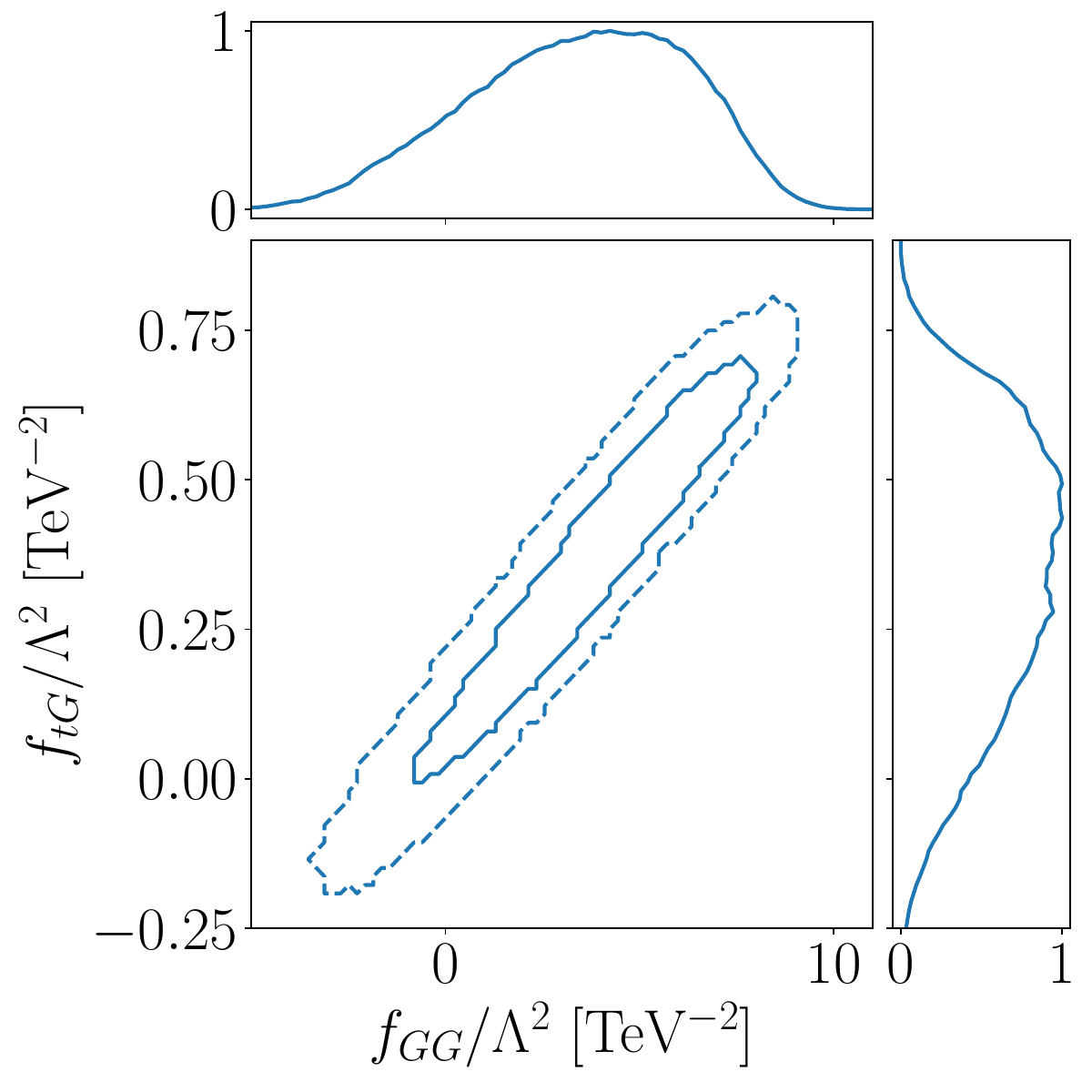}
  \caption{Left: marginalized likelihoods for the SFitter analysis
    with and without $f_{tG}$, using the same dataset; Right:
    marginalized correlation for the 21-dimensional SFitter analysis.}
  \label{fig:full_tG_corr}
\end{figure}

Even though completely justified, the only visible effect of including
$f_{tG}$ in the Higgs analysis is to wash out the limit on
$f_{GG}$. In Fig.~\ref{fig:full_tG_corr} we first show the change on
the 1-dimensional marginalized likelihood of $f_{GG}$ when we remove
$f_{tG}$ from the SFitter analysis. Indeed, the measurement of
$f_{GG}$ becomes much better. This is explained by the strong
correlation between $f_{GG}$ and $f_{tG}$ shown in the right
panel.\medskip

\begin{figure}[t]
  \includegraphics[width=.99\textwidth]{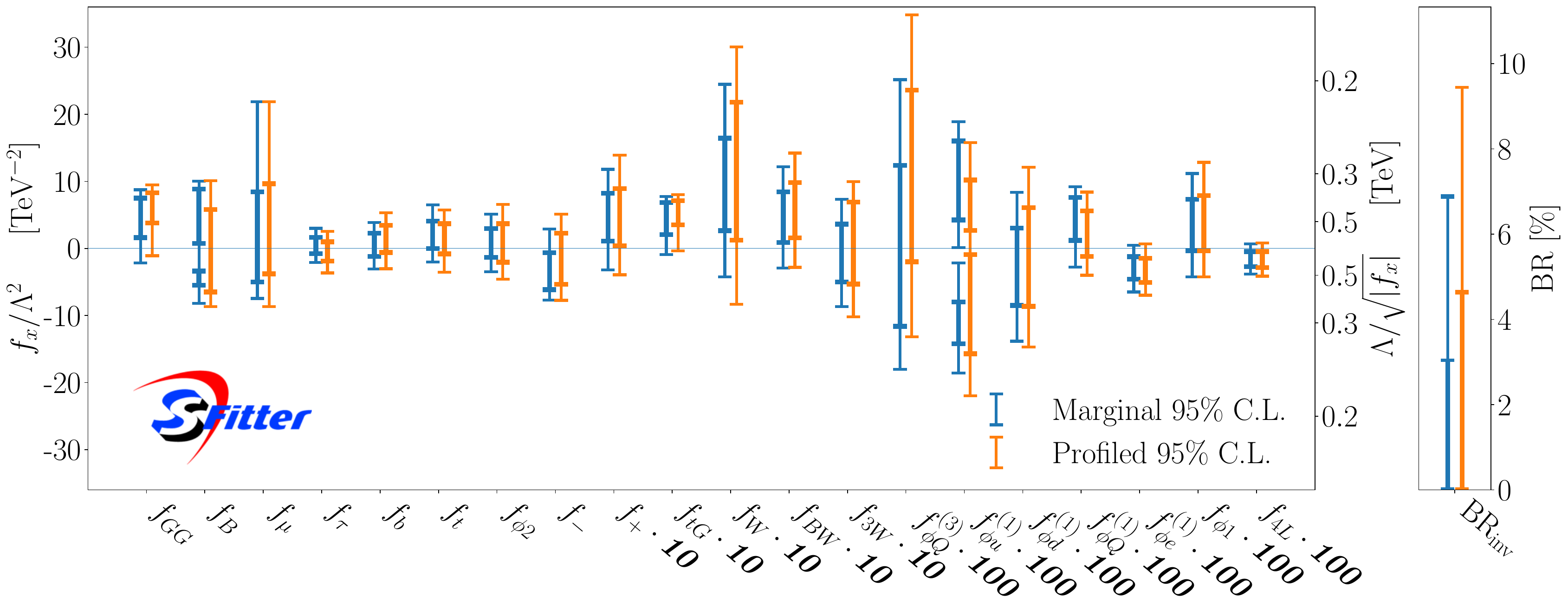}
  \caption{Comparison of 21-dimensional SFitter analysis with all
    updated measurements included. We show the 68\% and 95\%CL error
    bars from consistent marginalization and profile likelihood
    treatments of all nuisance parameters and Wilson coefficients.}
  \label{fig:bottomline}
\end{figure}

After the in-depth discussion of all features we show the 68\% and
95\%CL limits from the 21-dimensional SFitter analysis with the full
updated dataset in Fig.~\ref{fig:bottomline}. To extract these limits
we start with the respective 1-dimensional marginal or profile
likelihood, identify the maximum, and move outward keeping the
likelihood values on the left and the right border of the integral the
same. If there exists an additional peak, we compute the integral
under the likelihood for the part of the curve above a given
likelihood threshold. The 68\% and 95\%CL error bars are then defined
the same way for the marginal and profile likelihood.

The profile likelihood results in Fig.~\ref{fig:bottomline} provide an
update of the limits shown in
Fig.~\ref{fig:comparison_freq_vs_bayesian_summary}~\cite{Biekoetter:2018ypq}. We
emphasize that this update does not automatically mean an improvement
of the limits, because of our more comprehensive uncertainty
treatment, the added operator $\ope_{tG}$, and the now measured Yukawa
coupling $f_\mu$. Computing the uncertainties on the Wilson
coefficients which are all in agreement with the Standard Model at
least for the profile likelihood approach, we remove modes around
non-SM likelihood maxima. Those appear through sign flips in Yukawa
couplings and in $f_{+}$ and would require order-one effects from new
physics. We safely assume that new physics with this kind of effects
would have been observed somewhere already.

In Fig.~\ref{fig:bottomline} we see that all results from the
marginalization and profiling approach are consistent with each
other. The only kind-of-significant deviation appears in $f_B$ and the
correlated gauge-fermion operators like $f_{\phi u}^{(1)}$. The reason
for this discrepancy can be traced back to an under-fluctuation in the
$m_{WW}$ measurement and actual differences between the likelihood and
Bayesian approaches.

\section{Outlook}

Global SMEFT analyses are the first step into the direction of
interpreting all LHC data on hard scattering process in a common
framework. They allow us to combine rate and kinematic measurements
from the Higgs-gauge sector, the top sector, jet production, exotics
searches, even including parton densities and flavor physics. They can
be considered improved bin-wise analyses of LHC measurements, but
with a consistent effective theory framework. This framework allows us
to provide precision predictions matching the precision of the data we
analyze, and it ensures that their result is relevant fundamental
physics. Because any realistic effective theory description involves a
truncation in dimensionality, SMEFT results always have to be
considered in relation to the fundamental physics models they
represent.

From a brief look at the analyzed data we know that our SMEFT analysis
of the electroweak gauge and Higgs sector will not describe
established anomalies, but serve as a consistent, global limit-setting
tool. This makes it even more important to treat all uncertainties,
statistical, systematic, and theory, completely and
consistently. Technically, this leads us directly to the question if
we want to use a profile likelihood or a Bayesian marginalization
treatment. Because the two methods ask different questions, it is not
at all clear that technically correct analyses following the two
approaches lead to the same results. We have shown, for a first time,
what the current challenges in global LHC analyses are and how the two
methods do turn up slight differences.

We have started with an in-depth discussion of the current challenges
in the Higgs and electroweak data and the corresponding validation of
the marginalization in SFitter, in comparison to our classic profile
likelihoods. Using the established dataset of
Ref.~\cite{Biekoetter:2018ypq} we have shown that the two methods give
extremely similar results. We have also found that for this dataset
the exact treatment of the theory uncertainties is not a leading
problem, while a correct treatment of correlations of the measurements
and the uncertainties is crucial.

Next, we have updated this dataset, including a set of kinematic
di-boson measurements and boosted Higgs production. These measurements
allow us to constrain operators with a modified Lorentz structure
especially well. Kinematic distributions from di-boson resonance
searches probe the largest momentum transfers of our SFitter dataset,
but their interpretation in terms of SMEFT operators requires
significant effort. A systematic publication of the corresponding
likelihood by ATLAS and CMS would fundamentally change the
appreciation for these analyses, from failed resonance searches to the
most exciting SMEFT results.

Accidentally, the updated dataset also leads to differences in the
marginalization and profiling treatments of the same exclusive
likelihood. The measurement driving this difference is an
under-fluctuation in the tail of the kinematic $m_{WW}$
distribution. Under-fluctuations are difficult to reconcile with SMEFT
analyses, because they require a balance between linear and squared
operator contributions. To complicate things, a sizeable number of
kinematic distributions probes large momentum transfer, all consistent
with the Standard Model. For a small number of Wilson coefficients one
under-fluctuation will just lead to a poor log-likelihood value in the
SM-like likelihood maximum. A larger number of Wilson coefficients
defines a powerful model which accommodated this deviation. For the
final result, the complex correlations between Wilson coefficients
lead to volume effects in the marginalization, which, expectedly,
separated the final profile likelihood and marginalized results.
As a bonus in the Appendix we illustrate the potential of the High
Lumi LHC by scaling out Run 2 analysis to the corresponding luminosity.

\section*{Acknowledgments}

We would like to thank Anke Biek\"otter, Anja Butter and Tyler Corbett
for ongoing help with SFitter details. For advice on Bayesian methods
we thank Kevin Kr\"oninger; for details on the different analyses we
are grateful to Ines Ochoa, Pascal Baertschi, Alberto Zucchetta, and
Christian Sander; for the implementation of $f_{tG}$ limits we had
valuable support from Ken Mimasu and Eleni Vryonidou, and our Delphes
questions were answered by Michele Selvaggi.  Finally, we are grateful
to Wolfgang Kilian and Michael Kr\"amer for many discussions and to
Kevin Gauss for his contributions to an early phase of the project.
IB acknowledges funding from the Swiss National Science Foundation
(SNF) through the PRIMA grant no. 201508.  EG is supported by the
International Max Planck School \textsl{Precision Tests of Fundamental
  Symmetries}.  The research of all authors is supported by the Deutsche
Forschungsgemeinschaft (DFG, German Research Foundation) under grant
396021762 -- TRR~257 \textsl{Particle Physics Phenomenology after the
  Higgs Discovery}. Last, but not least, we acknowledge support by the
state of Baden-Württemberg through bwHPC and the German Research
Foundation (DFG) through grant no INST 39/963-1 FUGG (bwForCluster
NEMO).

\appendix

\section{High luminosity LHC}

\begin{figure}[b!]
  \includegraphics[width=.99\textwidth]{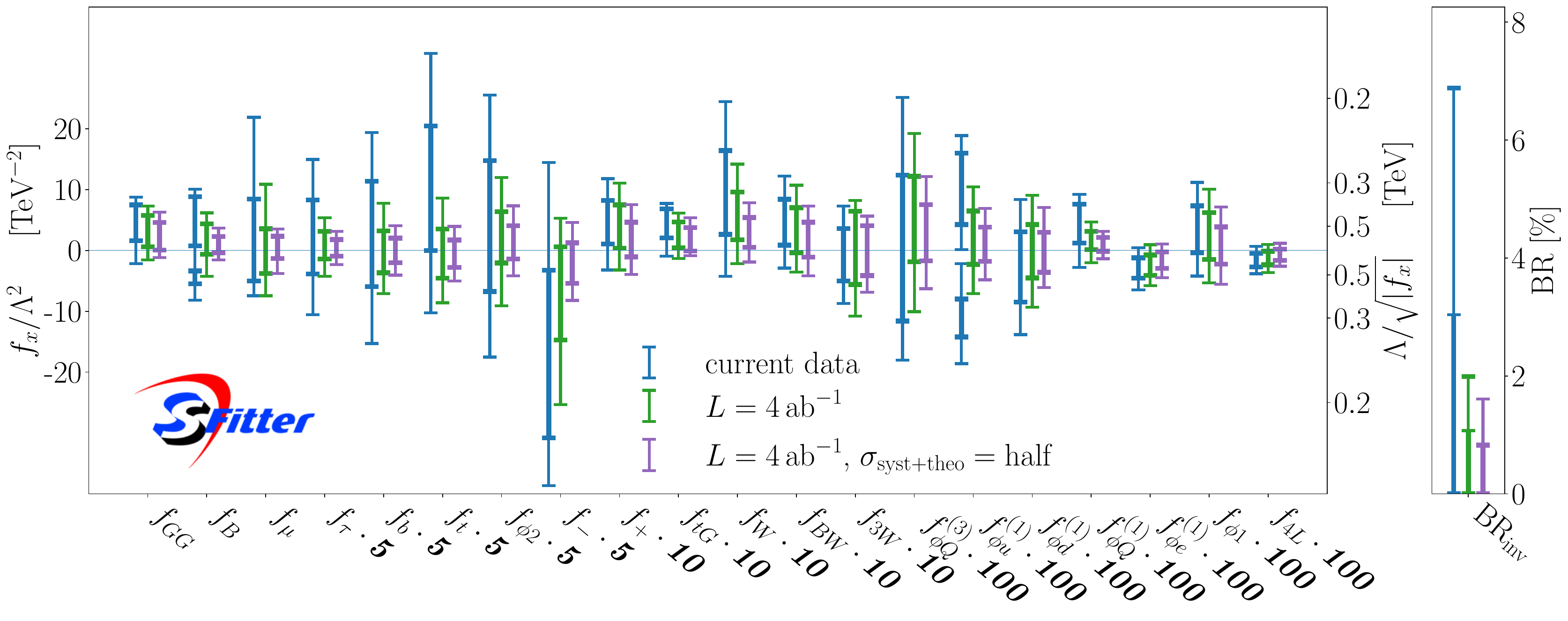}
  \caption{Green: 21-dimensional SFitter analysis with all updated measurements included, where all LHC measurements were scaled to a high-luminosity of $\unit[4]{ab^{-1}}$ and set to their background values. The electroweak precision data from LEP is kept the same as in previous fits. Purple: a second set of limits is derived setting all systematic and theoretical uncertainties to half their current values. We show the 68\% and 95\%CL error bars from a marginalization treatment of all the nuisance parameters and Wilson coefficients. Blue: current marginalized limits already presented in Fig.~\ref{fig:bottomline}.}
  \label{fig:highlumi}
\end{figure}

Figure~\ref{fig:highlumi} shows the projected limits obtained with a marginalized treatment for the high-luminosity LHC on the 21 SMEFT Wilson coefficients presented in Eq.~\ref{eq:ourlag}. We use the full set of measurements presented in Sec.~\ref{sec:fit}, where all the LHC measurements were set to their background values and their luminosity scaled to $\unit[4]{ab^{-1}}$. The electroweak precision data from LEP is kept the same as in previous fits. We also derived the projected high-luminosity limits assuming halved theory and systematic uncertainties.

A significant gain is expected on the invisible branching ratio $\text{BR}_{\text{inv}}$ as well as the Yukawa corrections $f_t$, $f_b$ and $f_\tau$. The constraints improve for most of the other Wilson coefficients, with a few exceptions. Limits on $f_{tG}$ and $f_{GG}$ do not change because the main constraint on these operators stems from the external measurement or prior described in Sec.~\ref{sec:new_top}. As this constraint is the result of a global fit to several top processes, it does not simply scale as a statistical uncertainty and we did not change it. $f_{\phi e}^{(1)}$, $f_{\phi 1}$, $f_{4L}$ and $f_{BW}$ also exhibit stable constraints, because they are mostly set by the electroweak precision data. Both the top constraints and the electroweak precision data are also the reason why the constraints are not always centred around zero. Finally, the only operator for which the constraints worsen is $f_{3W}$, mostly due to the change in the central value. Indeed, this particular operator is especially sensitive to high kinematic searches for $VV$. Thus, the pure background assumption may lead to a reduced constraining power, as there are no under-fluctuations in the higher bins of the distributions.

Once we also consider improved systematic and theoretical uncertainties, the limits on all Wilson coefficients improve---sometimes significantly, as for $f_{\mu}$ or $f_-$. In the case of $f_\mu$, this can be understood considering that the one analysis constraining this coefficient has a significance of only $3.0$ standard deviations and is dominated by large experimental uncertainties.

\section{Numerical results}
Table~\ref{tab:numerical_res} reports the numerical values of the boundaries of the 68\% and 95\% CL intervals shown in Fig.~\ref{fig:bottomline}.

\begin{table}[t!]\centering
\renewcommand{\arraystretch}{1.5}
 \begin{tabular}{|>{$}c<{$}|cc|cc|}
 \hline
 & \multicolumn{2}{c|}{Marginalized} & \multicolumn{2}{c|}{Profiled}\\
 \text{Coefficient} & 68\% CL & 95\% CL & 68\% CL & 95\% CL \\
 \hline
 
f_{GG} &  [1.61, 7.49] &  [-2.17, 8.75] &  [3.79, 8.28] & [-1.09, 9.50] \\
\multirow{2}{*}{$f_B$} &  [-5.49,  -3.38] & [-8.20, 10.05] &  [-6.49, 5.79] & [-8.69, 10.08] \\
 & [0.74, 8,84] & & & \\
f_{\phi 2} &  [-1.35, 2.95] &  [-3.51, 5.11] &  [-2.07, 3.68] & [-4.59, 6.55] \\
f_\mu &  [-5.01, 8.43] &  [-7.45, 21.88] &  [-3.79, 9.66] & [-8.68, 21.88] \\
f_t &  [-0.01, 4.09] &  [-2.05, 6.47] &  [-0.80, 3.68] & [-3.56, 5.75] \\
f_b &  [-1.19, 2.28] &  [-3.06, 3.88] &  [-0.60, 3.44] & [-3.03, 5.33] \\
f_\tau &  [-0.78, 1.66] &  [-2.11, 2.99] &  [-1.88, 1.00] & [-3.66, 2.55] \\
f_- &  [-6.16, -0.65] &  [-7.73, 2.89] &  [-5.34, 2.28] & [-7.75, 5.09] \\
f_+\times 10 &  [1.07, 8.21] &  [-3.21, 11.79] &  [0.36, 8.93] & [-3.93, 13.93] \\
f_{tG}\times 10 &  [2.05, 6.82] &  [-0.93, 7.72] &  [3.53, 7.12] & [-0.36, 8.02] \\
f_W\times 10 &  [2.64, 16.43] &  [-4.26, 24.47] &  [1.25, 21.80] & [-8.35, 30.03] \\
f_{BW}\times 10 &  [0.86, 8.42] &  [-2.91, 12.19] &  [1.57, 9.82] & [-2.83, 14.22] \\
f_{3W}\times 10 &  [-5.0, 3.62] &  [-8.7, 7.31] &  [-5.31, 6.89] & [-10.19, 9.94] \\
f_{\phi Q}^{(3)}\times 100 & [-11.61, 12.38] &  [-18.01, 25.17] &  [-1.99, 23.60] & [-13.19, 34.80] \\
\multirow{2}{*}{$f_{\phi u}^{(1)}\times 100$} &  [-14.22, -7.98]  & [-18.60, -2.17]  &  [-15.70, -0.90]  & [-22.00, 15.80] \\
 & [4.25, 16.01] & [0.14 18.87] & [2.70, 10.20] & \\
f_{\phi d}^{(1)}\times 100 &  [-8.51, 3.04] &  [-13.84, 8.37] &  [-8.64, 6.07] & [-14.69, 12.13] \\
f_{\phi 1}\times 100 &  [-0.37, 7.32] &  [-4.22, 11.17] &  [-0.37, 7.88] & [-4.22, 12.83] \\
f_{4L}\times 100 &  [-2.7, -0.46] &  [-3.82, 0.66] &  [-2.86, -0.46] & [-4.14, 0.82] \\
f_{\phi Q}^{(1)}\times 100 &  [1.2, 7.6] &  [-2.8, 9.2] &  [-1.20, 5.60] & [-4.00, 8.40] \\
f_{\phi e}^{(1)}\times 100 &  [-4.58, -1.22] &  [-6.5, 0.46] &  [-5.06, -1.46] & [-6.98, 0.67] \\
\text{BR}_{\text{inv}} &  [0, 3.04] &  [0, 6.88] & [0, 4.64] & [0, 9.44] \\

 \hline
 \end{tabular}
\caption{Numerical values for the results shown in Fig.~\ref{fig:bottomline}.}\label{tab:numerical_res}
\end{table}

Table~\ref{tab:numerical_res_high_lumi} report the numerical values of the boundaries of the 68\% and 95\% CL intervals shown in Fig.~\ref{fig:highlumi}.

\begin{table}[t!]\centering
\renewcommand{\arraystretch}{1.5}
 \begin{tabular}{|>{$}c<{$}|cc|cc|}
 \hline
 & \multicolumn{2}{c|}{$L=\unit[4]{ab^{-1}}$} & \multicolumn{2}{c|}{$L=\unit[4]{ab^{-1}}, \sigma_{\text{syst} + \text{theo}} = \text{half}$}\\
 \text{Coefficient} & 68\% CL & 95\% CL & 68\% CL & 95\% CL \\
 \hline
 
f_{GG} &  [0.42, 5.45] &  [-1.78, 7.02] & [-0.18, 4.03] & [-1.42, 6.01] \\
f_B &  [-0.41, 3.68] &  [-3.04, 6.02] & [-0.21, 2.2] & [-1.74, 3.3] \\
f_{\phi 2} \times 5 &  [-2.85, 4.64] &  [-9.09, 9.63] & [-1.17, 3.89] & [-4.2, 6.42] \\
f_\mu &  [-5.01, 5.99] &  [-9.9, 21.88] & [-1.34, 2.32] & [-3.79, 5.99] \\
f_t \times 5 &  [-4.77, 3.74] &  [-9.63, 8.61] & [-3.09, 2.17] & [-5.72, 4.8] \\
f_b \times 5 &  [-3.96, 2.57] &  [-9.19, 7.8] & [-1.91, 2.11] & [-4.32, 4.12] \\
f_\tau \times 5 &  [-1.45, 3.3] &  [-6.2, 7.1] & [-1.01, 1.88] & [-2.75, 3.03] \\
f_- \times 5 &  [-14.31, 0.84] &  [-22.39, 5.89] & [-4.88, 1.47] & [-8.06, 4.12] \\
f_+\times 10 &  [0.36, 7.5] &  [-3.93, 11.07] & [-1.07, 3.93] & [-3.93, 6.79] \\
f_{tG}\times 10 &  [-0.13, 4.13] &  [-1.46, 6.0] & [-0.14, 3.39] & [-1.18, 5.06] \\
f_W\times 10 &  [0.59, 7.28] &  [-2.75, 12.63] & [0.29, 4.55] & [-2.03, 6.48] \\
f_{BW}\times 10 &  [-0.05, 7.15] &  [-3.42, 11.0] & [-1.5, 3.88] & [-3.98, 7.19] \\
f_{3W}\times 10 &  [-4.44, 5.17] &  [-8.92, 7.09] & [-3.03, 3.29] & [-5.35, 4.62] \\
f_{\phi Q}^{(3)}\times 100 &  [-0.9, 11.55] &  [-8.37, 19.01] & [-1.44, 7.22] & [-5.78, 11.55] \\
f_{\phi u}^{(1)}\times 100 &  [-1.45, 6.44] &  [-6.3, 8.86] & [-1.63, 3.64] & [-4.5, 6.51] \\
f_{\phi d}^{(1)}\times 100 &  [-4.21, 3.46] &  [-8.4, 7.65] & [-2.89, 2.97] & [-5.6, 6.13] \\
f_{\phi 1}\times 100 &  [-1.1, 6.44] &  [-4.88, 11.3] & [-2.51, 3.29] & [-5.41, 6.67] \\
f_{4L}\times 100 &  [-2.37, 0.02] &  [-3.65, 1.14] & [-1.46, 0.21] & [-2.43, 1.04] \\
f_{\phi Q}^{(1)}\times 100 &  [-0.21, 2.65] &  [-2.11, 4.32] & [-0.32, 1.74] & [-1.45, 2.86] \\
f_{\phi e}^{(1)}\times 100 &  [-4.23, -0.69] &  [-5.89, 1.44] & [-2.68, -0.24] & [-4.1, 0.98] \\
\text{BR}_{\text{inv}} &  [0, 1.05] &  [0, 2.26] & [0, 0.85] & [0, 1.75] \\

 \hline
 \end{tabular}
\caption{Numerical values for the results shown in Fig.~\ref{fig:highlumi}.}\label{tab:numerical_res_high_lumi}
\end{table}

\clearpage

\bibliographystyle{SciPost-bibstyle-arxiv}
\bibliography{paper_final}

\begin{thebibliography}{100}
\providecommand{\url}[1]{\texttt{#1}}
\providecommand{\urlprefix}{URL }
\expandafter\ifx\csname urlstyle\endcsname\relax
  \providecommand{\doi}[1]{doi:\discretionary{}{}{}#1}\else
  \providecommand{\doi}{doi:\discretionary{}{}{}\begingroup
  \urlstyle{rm}\Url}\fi
\providecommand{\eprint}[2][]{\url{#2}}

\bibitem{Dawson:2018dcd}
S.~Dawson, C.~Englert and T.~Plehn,
\newblock \emph{{Higgs Physics: It ain't over till it's over}},
\newblock Phys. Rept. \textbf{816}, 1 (2019),
\newblock \doi{10.1016/j.physrep.2019.05.001},
\newblock \href{http://arxiv.org/abs/1808.01324}{{arXiv:1808.01324}}.

\bibitem{Brivio:2017vri}
I.~Brivio and M.~Trott,
\newblock \emph{{The Standard Model as an Effective Field Theory}},
\newblock Phys. Rept. \textbf{793}, 1 (2019),
\newblock \doi{10.1016/j.physrep.2018.11.002},
\newblock \href{http://arxiv.org/abs/1706.08945}{{arXiv:1706.08945}}.

\bibitem{Biekoetter:2018ypq}
A.~Biekoetter, T.~Corbett and T.~Plehn,
\newblock \emph{{The Gauge-Higgs Legacy of the LHC Run II}},
\newblock SciPost Phys. \textbf{6}(6), 064 (2019),
\newblock \doi{10.21468/SciPostPhys.6.6.064},
\newblock \href{http://arxiv.org/abs/1812.07587}{{arXiv:1812.07587}}.

\bibitem{daSilvaAlmeida:2018iqo}
E.~da~Silva~Almeida, A.~Alves, N.~Rosa~Agostinho, O.~J.~P. \'Eboli and M.~C.
  Gonzalez-Garcia,
\newblock \emph{{Electroweak Sector Under Scrutiny: A Combined Analysis of LHC
  and Electroweak Precision Data}},
\newblock Phys. Rev. D \textbf{99}(3), 033001 (2019),
\newblock \doi{10.1103/PhysRevD.99.033001},
\newblock \href{http://arxiv.org/abs/1812.01009}{{arXiv:1812.01009}}.

\bibitem{Kraml:2019sis}
S.~Kraml, T.~Q. Loc, D.~T. Nhung and L.~D. Ninh,
\newblock \emph{{Constraining new physics from Higgs measurements with Lilith:
  update to LHC Run 2 results}},
\newblock SciPost Phys. \textbf{7}(4), 052 (2019),
\newblock \doi{10.21468/SciPostPhys.7.4.052},
\newblock \href{http://arxiv.org/abs/1908.03952}{{arXiv:1908.03952}}.

\bibitem{vanBeek:2019evb}
S.~van Beek, E.~R. Nocera, J.~Rojo and E.~Slade,
\newblock \emph{{Constraining the SMEFT with Bayesian reweighting}},
\newblock SciPost Phys. \textbf{7}(5), 070 (2019),
\newblock \doi{10.21468/SciPostPhys.7.5.070},
\newblock \href{http://arxiv.org/abs/1906.05296}{{arXiv:1906.05296}}.

\bibitem{Dawson:2020oco}
S.~Dawson, S.~Homiller and S.~D. Lane,
\newblock \emph{{Putting standard model EFT fits to work}},
\newblock Phys. Rev. D \textbf{102}(5), 055012 (2020),
\newblock \doi{10.1103/PhysRevD.102.055012},
\newblock \href{http://arxiv.org/abs/2007.01296}{{arXiv:2007.01296}}.

\bibitem{Almeida:2021asy}
E.~d.~S. Almeida, A.~Alves, O.~J.~P. \'Eboli and M.~C. Gonzalez-Garcia,
\newblock \emph{{Electroweak legacy of the LHC run II}},
\newblock Phys. Rev. D \textbf{105}(1), 013006 (2022),
\newblock \doi{10.1103/PhysRevD.105.013006},
\newblock \href{http://arxiv.org/abs/2108.04828}{{arXiv:2108.04828}}.

\bibitem{Brown:2018gzb}
S.~Brown, A.~Buckley, C.~Englert, J.~Ferrando, P.~Galler, D.~J. Miller,
  L.~Moore, M.~Russell, C.~White and N.~Warrack,
\newblock \emph{{TopFitter: Fitting top-quark Wilson Coefficients to Run II
  data}},
\newblock PoS \textbf{ICHEP2018}, 293 (2019),
\newblock \doi{10.22323/1.340.0293},
\newblock \href{http://arxiv.org/abs/1901.03164}{{arXiv:1901.03164}}.

\bibitem{Hartland:2019bjb}
N.~P. Hartland, F.~Maltoni, E.~R. Nocera, J.~Rojo, E.~Slade, E.~Vryonidou and
  C.~Zhang,
\newblock \emph{{A Monte Carlo global analysis of the Standard Model Effective
  Field Theory: the top quark sector}},
\newblock JHEP \textbf{04}, 100 (2019),
\newblock \doi{10.1007/JHEP04(2019)100},
\newblock \href{http://arxiv.org/abs/1901.05965}{{arXiv:1901.05965}}.

\bibitem{Brivio:2019ius}
I.~Brivio, S.~Bruggisser, F.~Maltoni, R.~Moutafis, T.~Plehn, E.~Vryonidou,
  S.~Westhoff and C.~Zhang,
\newblock \emph{{O new physics, where art thou? A global search in the top
  sector}},
\newblock JHEP \textbf{02}, 131 (2020),
\newblock \doi{10.1007/JHEP02(2020)131},
\newblock \href{http://arxiv.org/abs/1910.03606}{{arXiv:1910.03606}}.

\bibitem{Ellis:2020unq}
J.~Ellis, M.~Madigan, K.~Mimasu, V.~Sanz and T.~You,
\newblock \emph{{Top, Higgs, Diboson and Electroweak Fit to the Standard Model
  Effective Field Theory}},
\newblock JHEP \textbf{04}, 279 (2021),
\newblock \doi{10.1007/JHEP04(2021)279},
\newblock \href{http://arxiv.org/abs/2012.02779}{{arXiv:2012.02779}}.

\bibitem{Ethier:2021bye}
J.~J. Ethier, G.~Magni, F.~Maltoni, L.~Mantani, E.~R. Nocera, J.~Rojo,
  E.~Slade, E.~Vryonidou and C.~Zhang,
\newblock \emph{{Combined SMEFT interpretation of Higgs, diboson, and top quark
  data from the LHC}},
\newblock JHEP \textbf{11}, 089 (2021),
\newblock \doi{10.1007/JHEP11(2021)089},
\newblock \href{http://arxiv.org/abs/2105.00006}{{arXiv:2105.00006}}.

\bibitem{Iranipour:2022iak}
S.~Iranipour and M.~Ubiali,
\newblock \emph{{A new generation of simultaneous fits to LHC data using deep
  learning}},
\newblock JHEP \textbf{05}, 032 (2022),
\newblock \doi{10.1007/JHEP05(2022)032},
\newblock \href{http://arxiv.org/abs/2201.07240}{{arXiv:2201.07240}}.

\bibitem{Carrazza_2019}
S.~Carrazza, C.~Degrande, S.~Iranipour, J.~Rojo and M.~Ubiali,
\newblock \emph{Can new physics hide inside the proton?},
\newblock Physical Review Letters \textbf{123}(13) (2019),
\newblock \doi{10.1103/physrevlett.123.132001},
\newblock \url{https://doi.org/10.1103%2Fphysrevlett.123.132001}.

\bibitem{Greljo_2021}
A.~Greljo, S.~Iranipour, Z.~Kassabov, M.~Madigan, J.~Moore, J.~Rojo, M.~Ubiali
  and C.~Voisey,
\newblock \emph{Parton distributions in the {SMEFT} from high-energy drell-yan
  tails},
\newblock Journal of High Energy Physics \textbf{2021}(7) (2021),
\newblock \doi{10.1007/jhep07(2021)122},
\newblock \url{https://doi.org/10.1007%2Fjhep07%282021%29122}.

\bibitem{Kassabov_2023}
Z.~Kassabov, M.~Madigan, L.~Mantani, J.~Moore, M.~M. Alvarado, J.~Rojo and
  M.~Ubiali,
\newblock \emph{The top quark legacy of the {LHC} run {II} for {PDF} and
  {SMEFT} analyses},
\newblock Journal of High Energy Physics \textbf{2023}(5) (2023),
\newblock \doi{10.1007/jhep05(2023)205},
\newblock \url{https://doi.org/10.1007%2Fjhep05%282023%29205}.

\bibitem{Bissmann:2019qcd}
S.~Bi\ss{}mann, J.~Erdmann, C.~Grunwald, G.~Hiller and K.~Kr\"oninger,
\newblock \emph{{Correlating uncertainties in global analyses within SMEFT
  matters}},
\newblock Phys. Rev. D \textbf{102}, 115019 (2020),
\newblock \doi{10.1103/PhysRevD.102.115019},
\newblock \href{http://arxiv.org/abs/1912.06090}{{arXiv:1912.06090}}.

\bibitem{Lafaye:2009vr}
R.~Lafaye, T.~Plehn, M.~Rauch, D.~Zerwas and M.~Duhrssen,
\newblock \emph{{Measuring the Higgs Sector}},
\newblock JHEP \textbf{08}, 009 (2009),
\newblock \doi{10.1088/1126-6708/2009/08/009},
\newblock \href{http://arxiv.org/abs/0904.3866}{{arXiv:0904.3866}}.

\bibitem{Klute:2012pu}
M.~Klute, R.~Lafaye, T.~Plehn, M.~Rauch and D.~Zerwas,
\newblock \emph{{Measuring Higgs Couplings from LHC Data}},
\newblock Phys. Rev. Lett. \textbf{109}, 101801 (2012),
\newblock \doi{10.1103/PhysRevLett.109.101801},
\newblock \href{http://arxiv.org/abs/1205.2699}{{arXiv:1205.2699}}.

\bibitem{Corbett:2015ksa}
T.~Corbett, O.~J.~P. Eboli, D.~Goncalves, J.~Gonzalez-Fraile, T.~Plehn and
  M.~Rauch,
\newblock \emph{{The Higgs Legacy of the LHC Run I}},
\newblock JHEP \textbf{08}, 156 (2015),
\newblock \doi{10.1007/JHEP08(2015)156},
\newblock \href{http://arxiv.org/abs/1505.05516}{{arXiv:1505.05516}}.

\bibitem{Butter:2016cvz}
A.~Butter, O.~J.~P. Eboli, J.~Gonzalez-Fraile, M.~C. Gonzalez-Garcia, T.~Plehn
  and M.~Rauch,
\newblock \emph{{The Gauge-Higgs Legacy of the LHC Run I}},
\newblock JHEP \textbf{07}, 152 (2016),
\newblock \doi{10.1007/JHEP07(2016)152},
\newblock \href{http://arxiv.org/abs/1604.03105}{{arXiv:1604.03105}}.

\bibitem{Lafaye:2007vs}
R.~Lafaye, T.~Plehn, M.~Rauch and D.~Zerwas,
\newblock \emph{{Measuring Supersymmetry}},
\newblock Eur. Phys. J. C \textbf{54}, 617 (2008),
\newblock \doi{10.1140/epjc/s10052-008-0548-z},
\newblock \href{http://arxiv.org/abs/0709.3985}{{arXiv:0709.3985}}.

\bibitem{Weinberg:1978kz}
S.~Weinberg,
\newblock \emph{{Phenomenological Lagrangians}},
\newblock Physica A \textbf{96}(1-2), 327 (1979),
\newblock \doi{10.1016/0378-4371(79)90223-1}.

\bibitem{Georgi:1984zwz}
H.~Georgi,
\newblock \emph{{Weak Interactions and Modern Particle Theory}},
\newblock ISBN 978-0-8053-3163-9 (1984).

\bibitem{Donoghue:1992dd}
J.~F. Donoghue, E.~Golowich and B.~R. Holstein,
\newblock \emph{{Dynamics of the standard model}}, vol.~2,
\newblock CUP,
\newblock \doi{10.1017/CBO9780511524370} (2014).

\bibitem{Leung:1984ni}
C.~N. Leung, S.~T. Love and S.~Rao,
\newblock \emph{{Low-Energy Manifestations of a New Interaction Scale: Operator
  Analysis}},
\newblock Z. Phys. C \textbf{31}, 433 (1986),
\newblock \doi{10.1007/BF01588041}.

\bibitem{Buchmuller:1985jz}
W.~Buchmuller and D.~Wyler,
\newblock \emph{{Effective Lagrangian Analysis of New Interactions and Flavor
  Conservation}},
\newblock Nucl. Phys. B \textbf{268}, 621 (1986),
\newblock \doi{10.1016/0550-3213(86)90262-2}.

\bibitem{Gonzalez-Garcia:1999ije}
M.~C. Gonzalez-Garcia,
\newblock \emph{{Anomalous Higgs couplings}},
\newblock Int. J. Mod. Phys. A \textbf{14}, 3121 (1999),
\newblock \doi{10.1142/S0217751X99001494},
\newblock \href{http://arxiv.org/abs/hep-ph/9902321}{{arXiv:hep-ph/9902321}}.

\bibitem{Grzadkowski:2010es}
B.~Grzadkowski, M.~Iskrzynski, M.~Misiak and J.~Rosiek,
\newblock \emph{{Dimension-Six Terms in the Standard Model Lagrangian}},
\newblock JHEP \textbf{10}, 085 (2010),
\newblock \doi{10.1007/JHEP10(2010)085},
\newblock \href{http://arxiv.org/abs/1008.4884}{{arXiv:1008.4884}}.

\bibitem{Passarino:2012cb}
G.~Passarino,
\newblock \emph{{NLO Inspired Effective Lagrangians for Higgs Physics}},
\newblock Nucl. Phys. B \textbf{868}, 416 (2013),
\newblock \doi{10.1016/j.nuclphysb.2012.11.018},
\newblock \href{http://arxiv.org/abs/1209.5538}{{arXiv:1209.5538}}.

\bibitem{Corbett:2012ja}
T.~Corbett, O.~J.~P. Eboli, J.~Gonzalez-Fraile and M.~C. Gonzalez-Garcia,
\newblock \emph{{Robust Determination of the Higgs Couplings: Power to the
  Data}},
\newblock Phys. Rev. D \textbf{87}, 015022 (2013),
\newblock \doi{10.1103/PhysRevD.87.015022},
\newblock \href{http://arxiv.org/abs/1211.4580}{{arXiv:1211.4580}}.

\bibitem{DiVita:2017eyz}
S.~Di~Vita, C.~Grojean, G.~Panico, M.~Riembau and T.~Vantalon,
\newblock \emph{{A global view on the Higgs self-coupling}},
\newblock JHEP \textbf{09}, 069 (2017),
\newblock \doi{10.1007/JHEP09(2017)069},
\newblock \href{http://arxiv.org/abs/1704.01953}{{arXiv:1704.01953}}.

\bibitem{Goncalves:2018qas}
D.~Gon\c{c}alves, T.~Han, F.~Kling, T.~Plehn and M.~Takeuchi,
\newblock \emph{{Higgs boson pair production at future hadron colliders: From
  kinematics to dynamics}},
\newblock Phys. Rev. D \textbf{97}(11), 113004 (2018),
\newblock \doi{10.1103/PhysRevD.97.113004},
\newblock \href{http://arxiv.org/abs/1802.04319}{{arXiv:1802.04319}}.

\bibitem{Chang:2018uwu}
J.~Chang, K.~Cheung, J.~S. Lee, C.-T. Lu and J.~Park,
\newblock \emph{{Higgs-boson-pair production H($\rightarrow b\bar
  b$)H($\rightarrow\gamma\gamma$) from gluon fusion at the HL-LHC and HL-100
  TeV hadron collider}},
\newblock Phys. Rev. D \textbf{100}(9), 096001 (2019),
\newblock \doi{10.1103/PhysRevD.100.096001},
\newblock \href{http://arxiv.org/abs/1804.07130}{{arXiv:1804.07130}}.

\bibitem{Biekotter:2018jzu}
A.~Biek\"otter, D.~Gon\c{c}alves, T.~Plehn, M.~Takeuchi and D.~Zerwas,
\newblock \emph{{The global Higgs picture at 27 TeV}},
\newblock SciPost Phys. \textbf{6}(2), 024 (2019),
\newblock \doi{10.21468/SciPostPhys.6.2.024},
\newblock \href{http://arxiv.org/abs/1811.08401}{{arXiv:1811.08401}}.

\bibitem{Borowka:2018pxx}
S.~Borowka, C.~Duhr, F.~Maltoni, D.~Pagani, A.~Shivaji and X.~Zhao,
\newblock \emph{{Probing the scalar potential via double Higgs boson production
  at hadron colliders}},
\newblock JHEP \textbf{04}, 016 (2019),
\newblock \doi{10.1007/JHEP04(2019)016},
\newblock \href{http://arxiv.org/abs/1811.12366}{{arXiv:1811.12366}}.

\bibitem{Krauss:2016ely}
F.~Krauss, S.~Kuttimalai and T.~Plehn,
\newblock \emph{{LHC multijet events as a probe for anomalous dimension-six
  gluon interactions}},
\newblock Phys. Rev. D \textbf{95}(3), 035024 (2017),
\newblock \doi{10.1103/PhysRevD.95.035024},
\newblock \href{http://arxiv.org/abs/1611.00767}{{arXiv:1611.00767}}.

\bibitem{Ellis:2018gqa}
J.~Ellis, C.~W. Murphy, V.~Sanz and T.~You,
\newblock \emph{{Updated Global SMEFT Fit to Higgs, Diboson and Electroweak
  Data}},
\newblock JHEP \textbf{06}, 146 (2018),
\newblock \doi{10.1007/JHEP06(2018)146},
\newblock \href{http://arxiv.org/abs/1803.03252}{{arXiv:1803.03252}}.

\bibitem{Zhang:2016zsp}
Z.~Zhang,
\newblock \emph{{Time to Go Beyond Triple-Gauge-Boson-Coupling Interpretation
  of $W$ Pair Production}},
\newblock Phys. Rev. Lett. \textbf{118}(1), 011803 (2017),
\newblock \doi{10.1103/PhysRevLett.118.011803},
\newblock \href{http://arxiv.org/abs/1610.01618}{{arXiv:1610.01618}}.

\bibitem{Corbett:2017qgl}
T.~Corbett, O.~J.~P. \'Eboli and M.~C. Gonzalez-Garcia,
\newblock \emph{{Unitarity Constraints on Dimension-six Operators II: Including
  Fermionic Operators}},
\newblock Phys. Rev. D \textbf{96}(3), 035006 (2017),
\newblock \doi{10.1103/PhysRevD.96.035006},
\newblock \href{http://arxiv.org/abs/1705.09294}{{arXiv:1705.09294}}.

\bibitem{Baglio:2017bfe}
J.~Baglio, S.~Dawson and I.~M. Lewis,
\newblock \emph{{An NLO QCD effective field theory analysis of $W^+W^-$
  production at the LHC including fermionic operators}},
\newblock Phys. Rev. D \textbf{96}(7), 073003 (2017),
\newblock \doi{10.1103/PhysRevD.96.073003},
\newblock \href{http://arxiv.org/abs/1708.03332}{{arXiv:1708.03332}}.

\bibitem{Baglio:2018bkm}
J.~Baglio, S.~Dawson and I.~M. Lewis,
\newblock \emph{{NLO effects in EFT fits to $W^+W^-$ production at the LHC}},
\newblock Phys. Rev. D \textbf{99}(3), 035029 (2019),
\newblock \doi{10.1103/PhysRevD.99.035029},
\newblock \href{http://arxiv.org/abs/1812.00214}{{arXiv:1812.00214}}.

\bibitem{Alves:2018nof}
A.~Alves, N.~Rosa-Agostinho, O.~J.~P. \'Eboli and M.~C. Gonzalez-Garcia,
\newblock \emph{{Effect of Fermionic Operators on the Gauge Legacy of the LHC
  Run I}},
\newblock Phys. Rev. D \textbf{98}(1), 013006 (2018),
\newblock \doi{10.1103/PhysRevD.98.013006},
\newblock \href{http://arxiv.org/abs/1805.11108}{{arXiv:1805.11108}}.

\bibitem{Dawson:2018jlg}
S.~Dawson and A.~Ismail,
\newblock \emph{{Standard model EFT corrections to Z boson decays}},
\newblock Phys. Rev. D \textbf{98}(9), 093003 (2018),
\newblock \doi{10.1103/PhysRevD.98.093003},
\newblock \href{http://arxiv.org/abs/1808.05948}{{arXiv:1808.05948}}.

\bibitem{Cirigliano:2009wk}
V.~Cirigliano, J.~Jenkins and M.~Gonzalez-Alonso,
\newblock \emph{{Semileptonic decays of light quarks beyond the Standard
  Model}},
\newblock Nucl. Phys. B \textbf{830}, 95 (2010),
\newblock \doi{10.1016/j.nuclphysb.2009.12.020},
\newblock \href{http://arxiv.org/abs/0908.1754}{{arXiv:0908.1754}}.

\bibitem{Falkowski:2017pss}
A.~Falkowski, M.~Gonz\'alez-Alonso and K.~Mimouni,
\newblock \emph{{Compilation of low-energy constraints on 4-fermion operators
  in the SMEFT}},
\newblock JHEP \textbf{08}, 123 (2017),
\newblock \doi{10.1007/JHEP08(2017)123},
\newblock \href{http://arxiv.org/abs/1706.03783}{{arXiv:1706.03783}}.

\bibitem{Alioli:2017ces}
S.~Alioli, V.~Cirigliano, W.~Dekens, J.~de~Vries and E.~Mereghetti,
\newblock \emph{{Right-handed charged currents in the era of the Large Hadron
  Collider}},
\newblock JHEP \textbf{05}, 086 (2017),
\newblock \doi{10.1007/JHEP05(2017)086},
\newblock \href{http://arxiv.org/abs/1703.04751}{{arXiv:1703.04751}}.

\bibitem{Maltoni:2016yxb}
F.~Maltoni, E.~Vryonidou and C.~Zhang,
\newblock \emph{{Higgs production in association with a top-antitop pair in the
  Standard Model Effective Field Theory at NLO in QCD}},
\newblock JHEP \textbf{10}, 123 (2016),
\newblock \doi{10.1007/JHEP10(2016)123},
\newblock \href{http://arxiv.org/abs/1607.05330}{{arXiv:1607.05330}}.

\bibitem{Grazzini:2016paz}
M.~Grazzini, A.~Ilnicka, M.~Spira and M.~Wiesemann,
\newblock \emph{{Modeling BSM effects on the Higgs transverse-momentum spectrum
  in an EFT approach}},
\newblock JHEP \textbf{03}, 115 (2017),
\newblock \doi{10.1007/JHEP03(2017)115},
\newblock \href{http://arxiv.org/abs/1612.00283}{{arXiv:1612.00283}}.

\bibitem{Deutschmann:2017qum}
N.~Deutschmann, C.~Duhr, F.~Maltoni and E.~Vryonidou,
\newblock \emph{{Gluon-fusion Higgs production in the Standard Model Effective
  Field Theory}},
\newblock JHEP \textbf{12}, 063 (2017),
\newblock \doi{10.1007/JHEP12(2017)063},
\newblock [Erratum: JHEP 02, 159 (2018)],
\newblock \href{http://arxiv.org/abs/1708.00460}{{arXiv:1708.00460}}.

\bibitem{Alonso:2013hga}
R.~Alonso, E.~E. Jenkins, A.~V. Manohar and M.~Trott,
\newblock \emph{{Renormalization Group Evolution of the Standard Model
  Dimension Six Operators III: Gauge Coupling Dependence and Phenomenology}},
\newblock JHEP \textbf{04}, 159 (2014),
\newblock \doi{10.1007/JHEP04(2014)159},
\newblock \href{http://arxiv.org/abs/1312.2014}{{arXiv:1312.2014}}.

\bibitem{Dawson:2021xei}
S.~Dawson, S.~Homiller and M.~Sullivan,
\newblock \emph{{Impact of dimension-eight SMEFT contributions: A case study}},
\newblock Phys. Rev. D \textbf{104}(11), 115013 (2021),
\newblock \doi{10.1103/PhysRevD.104.115013},
\newblock \href{http://arxiv.org/abs/2110.06929}{{arXiv:2110.06929}}.

\bibitem{Dawson:2022cmu}
S.~Dawson, D.~Fontes, S.~Homiller and M.~Sullivan,
\newblock \emph{{Beyond 6: the role of dimension-8 operators in an EFT for the
  2HDM}}  (2022),
\newblock \href{http://arxiv.org/abs/2205.01561}{{arXiv:2205.01561}}.

\bibitem{Biekotter:2016ecg}
A.~Biek{\"o}tter, J.~Brehmer and T.~Plehn,
\newblock \emph{{Extending the limits of Higgs effective theory}},
\newblock Phys. Rev. \textbf{D94}(5), 055032 (2016),
\newblock \doi{10.1103/PhysRevD.94.055032},
\newblock \href{http://arxiv.org/abs/1602.05202}{{arXiv:1602.05202}}.

\bibitem{Anisha:2020ggj}
Anisha, S.~Das~Bakshi, J.~Chakrabortty and S.~K. Patra,
\newblock \emph{{Connecting electroweak-scale observables to BSM physics
  through EFT and Bayesian statistics}},
\newblock Phys. Rev. D \textbf{103}(7), 076007 (2021),
\newblock \doi{10.1103/PhysRevD.103.076007},
\newblock \href{http://arxiv.org/abs/2010.04088}{{arXiv:2010.04088}}.

\bibitem{DasBakshi:2020pbf}
S.~Das~Bakshi, J.~Chakrabortty and M.~Spannowsky,
\newblock \emph{{Classifying Standard Model Extensions Effectively with
  Precision Observables}},
\newblock Phys. Rev. D \textbf{103}(5), 056019 (2021),
\newblock \doi{10.1103/PhysRevD.103.056019},
\newblock \href{http://arxiv.org/abs/2012.03839}{{arXiv:2012.03839}}.

\bibitem{DasBakshi:2021xbl}
S.~Das~Bakshi, J.~Chakrabortty, S.~Prakash, S.~U. Rahaman and M.~Spannowsky,
\newblock \emph{{EFT diagrammatica: UV roots of the CP-conserving SMEFT}},
\newblock JHEP \textbf{06}, 033 (2021),
\newblock \doi{10.1007/JHEP06(2021)033},
\newblock \href{http://arxiv.org/abs/2103.11593}{{arXiv:2103.11593}}.

\bibitem{Cepedello:2022pyx}
R.~Cepedello, F.~Esser, M.~Hirsch and V.~Sanz,
\newblock \emph{{Mapping the SMEFT to discoverable models}}  (2022),
\newblock \href{http://arxiv.org/abs/2207.13714}{{arXiv:2207.13714}}.

\bibitem{Brass:2018hfw}
S.~Brass, C.~Fleper, W.~Kilian, J.~Reuter and M.~Sekulla,
\newblock \emph{{Transversal Modes and Higgs Bosons in Electroweak Vector-Boson
  Scattering at the LHC}},
\newblock Eur. Phys. J. C \textbf{78}(11), 931 (2018),
\newblock \doi{10.1140/epjc/s10052-018-6398-4},
\newblock \href{http://arxiv.org/abs/1807.02512}{{arXiv:1807.02512}}.

\bibitem{Brehmer:2015rna}
J.~Brehmer, A.~Freitas, D.~Lopez-Val and T.~Plehn,
\newblock \emph{{Pushing Higgs Effective Theory to its Limits}},
\newblock Phys. Rev. D \textbf{93}(7), 075014 (2016),
\newblock \doi{10.1103/PhysRevD.93.075014},
\newblock \href{http://arxiv.org/abs/1510.03443}{{arXiv:1510.03443}}.

\bibitem{Drozd:2015rsp}
A.~Drozd, J.~Ellis, J.~Quevillon and T.~You,
\newblock \emph{{The Universal One-Loop Effective Action}},
\newblock JHEP \textbf{03}, 180 (2016),
\newblock \doi{10.1007/JHEP03(2016)180},
\newblock \href{http://arxiv.org/abs/1512.03003}{{arXiv:1512.03003}}.

\bibitem{Fuentes-Martin:2016uol}
J.~Fuentes-Martin, J.~Portoles and P.~Ruiz-Femenia,
\newblock \emph{{Integrating out heavy particles with functional methods: a
  simplified framework}},
\newblock JHEP \textbf{09}, 156 (2016),
\newblock \doi{10.1007/JHEP09(2016)156},
\newblock \href{http://arxiv.org/abs/1607.02142}{{arXiv:1607.02142}}.

\bibitem{Henning:2016lyp}
B.~Henning, X.~Lu and H.~Murayama,
\newblock \emph{{One-loop Matching and Running with Covariant Derivative
  Expansion}},
\newblock JHEP \textbf{01}, 123 (2018),
\newblock \doi{10.1007/JHEP01(2018)123},
\newblock \href{http://arxiv.org/abs/1604.01019}{{arXiv:1604.01019}}.

\bibitem{Kramer:2019fwz}
M.~Kr{\"a}mer, B.~Summ and A.~Voigt,
\newblock \emph{{Completing the scalar and fermionic Universal One-Loop
  Effective Action}},
\newblock JHEP \textbf{01}, 079 (2020),
\newblock \doi{10.1007/JHEP01(2020)079},
\newblock \href{http://arxiv.org/abs/1908.04798}{{arXiv:1908.04798}}.

\bibitem{Cohen:2020fcu}
T.~Cohen, X.~Lu and Z.~Zhang,
\newblock \emph{{Functional Prescription for EFT Matching}},
\newblock JHEP \textbf{02}, 228 (2021),
\newblock \doi{10.1007/JHEP02(2021)228},
\newblock \href{http://arxiv.org/abs/2011.02484}{{arXiv:2011.02484}}.

\bibitem{Brivio:2021alv}
I.~Brivio, S.~Bruggisser, E.~Geoffray, W.~Killian, M.~Kr\"amer, M.~Luchmann,
  T.~Plehn and B.~Summ,
\newblock \emph{{From models to SMEFT and back?}},
\newblock SciPost Phys. \textbf{12}(1), 036 (2022),
\newblock \doi{10.21468/SciPostPhys.12.1.036},
\newblock \href{http://arxiv.org/abs/2108.01094}{{arXiv:2108.01094}}.

\bibitem{Dumont:2013wma}
B.~Dumont, S.~Fichet and G.~von Gersdorff,
\newblock \emph{{A Bayesian view of the Higgs sector with higher dimensional
  operators}},
\newblock JHEP \textbf{07}, 065 (2013),
\newblock \doi{10.1007/JHEP07(2013)065},
\newblock \href{http://arxiv.org/abs/1304.3369}{{arXiv:1304.3369}}.

\bibitem{Fichet:2015xla}
S.~Fichet and G.~Moreau,
\newblock \emph{{Anatomy of the Higgs fits: a first guide to statistical
  treatments of the theoretical uncertainties}},
\newblock Nucl. Phys. B \textbf{905}, 391 (2016),
\newblock \doi{10.1016/j.nuclphysb.2016.02.019},
\newblock \href{http://arxiv.org/abs/1509.00472}{{arXiv:1509.00472}}.

\bibitem{DeBlas:2019ehy}
J.~De~Blas \emph{et~al.},
\newblock \emph{{$\texttt{HEPfit}$: a code for the combination of indirect and
  direct constraints on high energy physics models}},
\newblock Eur. Phys. J. C \textbf{80}(5), 456 (2020),
\newblock \doi{10.1140/epjc/s10052-020-7904-z},
\newblock \href{http://arxiv.org/abs/1910.14012}{{arXiv:1910.14012}}.

\bibitem{deBlas:2021wap}
J.~de~Blas, M.~Ciuchini, E.~Franco, A.~Goncalves, S.~Mishima, M.~Pierini,
  L.~Reina and L.~Silvestrini,
\newblock \emph{{Global analysis of electroweak data in the Standard Model}},
\newblock Phys. Rev. D \textbf{106}(3), 033003 (2022),
\newblock \doi{10.1103/PhysRevD.106.033003},
\newblock \href{http://arxiv.org/abs/2112.07274}{{arXiv:2112.07274}}.

\bibitem{Hocker:2001xe}
A.~Hocker, H.~Lacker, S.~Laplace and F.~Le~Diberder,
\newblock \emph{{A New approach to a global fit of the CKM matrix}},
\newblock Eur. Phys. J. C \textbf{21}, 225 (2001),
\newblock \doi{10.1007/s100520100729},
\newblock \href{http://arxiv.org/abs/hep-ph/0104062}{{arXiv:hep-ph/0104062}}.

\bibitem{Brehmer:2019gmn}
J.~Brehmer, S.~Dawson, S.~Homiller, F.~Kling and T.~Plehn,
\newblock \emph{{Benchmarking simplified template cross sections in $WH$
  production}},
\newblock JHEP \textbf{11}, 034 (2019),
\newblock \doi{10.1007/JHEP11(2019)034},
\newblock \href{http://arxiv.org/abs/1908.06980}{{arXiv:1908.06980}}.

\bibitem{Aad:2020ddw}
G.~Aad \emph{et~al.},
\newblock \emph{{Search for heavy diboson resonances in semileptonic final
  states in pp collisions at $\sqrt{s}=13$ TeV with the ATLAS detector}},
\newblock Eur. Phys. J. C \textbf{80}(12), 1165 (2020),
\newblock \doi{10.1140/epjc/s10052-020-08554-y},
\newblock \href{http://arxiv.org/abs/2004.14636}{{arXiv:2004.14636}}.

\bibitem{Brehmer:2016nyr}
J.~Brehmer, K.~Cranmer, F.~Kling and T.~Plehn,
\newblock \emph{{Better Higgs boson measurements through information
  geometry}},
\newblock Phys. Rev. D \textbf{95}(7), 073002 (2017),
\newblock \doi{10.1103/PhysRevD.95.073002},
\newblock \href{http://arxiv.org/abs/1612.05261}{{arXiv:1612.05261}}.

\bibitem{Alwall:2014hca}
J.~Alwall, R.~Frederix, S.~Frixione, V.~Hirschi, F.~Maltoni, O.~Mattelaer,
  H.~S. Shao, T.~Stelzer, P.~Torrielli and M.~Zaro,
\newblock \emph{{The automated computation of tree-level and next-to-leading
  order differential cross sections, and their matching to parton shower
  simulations}},
\newblock JHEP \textbf{07}, 079 (2014),
\newblock \doi{10.1007/JHEP07(2014)079},
\newblock \href{http://arxiv.org/abs/1405.0301}{{arXiv:1405.0301}}.

\bibitem{Sjostrand:2014zea}
T.~Sj\"ostrand, S.~Ask, J.~R. Christiansen, R.~Corke, N.~Desai, P.~Ilten,
  S.~Mrenna, S.~Prestel, C.~O. Rasmussen and P.~Z. Skands,
\newblock \emph{{An introduction to PYTHIA 8.2}},
\newblock Comput. Phys. Commun. \textbf{191}, 159 (2015),
\newblock \doi{10.1016/j.cpc.2015.01.024},
\newblock \href{http://arxiv.org/abs/1410.3012}{{arXiv:1410.3012}}.

\bibitem{Cacciari:2011ma}
M.~Cacciari, G.~P. Salam and G.~Soyez,
\newblock \emph{{FastJet User Manual}},
\newblock Eur. Phys. J. C \textbf{72}, 1896 (2012),
\newblock \doi{10.1140/epjc/s10052-012-1896-2},
\newblock \href{http://arxiv.org/abs/1111.6097}{{arXiv:1111.6097}}.

\bibitem{deFavereau:2013fsa}
J.~de~Favereau, C.~Delaere, P.~Demin, A.~Giammanco, V.~Lema\^\i{}tre,
  A.~Mertens and M.~Selvaggi,
\newblock \emph{{DELPHES 3, A modular framework for fast simulation of a
  generic collider experiment}},
\newblock JHEP \textbf{02}, 057 (2014),
\newblock \doi{10.1007/JHEP02(2014)057},
\newblock \href{http://arxiv.org/abs/1307.6346}{{arXiv:1307.6346}}.

\bibitem{Campbell:2011bn}
J.~M. Campbell, R.~K. Ellis and C.~Williams,
\newblock \emph{{Vector boson pair production at the LHC}},
\newblock JHEP \textbf{07}, 018 (2011),
\newblock \doi{10.1007/JHEP07(2011)018},
\newblock \href{http://arxiv.org/abs/1105.0020}{{arXiv:1105.0020}}.

\bibitem{ATLAS-CONF-2021-014}
{ATLAS Collaboration},
\newblock \emph{{Measurements of gluon fusion and vector-boson-fusion
  production of the Higgs boson in $H\rightarrow W W^* \rightarrow e\nu \mu\nu$
  decays using $pp$ collisions at $\sqrt{s}=13$ TeV with the ATLAS detector}}
  (2021),
\newblock \url{https://cds.cern.ch/record/2759651}.

\bibitem{CMS:2021fyk}
{CMS Collaboration},
\newblock \emph{{Search for a heavy vector resonance decaying to a
  ${\mathrm{Z}}_{\mathrm{}}^{\mathrm{}}$ ~boson and a Higgs boson in
  proton-proton collisions at $\sqrt{s} = 13\,\text {Te}\text {V} $}},
\newblock Eur. Phys. J. C \textbf{81}(8), 688 (2021),
\newblock \doi{10.1140/epjc/s10052-021-09348-6},
\newblock \href{http://arxiv.org/abs/2102.08198}{{arXiv:2102.08198}}.

\bibitem{ATLAS:2019jst}
{ATLAS Collaboration},
\newblock \emph{{Measurements and interpretations of Higgs-boson fiducial cross
  sections in the diphoton decay channel using 139 fb$^-1$ of $pp$ collision
  data at $\sqrt{s}$ = 13 TeV with the ATLAS detector}} ATLAS-CONF-2019-029
  (2019).

\bibitem{Ellis:1987xu}
R.~K. Ellis, I.~Hinchliffe, M.~Soldate and J.~J. van~der Bij,
\newblock \emph{{Higgs Decay to tau+ tau-: A Possible Signature of Intermediate
  Mass Higgs Bosons at the SSC}},
\newblock Nucl. Phys. B \textbf{297}, 221 (1988),
\newblock \doi{10.1016/0550-3213(88)90019-3}.

\bibitem{Baur:1989cm}
U.~Baur and E.~W.~N. Glover,
\newblock \emph{{Higgs Boson Production at Large Transverse Momentum in
  Hadronic Collisions}},
\newblock Nucl. Phys. B \textbf{339}, 38 (1990),
\newblock \doi{10.1016/0550-3213(90)90532-I}.

\bibitem{Banfi:2013yoa}
A.~Banfi, A.~Martin and V.~Sanz,
\newblock \emph{{Probing top-partners in Higgs+jets}},
\newblock JHEP \textbf{08}, 053 (2014),
\newblock \doi{10.1007/JHEP08(2014)053},
\newblock \href{http://arxiv.org/abs/1308.4771}{{arXiv:1308.4771}}.

\bibitem{Azatov:2013xha}
A.~Azatov and A.~Paul,
\newblock \emph{{Probing Higgs couplings with high $p_T$ Higgs production}},
\newblock JHEP \textbf{01}, 014 (2014),
\newblock \doi{10.1007/JHEP01(2014)014},
\newblock \href{http://arxiv.org/abs/1309.5273}{{arXiv:1309.5273}}.

\bibitem{Harlander:2013oja}
R.~V. Harlander and T.~Neumann,
\newblock \emph{{Probing the nature of the Higgs-gluon coupling}},
\newblock Phys. Rev. D \textbf{88}, 074015 (2013),
\newblock \doi{10.1103/PhysRevD.88.074015},
\newblock \href{http://arxiv.org/abs/1308.2225}{{arXiv:1308.2225}}.

\bibitem{Grojean:2013nya}
C.~Grojean, E.~Salvioni, M.~Schlaffer and A.~Weiler,
\newblock \emph{{Very boosted Higgs in gluon fusion}},
\newblock JHEP \textbf{05}, 022 (2014),
\newblock \doi{10.1007/JHEP05(2014)022},
\newblock \href{http://arxiv.org/abs/1312.3317}{{arXiv:1312.3317}}.

\bibitem{Buschmann:2014twa}
M.~Buschmann, C.~Englert, D.~Goncalves, T.~Plehn and M.~Spannowsky,
\newblock \emph{{Resolving the Higgs-Gluon Coupling with Jets}},
\newblock Phys. Rev. D \textbf{90}(1), 013010 (2014),
\newblock \doi{10.1103/PhysRevD.90.013010},
\newblock \href{http://arxiv.org/abs/1405.7651}{{arXiv:1405.7651}}.

\bibitem{Buschmann:2014sia}
M.~Buschmann, D.~Goncalves, S.~Kuttimalai, M.~Schonherr, F.~Krauss and
  T.~Plehn,
\newblock \emph{{Mass Effects in the Higgs-Gluon Coupling: Boosted vs Off-Shell
  Production}},
\newblock JHEP \textbf{02}, 038 (2015),
\newblock \doi{10.1007/JHEP02(2015)038},
\newblock \href{http://arxiv.org/abs/1410.5806}{{arXiv:1410.5806}}.

\bibitem{Degrande:2020evl}
C.~Degrande, G.~Durieux, F.~Maltoni, K.~Mimasu, E.~Vryonidou and C.~Zhang,
\newblock \emph{{Automated one-loop computations in the standard model
  effective field theory}},
\newblock Phys. Rev. D \textbf{103}(9), 096024 (2021),
\newblock \doi{10.1103/PhysRevD.103.096024},
\newblock \href{http://arxiv.org/abs/2008.11743}{{arXiv:2008.11743}}.

\bibitem{arxiv.2207.00348}
{ATLAS Collaboration},
\newblock \emph{Measurement of the properties of higgs boson production at
  $\sqrt{s} = 13$ tev in the $h\to\gamma\gamma$ channel using $139$ fb$^{-1}$
  of $pp$ collision data with the atlas experiment},
\newblock \doi{10.48550/ARXIV.2207.00348} (2022).

\bibitem{Sirunyan_2021}
{CMS Collaboration},
\newblock \emph{{Measurements of Higgs boson production cross sections and
  couplings in the diphoton decay channel at $\sqrt{s} = \unit[13]{TeV}$}},
\newblock Journal of High Energy Physics \textbf{2021}(7) (2021),
\newblock \doi{10.1007/jhep07(2021)027},
\newblock \url{https://doi.org/10.1007%2Fjhep07%282021%29027},
\newblock \href{http://arxiv.org/abs/2103.06956}{{arXiv:2103.06956}}.

\bibitem{Aad_2022}
{ATLAS Collaboration},
\newblock \emph{Search for associated production of a z boson with an invisibly
  decaying higgs boson or dark matter candidates at $\sqrt{s} = $ 13 tev with
  the {ATLAS} detector},
\newblock Physics Letters B \textbf{829}, 137066 (2022),
\newblock \doi{10.1016/j.physletb.2022.137066},
\newblock \url{https://doi.org/10.1016%2Fj.physletb.2022.137066}.

\bibitem{Sirunyan:2727805}
{CMS Collaboration},
\newblock \emph{{Search for dark matter produced in association with a
  leptonically decaying Z boson in proton-proton collisions at $\sqrt{s} = $ 13
  TeV}},
\newblock Eur. Phys. J. C \textbf{81}, 13. 33 p (2020),
\newblock \doi{10.1140/epjc/s10052-020-08739-5},
\newblock \url{https://cds.cern.ch/record/2727805},
\newblock \href{http://arxiv.org/abs/2008.04735}{{arXiv:2008.04735}}.

\bibitem{arxiv:2202.07953}
{ATLAS Collaboration},
\newblock \emph{Search for invisible higgs-boson decays in events with
  vector-boson fusion signatures using 139 $\text{fb}^{-1}$ of proton-proton
  data recorded by the atlas experiment},
\newblock \doi{10.48550/ARXIV.2202.07953} (2022).

\bibitem{CMS-HIG-20-003}
{CMS Collaboration},
\newblock \emph{Search for invisible decays of the higgs boson produced via
  vector boson fusion in proton-proton collisions at $\sqrt{s}$ = 13 tev},
\newblock \doi{10.48550/ARXIV.2201.11585} (2022).

\bibitem{arxiv.2201.08269}
{ATLAS Collaboration},
\newblock \emph{Measurements of higgs boson production cross-sections in the
  $h\to\tau^{+}\tau^{-}$ decay channel in $pp$ collisions at
  $\sqrt{s}=13\,\text{TeV}$ with the atlas detector},
\newblock \doi{10.48550/ARXIV.2201.08269} (2022).

\bibitem{arxiv.2204.12957}
{CMS Collaboration},
\newblock \emph{Measurements of higgs boson production in the decay channel
  with a pair of $\tau$ leptons in proton-proton collisions at $\sqrt{s}$ = 13
  tev},
\newblock \doi{10.48550/ARXIV.2204.12957} (2022).

\bibitem{arxiv.2206.09466}
{CMS Collaboration},
\newblock \emph{Measurements of the higgs boson production cross section and
  couplings in the w boson pair decay channel in proton-proton collisions at
  $\sqrt{s}$ = 13 tev},
\newblock \doi{10.48550/ARXIV.2206.09466} (2022).

\bibitem{Aad:2723187}
{ATLAS Collaboration},
\newblock \emph{{Measurements of $WH$ and $ZH$ production in the $H \rightarrow
  b\bar{b}$ decay channel in $pp$ collisions at 13 TeV with the ATLAS
  detector}},
\newblock Eur. Phys. J. C \textbf{81}, 178. 41 p (2020),
\newblock \doi{10.1140/epjc/s10052-020-08677-2},
\newblock \url{https://cds.cern.ch/record/2723187},
\newblock \href{http://arxiv.org/abs/2007.02873}{{arXiv:2007.02873}}.

\bibitem{Tumasyan_2021}
{CMS Collaboration},
\newblock \emph{{Evidence for Higgs boson decay to a pair of muons}},
\newblock Journal of High Energy Physics \textbf{2021}(1) (2021),
\newblock \doi{10.1007/jhep01(2021)148},
\newblock \url{https://doi.org/10.1007%2Fjhep01%282021%29148}.

\bibitem{HXSWGReport}
B.~Mellado~Garcia, P.~Musella, M.~Grazzini and R.~Harlander,
\newblock \emph{{CERN Report 4: Part I Standard Model Predictions}}  (2016),
\newblock \url{https://cds.cern.ch/record/2150771}.

\bibitem{HXSWGWebXS}
{LHC Higgs Working Group},
\newblock \emph{{SM Higgs production cross sections at $\sqrt{s} =
  \unit[13]{TeV}$ (update in CERN Report 4)}}  (2016),
\newblock
  \url{https://twiki.cern.ch/twiki/bin/view/LHCPhysics/CERNYellowReportPageAt13TeV}.

\bibitem{HXSWGWebBR}
{LHC Higgs Working Group},
\newblock \emph{{SM Higgs Branching Ratios and Total Decay Widths (update in
  CERN Report 4)}}  (2016),
\newblock
  \url{https://twiki.cern.ch/twiki/bin/view/LHCPhysics/CERNYellowReportPageBR}.

\bibitem{Aad:2020tps}
{ATLAS Collaboration},
\newblock \emph{{Search for resonances decaying into a weak vector boson and a
  Higgs boson in the fully hadronic final state produced in proton$-$proton
  collisions at $\sqrt{s} = 13$ TeV with the ATLAS detector}},
\newblock Phys. Rev. D \textbf{102}(11), 112008 (2020),
\newblock \doi{10.1103/PhysRevD.102.112008},
\newblock \href{http://arxiv.org/abs/2007.05293}{{arXiv:2007.05293}}.

\end{thebibliography}
\end{document}